\documentclass[12pt]{iopart}

\usepackage{iopams} 
\usepackage{graphicx} 
\begin{document}

\title[]{Multilayer coating for higher accelerating fields in superconducting radio-frequency cavities: a review of theoretical aspects}

\author{Takayuki Kubo}

\address{KEK, High Energy Accelerator Research Organization, Tsukuba, Ibaraki, Japan\\
SOKENDAI (the Graduate University for Advanced Studies), Hayama, Kanagawa, Japan}
\ead{kubotaka@post.kek.jp}
\vspace{10pt}
\begin{indented}
\item[]%February 2014??
\end{indented}

%%%%%%%%%%%%%%
%%%%%%%%%%%%%%
\begin{abstract}
%%%%%%%%%%%%%%
%%%%%%%%%%%%%%

Theory of the superconductor-insulator-superconductor (S-I-S) multilayer structure in superconducting accelerating cavity application is reviewed. 
The theoretical field limit, optimum layer thicknesses and material combination, and surface resistance are discussed. 
Those for the S-S bilayer structure are also reviewed.

\end{abstract}

%%%%%%%%%%%%%%%%%%
%%%%%%%%%%%%%%%%%%
\section{Introduction}
%%%%%%%%%%%%%%%%%%
%%%%%%%%%%%%%%%%%%

Science and technology of the superconducting radio-frequency (SRF) cavity made of niobium (Nb) have been studied strenuously over the last decades~\cite{hasan}. 
Improvements in fabrication and processing technologies combined with progresses in understanding of SRF physics~\cite{gurevich_review} have pushed up the frontier of the accelerating field. 
In the present day, the peak surface magnetic field around $B_{0} \simeq 150\,{\rm mT}$ has been commonly achieved by using the set of modern surface-preparation techniques: 
electropolishing followed by a heat treatment for hydrogen degassing~\cite{saitoEP, furuyaEP}, high-pressure rinsing~\cite{bernerdHPR, saitoHPR, kneiselHPR}, clean assembly~\cite{kojimaCA}, low temperature baking~\cite{kako_bake, kneisel_40, ono_bake, lilje_bake}, and local grind combined with optical inspection technique~\cite{iwashita, champion, yamamoto2010, ge, yamamoto2013, kubo_PTEP_pit}. 
Some laboratories have achieved $B_{0} \simeq 200\,{\rm mT} \sim B_{c1}^{\rm (Nb)} \sim B_{c}^{\rm (Nb)}$~\cite{geng, watanabe}, 
where $B_{c1}^{\rm (Nb)}$ and $B_c^{\rm (Nb)}$ are the lower critical field and the thermodynamic critical field, respectively.  
Further high fields, however, would not be expected 
because the present record field is thought to be close to the theoretical field limit, namely, the superheating field $B_s^{\rm (Nb)}(\sim B_c^{\rm (Nb)})$.

The superheating field $B_s$ is the field at which the Meissner state becomes absolutely unstable. 
When $B_0<B_{c1}$, 
the Meissner state of the type II superconductor corresponds to the global minimum of the free energy. 
For $B_0>B_{c1}$, 
the vortex state, instead of the Meissner state, becomes the global minimum. 
However, transition from the Meissner state to the vortex state does not necessarily take place, 
because these two states are connected with a finite change of the order parameter, 
and all the intermediate states have higher free energies than the Meissner states, 
which act as the energy barrier preventing the transition~\cite{galaiko, kramer}. 
The Meissner state may continue even at $B_0 > B_{c1}$ as a metastable state. 
At $B_0=B_s\, (> B_{c1})$, 
the free energy of all possible intermediates states achieved by perturbations to the Meissner state becomes smaller than that of the Meissner state: 
the Meissner state is unstable with respect to any small perturbation. 
Bean and Livingston~\cite{bean} examined a specific and crucial intermediate state within the London theory: 
a vortex near the surface. 
They showed there exists the energy barrier for penetration of vortex that originates in the attraction force between the surface and a single vortex (the Bean-Livingston barrier), 
and obtained the rough estimate of $B_s$ by finding the field at which the Bean-Livingston barrier disappears, 
which we call the vortex penetration field to distinguish the rough estimate from the true value of $B_s$. 
Rigorous calculations of $B_s$ have also been carried out so far within the Ginzburg-Landau (GL) theory~\cite{kramer, christiansen, chapman, transtrum} and the quasiclassical theory~\cite{catelani, lin}, 
which are valid at the vicinity of the critical temperature $T_c$ and at an arbitrary temperature $0<T<T_c$, respectively.

\begin{figure}[tb]
   \begin{center}
   \includegraphics[width=0.7\linewidth]{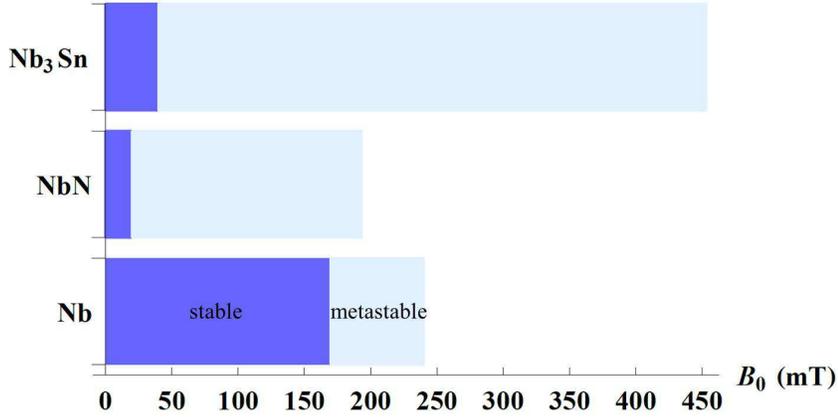}
   \end{center}\vspace{0cm}
   \caption{
The applied magnetic field $B_0$ and the stability of the Meissner state at $T\simeq 0$. 
The deep blue regions correspond to $B_0 < B_{c1}$ and represent the stable Meissner state.  
The light blue regions correspond to $B_{c1} < B_0 < B_s$, 
where the Meissner state is not stable but metastable and can transition to the more stable vortex state. 
Here we assumed the following material parameters~\cite{gurevich_review}: 
$B_{c1}^{\rm (Nb)}=170\,{\rm mT}$, $B_c^{\rm (Nb)}=200\,{\rm mT}$, and $B_s^{\rm (Nb)}=1.2 B_c^{\rm (Nb)}=240\,{\rm mT}$ for Nb; 
$B_{c1}^{\rm (NbN)}=20\,{\rm mT}$, $B_c^{\rm (NbN)}=230\,{\rm mT}$, and $B_s^{\rm (NbN)}=0.84 B_c^{\rm (NbN)}=190\,{\rm mT}$ for NbN; 
$B_{c1}^{\rm (Nb_3 Sn)}=40\,{\rm mT}$, $B_c^{\rm (Nb_3 Sn)}=540\,{\rm mT}$, and $B_s^{\rm (Nb_3 Sn)}=0.84 B_c^{\rm (Nb_3 Sn)}=450\,{\rm mT}$ for ${\rm Nb_3 Sn}$.
   }\label{fig1}
\end{figure}

Above $B_s$, only the highly dissipative vortex state, 
which yields much stronger dissipation than an acceptable level in SRF applications,
can exist. 
The superheating field $B_s$ at GHz frequencies defines the theoretical field limit of the SRF cavity. 
Then we may consider use of an alternative material that has a higher $B_s\, (\sim B_c)$ may push up the ultimate limit (see the light blue regions of Fig.~\ref{fig1}). 
Such a material, however, tends to have a small lower critical field $B_{c1}$ (see the deep blue regions of Fig.~\ref{fig1}), 
above which the Meissner state ceases to be stable and can transition to the vortex state. 
The energy barrier may protect the material against penetration of vortices as mentioned above, 
but it would not provides adequate protection: 
the actual cavity surface involves a tremendous number of materials and topographic defects which reduce the energy barrier, 
causing local penetration of vortices at $B_0 \sim B_{c1}$. 
In particular, at a temperature as low as that for SRF operations, 
vortices that locally penetrate at such a weak spot would develop into the thermomagnetic flux avalanche and cause a quench~\cite{aranson2001, aranson2005, duran_Nb, rudnev_NbN, rudnev_Nb3Sn, johansen_MgB2}. 
Eventually, use of an alternative material (simply in a homogeneous bulk form) is expected to restrict an achievable field to a region not much far from the deep blue region of Fig.~\ref{fig1}.

\begin{figure}[tb]
   \begin{center}
   \includegraphics[width=0.5\linewidth]{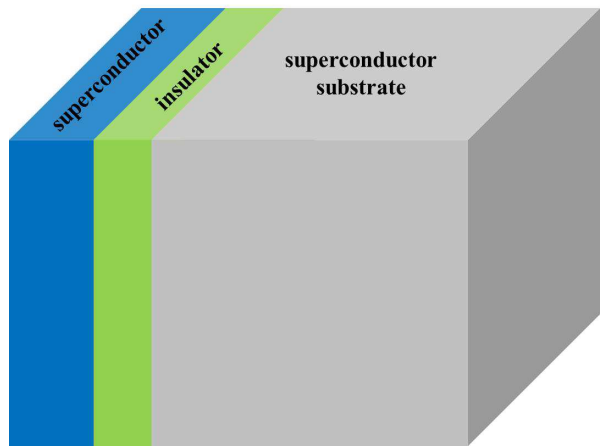}
   \end{center}\vspace{0cm}
   \caption{
The simplest multilayer superconductor: S-I-S structure. 
The blue, green, and gray regions correspond to a superconductor (${\mathcal S}$) layer, 
an insulator (${\mathcal I}$) layer, and a superconductor substrate, respectively 
   }\label{fig2}
\end{figure}

The multilayer approach~\cite{gurevich_APL} was proposed to address this problem and to push up the achievable field from the deep blue regions to the light blue regions in Fig.~\ref{fig1}. 
Its main idea is to arrest thermomagnetic avalanches caused by a local penetration of vortices at defects and not to allow them to develop into avalanches. 
Fig.~\ref{fig2} shows the simplest one (i. e., S-I-S structure). 
The bulk Nb substrate is coated with an insulator (${\mathcal I}$) layer and a superconductor (${\mathcal S}$) layer. 
The ${\mathcal I}$ layer is the essential gimmick, 
which intercepts the propagating vortex and localize the dissipation in the ${\mathcal S}$ layer. 
The ${\mathcal S}$ layer must be as thin as the penetration depth $\lambda$; 
otherwise the ${\mathcal S}$ layer may be regarded as just a bulk material and then lead to a thermal quench in the same manner as mentioned in the last paragraph. 
On the other hand, the ${\mathcal S}$ layer partly screens the surface magnetic field down to a level that the bulk Nb can withstand (i. e., $\sim B_{c1}^{\rm (Nb)} \sim B_{c}^{\rm (Nb)}$). 
Thus it should be thick enough to protect the Nb substrate. 
Now a question arise: 
how can we fix the thickness of the layers and a combination of materials?  
The recent main progress in the study of the multilayer coating is the finding of an answer to this question~\cite{kubo_APL, gurevich_AIP, posen_PRAppl, kubo_SRF2015}. 
The main topics of this article is to review how this question is solved.

\begin{figure}[tb]
   \begin{center}
   \includegraphics[width=0.5\linewidth]{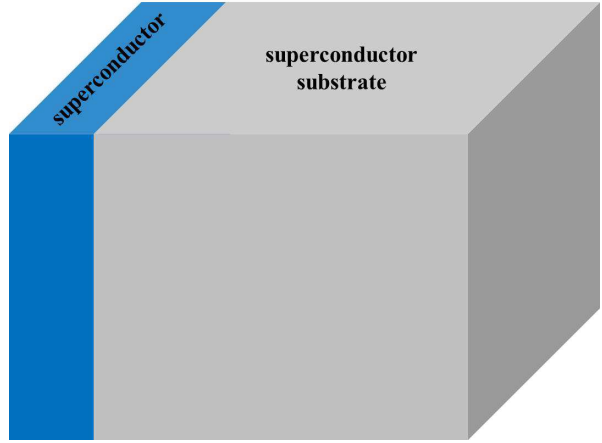}
   \end{center}\vspace{0cm}
   \caption{
The multilayer superconductor without insulator layers. 
When the superconductor layer (blue region) and the superconductor substrate (gray region) are a dirty Nb and a clean Nb, respectively, 
this can be regarded as a model of the Nb surface after the low temperature baking. 
   }\label{fig3}
\end{figure}

As mentioned above, the ${\mathcal I}$ layer is the essential constituent in the multilayer approach. 
However, the multilayer structure without insulator layers as shown in Fig.~\ref{fig3} is also worth studying because of the following two reasons. 
First, it can be regarded as a model of the surface of a superconductor that consists of superconductors with different penetration depths. 
As briefly mentioned in the discussion section of Ref.~\cite{kubo_PTEP_nano}, 
the Nb surface after the low temperature baking~\cite{kako_bake, kneisel_40, ono_bake, lilje_bake}, 
which has a depth dependent mean-free path~\cite{ciovati_bake, romanenko_bake} and then a depth dependent penetration depth, 
can be described by an S-S bilayer~\cite{kubo_LINAC14} with a thin dirty Nb and a clean Nb substrate as the simplest model. 
The same would be true for the modified baking~\cite{grassellino_bake}. 
Note here the present approach cannot incorporate the impurity-concentration dependence of the density of state in the current-carrying state~\cite{gurevich_review, lin}. 
The recent work on the Nb surface after the low temperature baking~\cite{checchin} is also the similar approach as the above. 
Second, some researchers have made S-S bilayer structures such as ${\rm Mg B_2}$-Nb and ${\rm Nb_3 Sn}$-Nb, and have carried out sample testing~\cite{tan_SRF2015, laxdal_TTC2016}, 
which should also be understood theoretically. 
In the last part of the present article, 
some features of the S-S bilayer structure are reviewed, 
which have already been known through the studies of the S-I-S structure so far~\cite{kubo_APL, gurevich_AIP, posen_PRAppl, kubo_SRF2015}.

The main purpose of this article is to summarize important formulae necessary for planning proof-of-concept experiments of the multilayer approach and to introduce some formulae for the S-S bilayer structure obtained as bi-products of studies on the S-I-S structure. 
The article is organized as follows. 
In Section~\ref{section:Bs}, the vortex penetration field and the superheating field are briefly reviewed, 
which are a necessary input parameter for calculating the field limit of the multilayer superconductor. 
In Sec.~\ref{section:multilayer}, 
we review how to optimize thicknesses of layers and a combination of materials of the S-I-S structure. 
First, the S-I-S structure with the ideal surface and negligibly thin ${\mathcal I}$ layer is studied. 
The results are expressed by using the vortex penetration field from the London theory and the superheating fields of the GL and quasiclassical theories step by step. 
The last one is valid at an arbitrary temperature ($0<T<T_c$). 
Then the theory that contains effects of a finite ${\mathcal I}$ layer thickness is also investigated. 
Finally, effects of surface defects are taken into account. 
The surface resistance of the S-I-S structure is also evaluated. 
In Sec.~\ref{section:multilayer_without_I}, 
some known results of the S-S bilayer structure are reviewed, 
where the similar techniques as those used in Sec.~\ref{section:multilayer} are used. 
First the optimization procedure of the layer thickness and material combination to maximize the theoretical field limit is reviewed. 
Then a barrier structure in the surface layer is examined: 
we see the S-S boundary has a role of barrier to prevent penetration of vortices. 
The surface resistance of the S-S bilayer structure is also derived in much the same way as the S-I-S structure. 
All the calculations are explained in detail for readers who want to follow derivation processes of the formulae.

%%%%%%%%%%%%%%%%%%%%%%%%%%%%%%%%%%%%
%%%%%%%%%%%%%%%%%%%%%%%%%%%%%%%%%%%%
%%%%%%%%%%%%%%%%%%%%%%%%%%%%%%%%%%%%
%%%%%%%%%%%%%%%%%%%%%%%%%%%%%%%%%%%%
%%%%%%%%%%%%%%%%%%%%%%%%%%%%%%%%%%%%
%%%%%%%%%%%%%%%%%%%%%%%%%%%%%%%%%%%%
%%%%%%%%%%%%%%%%%%%%%%%%%%%%%%%%%%%%
%%%%%%%%%%%%%%%%%%%%%%%%%%%%%%%%%%%%
%%%%%%%%%%%%%%%%%%%%%%%%%%%%%%%%%%%%
\section{Brief review of the superheating field} \label{section:Bs}
%%%%%%%%%%%%%%%%%%%%%%%%%%%%%%%%%%%%
%%%%%%%%%%%%%%%%%%%%%%%%%%%%%%%%%%%%
%%%%%%%%%%%%%%%%%%%%%%%%%%%%%%%%%%%%
%%%%%%%%%%%%%%%%%%%%%%%%%%%%%%%%%%%%
%%%%%%%%%%%%%%%%%%%%%%%%%%%%%%%%%%%%
%%%%%%%%%%%%%%%%%%%%%%%%%%%%%%%%%%%%
%%%%%%%%%%%%%%%%%%%%%%%%%%%%%%%%%%%%
%%%%%%%%%%%%%%%%%%%%%%%%%%%%%%%%%%%%
%%%%%%%%%%%%%%%%%%%%%%%%%%%%%%%%%%%%

\begin{figure}[tb]
   \begin{center}
   \includegraphics[width=0.7\linewidth]{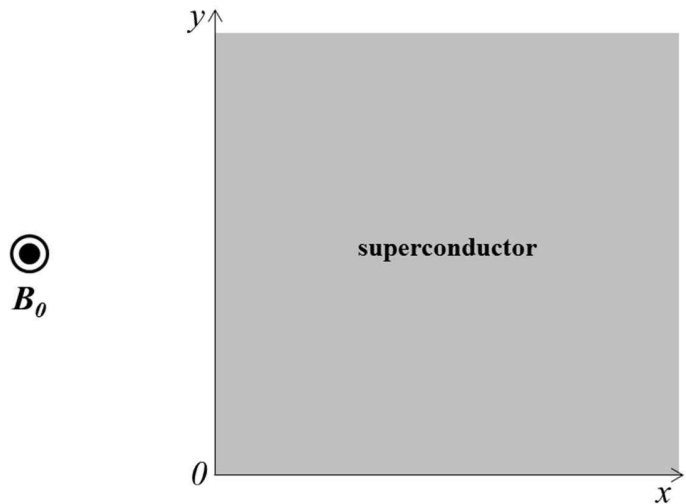}
   \end{center}\vspace{0cm}
   \caption{
Model of a semi-infinite superconductor. 
The surface is parallel to the $y$-$z$ plane and then perpendicular to the $x$-axis. 
The applied magnetic field is given by ${\bf B}_0=(0,0,B_0)$. 
   }\label{fig4}
\end{figure}

Let us begin with a brief review of the basics of the superheating field. 
We treat a semi-infinite superconductor shown in Fig.~\ref{fig4} through out this section. 
The surfaces of materials are assumed to be flat and parallel to the $y$-$z$ plane. 
The applied magnetic field is parallel to the $z$-axis and is given by ${\bf B}_0=(0,\,0,\,B_0)$. 

%

%%%%%%%%%%%%%%%%
\subsection{Vortex penetration field from the London theory}
%%%%%%%%%%%%%%%%

As mentioned in the last section, the transition from the Meissner state to the vortex state is prevented by the existence of intermediate states with higher free energies than the Meissner state even when $B_0 >B_{c1}$. 
The purpose of this subsection is to estimate the superheating field by examining a specific intermediate states in the framework of the London theory as ita was done by Bean and Livingston~\cite{bean}. 
We use the term ``vortex penetration field" instead of the superheating field in order to distinguish the rough estimate of the superheating field from the true one.

We assume the superconductor is made of an extreme type II material with a penetration depth $\lambda$ and a coherence length $\xi$ ($\xi \ll \lambda$). 
Let us put a vortex with the flux quantum $\phi_0=2.07\times10^{-15}{\rm Wb}$ parallel to ${\bf \hat{z}}$ at ${\bf r}_0=(x_0,0)$. 
Then the vortex feels two distinct forces ${\bf f}_{\rm M}(x_0)$ and ${\bf f}_{\rm B}(x_0)$. 
The former is the force from a Meissner screening current ${\bf J} =  J  {\bf \hat{y}}$ and is given by 
${\bf f}_{\rm M}(x_0)={\bf J}(x_0) \times \phi_0 {\bf \hat{z}} = \phi_0 J(x_0) {\bf \hat{x}}$. 
When the vortex is at the surface ($x_0=\xi$), we have
\begin{eqnarray}
{\bf f}_{\rm M}(x_0) \Bigr|_{x_0=\xi} \simeq \phi_0 J(0) {\bf \hat{x}} ,
\end{eqnarray}
which pushes the vortex into the inside. 
The latter, ${\bf f}_{\rm B}$, is a force due to an interaction between the vortex and the boundary. 
The simplest way to calculate ${\bf f}_{\rm B}$ is use of the method of images: 
remove the boundary, regard all the space as the superconductor, put an image vortex to satisfy the boundary condition, 
and evaluate the force due to the image. 
In this problem, the appropriate image is an antivortex at $(x,y)=(-x_0,0)$, 
by which the boundary condition of zero current normal to the surface is satisfied.  
Then ${\bf f}_{\rm B}(x_0)$ is given by ${\bf f}_{\rm B}(x_0)={\bf j}_{\rm img}(x_0) \times \phi_0 {\bf \hat{z}} = - (\phi_0^2/4\pi\mu_0 \lambda^2 x_0) {\bf \hat{x}}$, 
where ${\bf j}_{\rm img}(x_0) = -\phi_0/(2\pi\mu_0 \lambda^2 \cdot 2x_0){\bf \hat{y}}$ is the current circulating the image antivortex for $x_0 < \lambda$ (see \ref{appendix_infinite}). 
When the vortex is at the surface ($x_0=\xi$), we have
\begin{eqnarray}
{\bf f}_{\rm B}(x_0)\Bigr|_{x_0=\xi}
= - \frac{\phi_0^2}{4\pi\mu_0 \lambda^2 \xi} {\bf \hat{x}} ,
\label{eq:fB_semi_infinite}
\end{eqnarray}
which attracts the vortex to the surface. 
Instead of the method of images, 
${\bf f}_{\rm B}$ can be evaluated by a brute-force approach: 
solve the London equation $-\lambda^2 \nabla^2 {\bf B} + {\bf B} = \phi_0 \delta^{(2)}({\bf r}-{\bf r}_0)$ at the domain $x \ge 0$ with the boundary condition given by the zero current normal to the surface, 
evaluate the energy of the vortex interacting with the boundary, 
and differentiate the energy over the position of the vortex (see \ref{appendix_semi_infinite}).

When the screening current density $J(0)$ is so small that $|{\bf f}_{\rm M}| < |{\bf f}_{\rm B}|$, 
the total force directs the negative direction of the $x$-axis, 
which acts as a barrier that prevents penetration of vortices. 
This barrier is called the Bean-Livingston surface barrier~\cite{bean}. 
When $J(0)$ is large enough and $|{\bf f}_{\rm M}(\xi)| > |{\bf f}_{\rm B}(\xi)|$, 
the barrier disappears and the vortex is drawn into the material. 
Then the maximum current that the material can withstand against vortex penetration is derived from the condition $|{\bf f}_{\rm M}(\xi)|=|{\bf f}_{\rm B}(\xi)|$ and is given by
\begin{eqnarray}
J_{\rm max,L} = \frac{\phi_0}{4\pi \mu_0 \lambda^2 \xi}. 
\label{eq:London_superj}
\end{eqnarray}
where the subscript L represents the London theory. 
By using the London equation $J(0) = -A(0)/\mu_0 \lambda^2$, 
Eq.~(\ref{eq:London_superj}) can be expressed as $A_{\rm max,L} = |-\mu_0 \lambda^2 J_{\rm max,L}|$ or
\begin{eqnarray}
A_{\rm max,L} =\frac{\phi_0}{4\pi \xi} . 
\label{eq:London_superA}
\end{eqnarray}
The applied field corresponding to Eq.~(\ref{eq:London_superj}) or (\ref{eq:London_superA}) is the vortex penetration field, $B_{\rm v}$.  
In order to obtain $B_{\rm v}$, 
we need to know the relation between $B_0$ and $J$ (or $A$).  
Then the next task is to solve the London equation, 
\begin{eqnarray}
A'' -\frac{1}{\lambda^2}A = 0,  \label{eq:London_bulk}
\end{eqnarray}
where the prime denotes the derivative over $x$.  
The solution of Eq.~(\ref{eq:London_bulk}) under the boundary condition $B_0 = A'(0)$ is given by $A(x)=-\lambda B_0 e^{-x/\lambda}$ or $J(x)=(B_0/\mu_0 \lambda) e^{-x/\lambda}$. 
Since $B_0 = \mu_0 \lambda J(0)$,  
$B_{\rm v}$ is given by $B_{\rm v} = \mu_0 \lambda J_{\rm max,L}$ or~\cite{bean} 
\begin{eqnarray}
B_{\rm v} 
= \frac{\phi_0}{4\pi \lambda \xi} 
= \frac{1}{\sqrt{2}} B_c 
\simeq 0.71 B_c. 
\label{eq:BvBulk}  
\end{eqnarray}
It should be noted that the balance of forces at the surface means the flatness of the free energy at the surface: 
the disappearance of the energy barrier. 
The force approach is equivalent to the free energy approach~\cite{bean} in the evaluation of the vortex penetration field in the London theory~\cite{kubo_SRF2013, kubo_IPAC14}.

Clearly, the definition of the vortex penetration field is unsatisfactory. 
The London theory ignores the pair-breaking effect due to the current density, 
and the vortex core is replaced by the normal conducting filament with radius $\sim \xi$. 
In the above, we put a vortex at $x_0=\xi$ by hand and examine how large field is necessary to make it penetrate into the inside, 
where we necessarily introduce an ambiguity resulting from the short distance cutoff $\sim \xi$. 
The vortex penetration field only gives the order of magnitude of the true superheating field.  
For a rigorous discussions, 
at least the GL theory is necessary.

%%%%%%%%%%%%%%%%
\subsection{Superheating field at $T\simeq T_c$}
%%%%%%%%%%%%%%%%

Let us examine the superheating field within the GL theory, 
which is valid only at $T\simeq T_c$~\cite{kramer, christiansen, chapman, transtrum}. 
We use the same unit as Ref.~\cite{kramer}: 
$\widetilde{\nabla} \equiv \lambda \nabla$, 
$\widetilde{\bf A} \equiv {\bf A}/\sqrt{2} B_c \lambda$, 
$\widetilde{\bf B}=\widetilde{\nabla}\times \widetilde{\bf A} = {\bf B}/(\sqrt{2}B_c)$. 
In the follwing, we omit all the tildes for brevity. 
Then the free energy of a semi-infinite superconductor is given by
\begin{eqnarray}
\Omega = 
\int\! d^3 r \biggl[ \frac{1}{\kappa^2} (\nabla f)^2 + \frac{1}{2} (1-f^2)^2 + f^2{\bf A}^2 + ({\bf B}_0 - \nabla \times {\bf A})^2 \biggr], 
\label{eq:GL_free_energy}
\end{eqnarray}
where $\kappa=\lambda/\xi$ is the GL parameter, 
$f$ represents the real and dimensionless order parameter, 
and ${\bf B}_0$ is the applied magnetic field. 
In the absence of vortices, it is possible to choose the gauge in which $f$ is real, 
and the superfluid velocity is simply proportional to ${\bf A}$. 
The GL equations are given by
\begin{eqnarray}
\frac{1}{\kappa^2} \nabla^2 f = f (f^2 + {\bf A}^2 -1) , 
\hspace{1.5cm}
\nabla \times \nabla \times {\bf A} = -f^2 {\bf A} .
\label{eq:GL_equation_q}
\end{eqnarray}
Stability of the Meissner state can be discussed by considering the second variation of the free energy under small perturbations $f+\delta f$ and ${\bf A}+ {\bf \delta A}$, namely, 
\begin{eqnarray}
\fl
\delta^2 \Omega = 
\int\! d^3 r \biggl[ \frac{1}{\kappa^2} (\nabla \delta f)^2 + (3f^2 + {\bf A}^2 -1) \delta f^2 + 4 f  {\bf A}\cdot {\bf \delta A} \delta f + f^2 {\bf \delta A}^2 + (\nabla \times {\bf \delta A})^2 \biggr]. 
\label{eq:GL_free_energy_second_variation}
\end{eqnarray}
As long as $\delta^2 \Omega$ is positive definite, the Meissner state corresponds to the global minimum or a metastable local minimum~\cite{kramer}. 
The perturbations are generally given by $\delta f=\delta f (x,y)$ and ${\bf \delta A} = (\delta A_x (x,y), \delta A_y (x,y), 0)$ and can be expanded as $\delta f (x,y) =  \widetilde{\delta f}(x) \cos k y$, $\delta A_x (x,y) = \widetilde{\delta A_x}(x) \sin k y$, and $\delta A_y (x,y) = \widetilde{\delta A_y}(x) \cos k y$.

Let us consider the case $\kappa\to \infty$ for simplicity.  
Then, after some calculations, 
we find $\delta^2 \Omega$ is positive definite as long as $A^2 \le 1/3$, 
and the Meissner state becomes absolutely unstable when $|A|=1/\sqrt{3} \equiv A_{\rm max, GL}$. 
The subscript expresses the result is obtained by the GL theory. 
Restoring the dimensional units, we obtain
\begin{eqnarray}
A_{\rm max,GL} 
= \sqrt{2} B_c \lambda \frac{1}{\sqrt{3}} 
= \frac{\phi_0}{2\sqrt{3} \pi\xi} . 
\label{eq:superheating_GL_1}
\end{eqnarray}
The applied field corresponding to Eq.~(\ref{eq:superheating_GL_1}) is the superheating field. 
The applied field is related to $A$ through the relation
$B_0 = ({\rm rot}\,{\bf A}(0))_z = A'(0)$, 
where $A'$ can be obtained by solving the GL equation $A''=A-A^3$, 
corresponding to Eq.~(\ref{eq:GL_equation_q}) with $\kappa\to\infty$. 
Multiplying $A'$ on the both sides and integrating $A'A''=A A'-A^3 A'$, 
we obtain $A'^2=A^2-(1/2)A^4$, 
where the boundary conditions $A' =0$ and $A = 0$ at $x \to \infty$ are used. 
Then we find $B_0=\sqrt{A(0)^2-(1/2)A(0)^4}$. 
Subsituting $A(0)=A_{\rm max, GL}=1/\sqrt{3}$, 
we obtain $B_{s,{\rm GL}}=\sqrt{5/18}$. 
Restoring the dimensional units, it becomes~\cite{kramer, christiansen, chapman, transtrum}
\begin{eqnarray}
B_{s,{\rm GL}} = \sqrt{2} B_c \sqrt{\frac{5}{18}} = \frac{\sqrt{5}}{3} B_c \simeq 0.745 B_c,  
\label{eq:superheating_GL_2}
\end{eqnarray}
which is the superheating field of the superconductor with $\kappa\to \infty$ at $T\simeq T_c$.  
Note that Eq.~(\ref{eq:superheating_GL_2}) is modified for a finite $\kappa$~\cite{christiansen, transtrum}. 
For example, the superheating field of Nb ($\kappa \simeq 1$) is given by $B_{s}^{\rm (Nb)} \simeq 1.2 B_c^{\rm (Nb)}$ at $T\simeq T_c$. 
See Ref.~\cite{transtrum} for $B_{s,{\rm GL}}$ for an arbitrary $\kappa$.

%%%%%%%%%%%%%%%%
\subsection{Superheating field at $T = 0$}
%%%%%%%%%%%%%%%%

The superheating field evaluated in the GL theory, 
which is valid only at $T\simeq T_c$, 
is not applicable to the SRF cavity operated at $T\ll T_c$ in accelerator applications. 
The quasiclassical formalism~\cite{eilenberger}, 
which is applicable to an arbitrary temperature, 
is available for calculations of the superheating field at $T\ll T_c$. 
The superheating field for a clean superconductor with $\kappa\to\infty$ at $T\to 0$ is given by~\cite{galaiko, catelani, lin}
\begin{eqnarray}
B_s(0)
= \sqrt{1- (2^{\frac{5}{3}}-3)\exp(2^{\frac{4}{3}}-2) } \, B_c(0) 
\simeq 0.84 B_c(0) .
\label{eq:QC_Bs}
\end{eqnarray}
See also \ref{appendix_superheating} for the derivation process of Eq.~(\ref{eq:QC_Bs}). 
Extended results for $T\ne 0$ are seen in Ref.~\cite{catelani} and those for superconductor with impurities are in Ref.~\cite{lin}. 
Eq.~(\ref{eq:QC_Bs}) is approximately applicable to a superconductor with $\kappa \to \infty$ containing non-magnetic impurities~\cite{lin}. 
Note here the quasiclassical theory is valid at all temperature range in $0<T<T_c$, 
and Eq.~(\ref{eq:superheating_GL_2}) can also be derived by using the quasiclassical formalism by considering the case that $T\simeq T_c$.

%%%%%%%%%%%%%%%%%%%%%%%%%%%%%%%%%%%%
%%%%%%%%%%%%%%%%%%%%%%%%%%%%%%%%%%%%
%%%%%%%%%%%%%%%%%%%%%%%%%%%%%%%%%%%%
%%%%%%%%%%%%%%%%%%%%%%%%%%%%%%%%%%%%
%%%%%%%%%%%%%%%%%%%%%%%%%%%%%%%%%%%%
%%%%%%%%%%%%%%%%%%%%%%%%%%%%%%%%%%%%
%%%%%%%%%%%%%%%%%%%%%%%%%%%%%%%%%%%%
%%%%%%%%%%%%%%%%%%%%%%%%%%%%%%%%%%%%
%%%%%%%%%%%%%%%%%%%%%%%%%%%%%%%%%%%%
%%%%%%%%%%%%%%%%%%%%%%%%%%%%%%%%%%%%
%%%%%%%%%%%%%%%%%%%%%%%%%%%%%%%%%%%%
%%%%%%%%%%%%%%%%%%%%%%%%%%%%%%%%%%%%
%%%%%%%%%%%%%%%%%%%%%%%%%%%%%%%%%%%%
%%%%%%%%%%%%%%%%%%%%%%%%%%%%%%%%%%%%
%%%%%%%%%%%%%%%%%%%%%%%%%%%%%%%%%%%%
%%%%%%%%%%%%%%%%%%%%%%%%%%%%%%%%%%%%
\section{Multilayer superconductor}\label{section:multilayer}
%%%%%%%%%%%%%%%%%%%%%%%%%%%%%%%%%%%%
%%%%%%%%%%%%%%%%%%%%%%%%%%%%%%%%%%%%
%%%%%%%%%%%%%%%%%%%%%%%%%%%%%%%%%%%%
%%%%%%%%%%%%%%%%%%%%%%%%%%%%%%%%%%%%
%%%%%%%%%%%%%%%%%%%%%%%%%%%%%%%%%%%%
%%%%%%%%%%%%%%%%%%%%%%%%%%%%%%%%%%%%
%%%%%%%%%%%%%%%%%%%%%%%%%%%%%%%%%%%%
%%%%%%%%%%%%%%%%%%%%%%%%%%%%%%%%%%%%
%%%%%%%%%%%%%%%%%%%%%%%%%%%%%%%%%%%%
%%%%%%%%%%%%%%%%%%%%%%%%%%%%%%%%%%%%
%%%%%%%%%%%%%%%%%%%%%%%%%%%%%%%%%%%%
%%%%%%%%%%%%%%%%%%%%%%%%%%%%%%%%%%%%
%%%%%%%%%%%%%%%%%%%%%%%%%%%%%%%%%%%%
%%%%%%%%%%%%%%%%%%%%%%%%%%%%%%%%%%%%
%%%%%%%%%%%%%%%%%%%%%%%%%%%%%%%%%%%%
%%%%%%%%%%%%%%%%%%%%%%%%%%%%%%%%%%%%

%
\begin{figure}[tb]
   \begin{center}
   \includegraphics[width=0.8\linewidth]{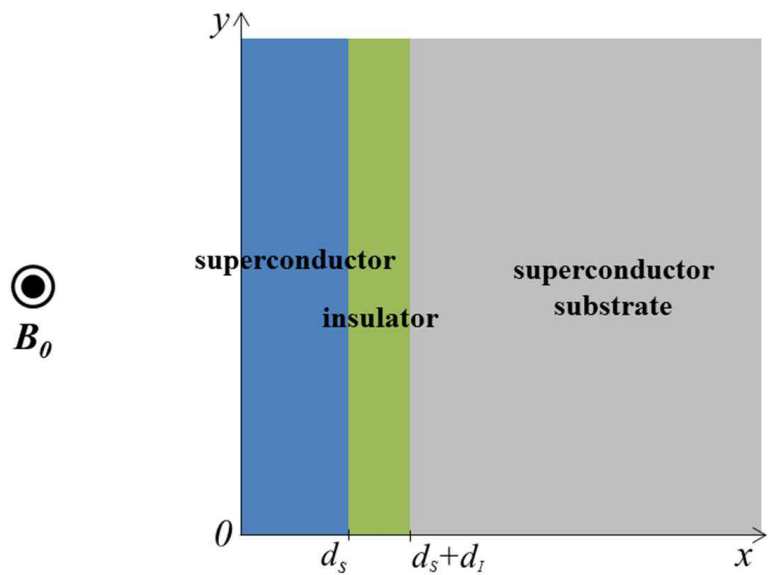}
   \end{center}\vspace{0cm}
   \caption{
Model of the S-I-S structure. 
The layers are parallel to the $y$-$z$ plane and then perpendicular to the $x$-axis. 
The thicknesses of $\mathcal{S}$ and $\mathcal{I}$ layers are $d_{\mathcal S}$ and $d_{\mathcal I}$, respectively. 
The applied magnetic field is given by ${\bf B}_0=(0,0,B_0)$. 
   }\label{fig5}
\end{figure}

Now we start to examine the S-I-S multilayer superconductor. 
The theoretical field limit of the S-I-S structure $B_{\rm max}$ and the optimum layer thicknesses and material combination to maximize $B_{\rm max}$ are discussed. 
We start from an investigation of a model with an ideally flat surface and a negligibly thin insulator in the London theory. 
Then we develop it towards a more quantitative model step by step. 
In the end of this section, we arrive at a realistic model with an imperfect surface and a finite insulator thickness; 
its field limit and the optimum parameters are expressed by using the superheating field of the quasiclassical theory, 
which is valid at an arbitrary temperature $0<T<T_c$. 
The surface resistance of the S-I-S structure is also derived. 
This step-by-step approach seems to be redundant, 
but would be beneficial for readers who want to follow all the calculations. 
Through out this section, we consider the model shown in Fig.~\ref{fig5}. 

%%%%%%%%%%%%%%%%
\subsection{S-I-S structure with a thin ${\mathcal I}$ layer in the London theory}
%%%%%%%%%%%%%%%%

While the London theory provides only a rough estimate of the field limit of the S-I-S structure, 
the analysis based on the London theory contains the essence of the optimization procedure of layer thicknesses and a material combination~\cite{kubo_APL}. 

\begin{figure*}[tb]
   \begin{center}
   \includegraphics[width=1\linewidth]{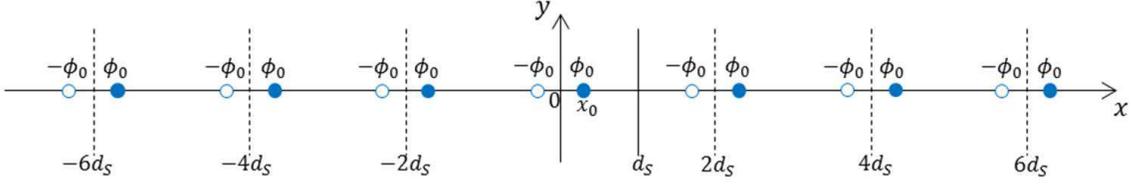}
   \end{center}\vspace{0cm}
   \caption{
Vortex at $x=x_0$ and images necessary for satisfying the boundary conditions at $x=0$ and $x=d_{\mathcal S}$.  
An infinite number of images are introduced. 
   }\label{fig6}
\end{figure*}

As mentioned in the last section, 
the vortex penetration field is defined by the balance of the two forces acting on a vortex at the surface: 
the force from the screening current, ${\bf f}_{\rm M}$, and that from the boundary, ${\bf f}_{\rm B}$. 
As seen in the last section, the former is given by
\begin{eqnarray}
{\bf f}_{\rm M}(x_0)\Bigr|_{x_0=\xi}
={\bf J}(\xi) \times \phi_0 {\bf \hat{z}} 
= \phi_0 J(\xi){\bf \hat{x}} 
\simeq  \phi_0 J(0) {\bf \hat{x}} , 
\end{eqnarray}
where ${\bf J}=J{\bf \hat{y}}$ is the screening current density. 
The later, ${\bf f}_{\rm B}$, can be evaluated by the method of images: 
remove both the boundaries at $x=0$ and $x=d_{\mathcal S}$, 
extend the ${\mathcal S}$ layer material to all the space, 
put appropriate images to satisfy the boundary conditions (zero current normal to the boundaries at $x=0$ and $x=d_{\mathcal S}$), 
and evaluate the force due to all the images. 
This time, unlike the last section, 
an infinite number of image are necessary to satisfy the boundary conditions. 
Suppose a vortex is placed at an arbitrary position $x_0$ in the ${\mathcal S}$ layer. 
Then we need to introduce
(i) an antivortex at $x=-x_0$ to satisfy the condition at $x=0$, 
(ii) an antivortex at $x=2d_{\mathcal S}-x_0$ and a vortex at $x=2d_{\mathcal S}+x_0$ to satisfy the condition at $x=d_{\mathcal S}$, 
which violate the condition at $x=0$, 
(iii) a vortex at $x=-2d_{\mathcal S}+x_0$ and an antivortex at $x=-2d_{\mathcal S}-x_0$ to satisfy the condition at $x=0$ again, 
which violate the condition at $x=d_{\mathcal S}$, 
(iv) an antivortex at $x=4d_{\mathcal S}-x_0$ and a vortex at $x=4d_{\mathcal S}+x_0$ to satisfy the condition at $x=d_{\mathcal S}$, and so on. 
Finally, an infinite number of image vortices are introduced as shown in Fig.~\ref{fig6}. 
All the images act on the vortex at $x=x_0$. 
When $d_{\mathcal S} \lesssim \lambda_1$ the total force can be calculated as (see \ref{appendix_infinite_fI_multi})
\begin{eqnarray}
\fl
{\bf f}_{\rm B}(x_0)
= \frac{\phi_0^2}{4\pi\mu_0 \lambda_1^2} 
\biggl[ -\frac{1}{x_0}+\sum_{n=1}^{\infty} \biggl( \frac{1}{n d_{\mathcal S}-x_0} -\frac{1}{n d_{\mathcal S}+x_0} \biggr) \biggr] {\bf \hat{x}} %\nonumber \\
= -\frac{\phi_0^2}{4\pi\mu_0 \lambda_1^2 d_{\mathcal S}} \pi \cot \frac{\pi x_0}{d_{\mathcal S}} {\bf \hat{x}} . 
\label{eq:infinite_fI_multi}
\end{eqnarray}
When the vortex is placed at the surface ($x_0=\xi$) and $\xi\ll d_{\mathcal S}$, 
Eq.~(\ref{eq:infinite_fI_multi}) is reduced to 
\begin{eqnarray}
{\bf f}_{\rm B}(x_0)\Bigr|_{x_0=\xi}
= -\frac{\phi_0^2}{4\pi\mu_0 \lambda_1^2 \xi}  {\bf \hat{x}} , 
\label{eq:infinite_fI_multi_reduced}
\end{eqnarray}
which corresponds with Eq.~(\ref{eq:fB_semi_infinite}) obtained for a semi-infinite superconductor in the last section. 
Eqs.~(\ref{eq:infinite_fI_multi}) and (\ref{eq:infinite_fI_multi_reduced}) can be derived by directly solving the London equation (see Ref.~\cite{stejic} and \ref{appendix_thin_film}).

When $J(0)$ is so large that $|{\bf f}_{\rm M}(\xi)| > |{\bf f}_{\rm B}(\xi)|$, 
the barrier disappears and the vortex is drawn into the material. 
The maximum current can be obtained by balancing ${\bf f}_{\rm M}$ and ${\bf f}_{\rm B}$ and is given by
\begin{eqnarray}
J_{\rm max,L}^{(\mathcal{S})} = \frac{\phi_0}{4\pi\mu_0 \lambda_1^2 \xi_1} . 
\label{eq:multi_London_jmaxS}
\end{eqnarray}
By using the London equation, this can be written as $A_{\rm max,L}^{(\mathcal{S})}=|-\mu_0\lambda_1^2 J_{\rm max,L}^{(\mathcal{S})}|$ or
\begin{eqnarray}
A_{\rm max,L}^{(\mathcal{S})} =\frac{\phi_0}{4\pi \xi_1} . 
\label{eq:multi_London_AmaxS}
\end{eqnarray}
Eqs.~(\ref{eq:multi_London_jmaxS}) and (\ref{eq:multi_London_AmaxS}) also correspond with those obtained for the semi-infinite superconductor in the last section.

\begin{figure}[tb]
   \begin{center}
   \includegraphics[width=1\linewidth]{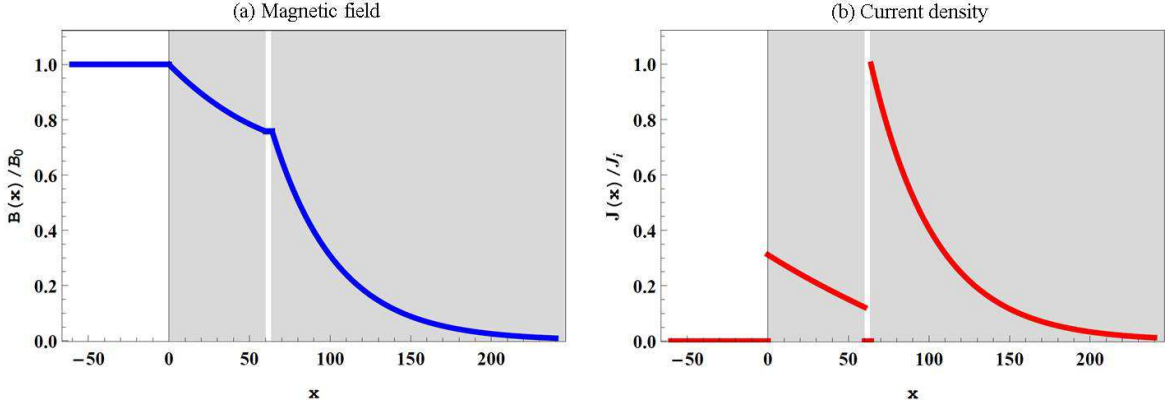}
   \end{center}\vspace{0cm}
   \caption{
Examples of the magnetic field and current density distributions in the S-I-S structure, 
where the magnetic field is normalized by $B_0$, and the current density is normalized by that at the interface $J_i$. 
Assumed parameters are $d_{\mathcal S}=60\,{\rm nm}$, $d_{\mathcal I}=4\,{\rm nm}$, $\lambda_1=120\,{\rm nm}$, and $\lambda_2=40\,{\rm nm}$. 
   }\label{fig7}
\end{figure}

In order to evaluate the maximum field that the $\mathcal{S}$ layer can withstand, 
we need to know the relation between $B_0$ and $J$ (or $A$).  
Here, for simplicity, we consider the case that $d_{\mathcal I}$ is negligibly small and solve the London equation, 
\begin{eqnarray}
A''=\frac{1}{\lambda^2}A , 
\label{eq:London_multi}
\end{eqnarray}
where $\lambda=\lambda_1$ at $0\le x \le d_{\mathcal S}$ and $\lambda=\lambda_2$ at $x> d_{\mathcal S}+d_{\mathcal I}\simeq d_{\mathcal S}$. 
The general solution is written as $A=C_1 e^{-\frac{x}{\lambda_1}} + C_2 e^{\frac{x}{\lambda_1}}$ at $0\le x \le d_{\mathcal S}$ and $A=C_3 e^{-\frac{x-d_{\mathcal S}}{\lambda_2}}$ at $x> d_{\mathcal S}$, 
where $C_i$ ($i=1,2,3$) are constants determined by boundary conditions. 
The boundary conditions are given by $B_0=({\rm rot}\,{\bf A}(0))_z=A'(0)=-C_1/\lambda_1 + C_2/\lambda_1$ and the continuity conditions of $B$ and $A$ at $x=d_{\mathcal S}$, namely, $C_1 e^{-\frac{d_{\mathcal S}}{\lambda_1}} + C_2 e^{\frac{d_{\mathcal S}}{\lambda_1}}=C_3$ 
and $-C_1 e^{-\frac{d_{\mathcal S}}{\lambda_1}} + C_2 e^{\frac{d_{\mathcal S}}{\lambda_1}}=-(\lambda_1/\lambda_2)C_3$. 
The solution is given by
\begin{eqnarray}
A = -\lambda_1 B_0
\frac{  \sinh \frac{d_{\mathcal{S}}-x}{\lambda_1} + \frac{\lambda_2}{\lambda_1} \cosh \frac{d_{\mathcal{S}}-x}{\lambda_1}}
     { \cosh \frac{d_{\mathcal{S}}}{\lambda_1} + \frac{\lambda_2}{\lambda_1}  \sinh \frac{d_{\mathcal{S}}}{\lambda_1}}  \hspace{1cm} (0\le x \le d_{\mathcal S}) ,
\label{eq:multi_London_A1}\\
A=-\lambda_2 B_0 \frac{ e^{-\frac{x-d_{\mathcal{S}} }{\lambda_2}}}
         { \cosh \frac{d_{\mathcal{S}}}{\lambda_1} + \frac{\lambda_2}{\lambda_1} \sinh \frac{d_{\mathcal{S}}}{\lambda_1}}  \hspace{1cm} ( d_{\mathcal S} < x < \infty) .
\label{eq:multi_London_A2}
\end{eqnarray}
The magnetic field distribution~\cite{kubo_APL} is given by $B(x)=A'(x)$ or 
\begin{eqnarray}
B = B_0 \frac{\cosh \frac{d_{\mathcal{S}}-x}{\lambda_1}  + \frac{\lambda_2}{\lambda_1} \sinh \frac{d_{\mathcal{S}}-x}{\lambda_1} }{ \cosh \frac{d_{\mathcal{S}}}{\lambda_1} + \frac{\lambda_2}{\lambda_1} \sinh \frac{d_{\mathcal{S}}}{\lambda_1} }
\hspace{1.5cm}  (0\le x \le d_{\mathcal S}) \\
B= B_0 \frac{ e^{-\frac{x-d_{\mathcal{S}} }{\lambda_2}}}
         { \cosh \frac{d_{\mathcal{S}}}{\lambda_1} + \frac{\lambda_2}{\lambda_1} \sinh \frac{d_{\mathcal{S}}}{\lambda_1}}  \hspace{2cm} ( d_{\mathcal S} < x < \infty) . 
\end{eqnarray}
The current density distribution~\cite{kubo_APL} is given by $J(x)=-B'(x)/\mu_0$ or
\begin{eqnarray}
J = \frac{B_0}{\mu_0 \lambda_1}
\frac{  \sinh \frac{d_{\mathcal{S}}-x}{\lambda_1} + \frac{\lambda_2}{\lambda_1} \cosh \frac{d_{\mathcal{S}}-x}{\lambda_1}}
     { \cosh \frac{d_{\mathcal{S}}}{\lambda_1} + \frac{\lambda_2}{\lambda_1}  \sinh \frac{d_{\mathcal{S}}}{\lambda_1}}  \hspace{1cm} (0\le x \le d_{\mathcal S}) ,
\label{eq:multi_London_J1}\\
J=\frac{B_0}{\mu_0 \lambda_2} \frac{ e^{-\frac{x-d_{\mathcal{S}} }{\lambda_2}}}
         { \cosh \frac{d_{\mathcal{S}}}{\lambda_1} + \frac{\lambda_2}{\lambda_1} \sinh \frac{d_{\mathcal{S}}}{\lambda_1}}  \hspace{1cm} ( d_{\mathcal S} < x < \infty) .
\label{eq:multi_London_J2}
\end{eqnarray}
Examples of the magnetic field and current density distributions are shown in Fig.~\ref{fig7}. 
Then, at the surface, we have~\cite{kubo_APL}  
\begin{eqnarray}
J(0) 
= \gamma_1 \frac{B_0}{\mu_0 \lambda_1}, 
\label{eq:multi_London_j0}
\end{eqnarray}
where the factor $\gamma_1$ defined by
\begin{eqnarray}
\gamma_1 
\equiv \frac{  \sinh \frac{d_{\mathcal{S}}}{\lambda_1} + \frac{\lambda_2}{\lambda_1} \cosh \frac{d_{\mathcal{S}}}{\lambda_1}}
     { \cosh \frac{d_{\mathcal{S}}}{\lambda_1} + \frac{\lambda_2}{\lambda_1}  \sinh \frac{d_{\mathcal{S}}}{\lambda_1}} ,
\label{eq:multi_London_gamma}
\end{eqnarray}
represents the difference of the surface current density between the S-I-S and a simple semi-infinite superconductor. 
This factor comes from the counterflow induced by the substrate~\cite{kubo_APL}. 
An intuitive explanation is as follows. 
Let us consider the magnetic field at the interface of the ${\mathcal S}$ layer and the substrate, $B_i$. 
The magnetic fields generated by the ${\mathcal S}$ layer current is parallel to $-{\bf \hat{z}}$ at the interface and negatively contributes to $B_i$;  
on the other hand, one due to the substrate current is parallel to $+{\bf \hat{z}}$ at the interface and positively contributes to $B_i$;  
these two contributions determines $B_i$. 
When the substrate is made of the same material as the ${\mathcal S}$ layer ($\lambda_2 = \lambda_1$), 
the magnetic field distribution becomes the well-known exponential decay for a simple semi-infinite superconductor: 
$B_i=B_0 e^{-d_{\mathcal S}/\lambda_1}$. 
If we replace the substrate material by a material with a smaller penetration depth $\lambda_2 \, (< \lambda_1)$, 
the magnetic field generated by the substrate increases, 
and the magnetic field at the interface also. 
Thus we have $B_i > B_0 e^{-d_{\mathcal S}/\lambda_1}$: 
the magnetic field attenuation in the ${\mathcal S}$ layer is prevented by the counterflow induced by the substrate with a smaller penetration depth [see Fig.~\ref{fig7}(a)]. 
Since the current density is given by the slope of the magnetic field attenuation, 
a prevention of the field attenuation means a suppression of the current density [see Fig.~\ref{fig7}(b)]. 
Then we have $\gamma_1 < 1$. 
Conversely, when the substrate is made of a material with $\lambda_2 > \lambda_1$, 
the positive contribution from the substrate current decreases, 
and $B_i < B_0 e^{-d_{\mathcal S}/\lambda_1}$: 
the magnetic field attenuation in the ${\mathcal S}$ layer is promoted. 
This means the surface current is enhanced and $\gamma_1 > 1$.

\begin{figure}[tb]
   \begin{center}
   \includegraphics[width=0.7\linewidth]{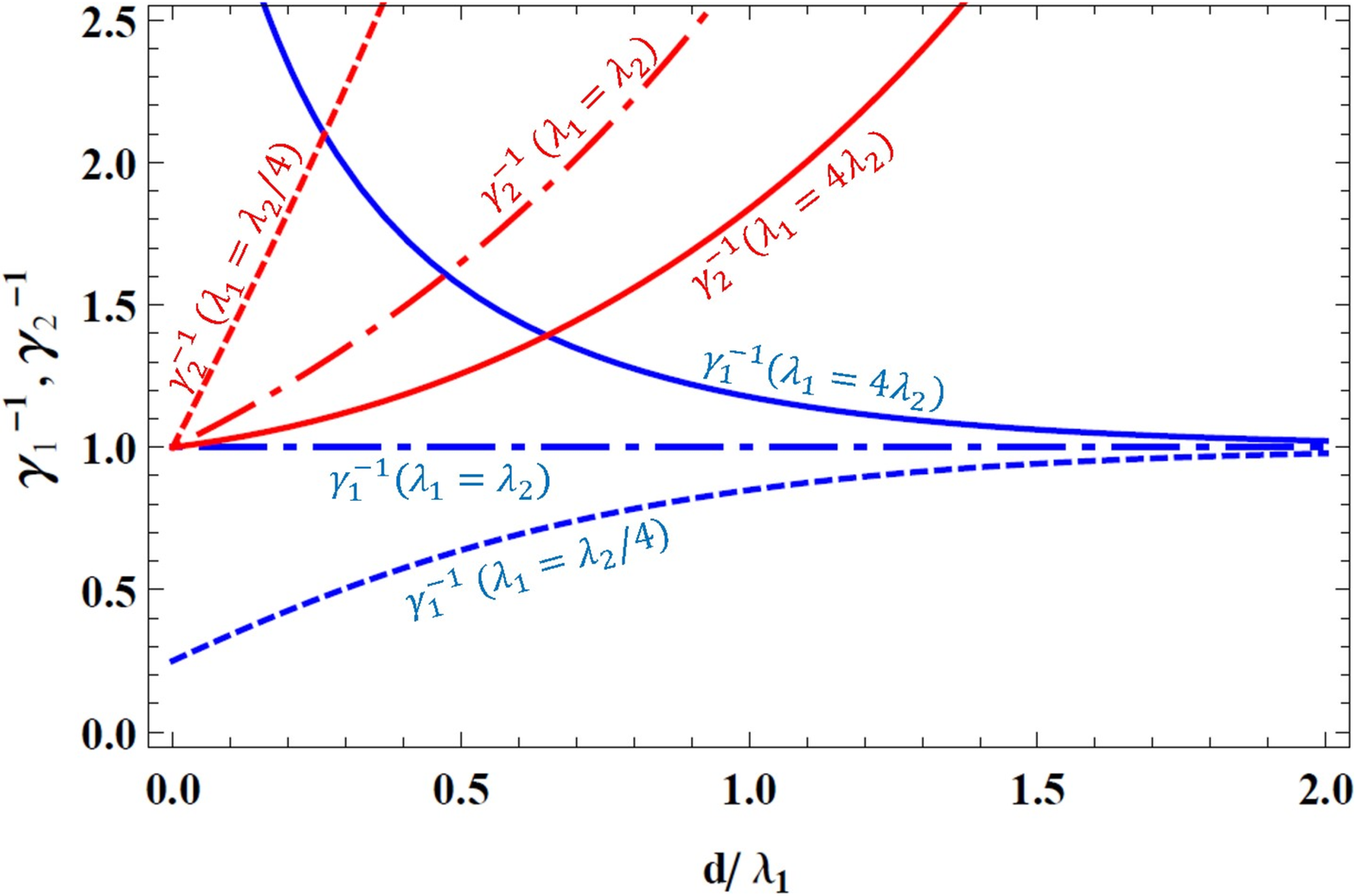}
   \end{center}\vspace{0cm}
   \caption{
$\gamma_1^{-1}$ and $\gamma_2^{-1}$ as functions of the ${\mathcal S}$ layer thickness $d_{\mathcal S}/\lambda_1$. 
   }\label{fig8}
\end{figure}

By using Eq.~(\ref{eq:multi_London_j0}) or $B_0 = \gamma_1^{-1} \mu_0 \lambda_1 J(0)$, 
the applied magnetic field corresponding to $J_{\rm max,L}^{\mathcal (S)}$ or $A_{\rm max,L}^{\mathcal (S)}$ is given by~\cite{kubo_APL} 
\begin{eqnarray}
B_{\rm max,L}^{(\mathcal{S})} 
= \gamma_1^{-1} \mu_0 \lambda_1 J_{\rm max,L}^{(\mathcal{S})} 
= \gamma_1^{-1} \frac{\phi_0}{4\pi \lambda_1 \xi_1}
= \gamma_1^{-1} \frac{B_c^{(\mathcal{S})} }{\sqrt{2}}
= \gamma_1^{-1} B_{\rm v}^{(\mathcal{S})} ,
\label{eq:BmaxS_multi}
\end{eqnarray}
where $B_c^{(\mathcal{S})}$ and $B_{\rm v}^{(\mathcal{S})}$ are the thermodynamic critical field and the vortex penetration field of the ${\mathcal S}$ layer material, respectively. 
$B_{\rm max,L}^{(\mathcal{S})}$ is the maximum field that the ${\mathcal S}$ layer can withstand. 
As mentioned in the above, 
$\gamma_1 <1$ or $\gamma_1^{-1} > 1$ when the condition~\cite{kubo_APL}  
\begin{eqnarray}
\lambda_1 > \lambda_2 
\label{eq:multilayer_condition}
\end{eqnarray}
is satisfied. 
Then $B_{\rm max,L}^{(\mathcal{S})}$ can exceed the vortex penetration field of the ${\mathcal S}$ layer material $B_{\rm v}^{(\mathcal{S})}$ by the factor $\gamma_1^{-1}$. 
This enhancement comes from the suppression of surface current by $\gamma_1$. 
Conversely, when $\lambda_1 < \lambda_2$,  
the surface current is enhanced by $\gamma_1 \,(>1)$, and $B_{\rm max,L}^{(\mathcal{S})}$ is suppressed by $\gamma_1^{-1} \,(<1)$. 
Fig.~\ref{fig8} shows $\gamma_1^{-1}$ as functions of the ${\mathcal S}$ layer thickness. 
When $\lambda_1 > \lambda_2$, 
the factor $\gamma_1^{-1}$ increases as $d_{\mathcal S}$ decreases: 
the thinner the ${\mathcal S}$ layer the larger the $B_{\rm max,L}^{(\mathcal{S})}$ (see the solid blue curve).

In the following discussion, Eq.~(\ref{eq:multilayer_condition}) is assumed to be satisfied.
While a thin ${\mathcal S}$ layer pushes up $B_{\rm max,L}^{(\mathcal{S})}$, 
an extremely thin $d_{\mathcal S}$ cannot protect the substrate. 
When the magnetic field at the interface of the substrate~\cite{kubo_APL}, 
\begin{eqnarray}
B_i \equiv B(d_{\mathcal S})=\gamma_2 B_0 ,
\hspace{1.5cm}
\gamma_2
\equiv \frac{1}{ \cosh \frac{d_{\mathcal{S}}}{\lambda_1} + \frac{\lambda_2}{\lambda_1}  \sinh \frac{d_{\mathcal{S}}}{\lambda_1}} , 
\label{eq:multi_London_gamma2}
\end{eqnarray}
exceeds the field limit of the substrate $B_{\rm max}^{\rm (sub)}$, 
it causes a breakdown, 
where $B_{\rm max}^{\rm (sub)}$ is an empirical field limit of the substrate material (e. g.,  $B_{\rm max}^{\rm (sub)}\sim B_{c1}^{\rm (Nb)} \sim B_{c}^{\rm (Nb)}$ for a bulk Nb). 
Thus, in order to improve the field limit of the whole S-I-S structure $B_{\rm max,L}$, 
we need to optimize $d_{\mathcal S}$ so as to simultaneously increase $B_{\rm max,L}^{({\mathcal S})}$ and suppress $B_i$.
For a given $d_{\mathcal S}$, 
$B_{\rm max,L}$ is given by $B_0$ that satisfies $B_0< B_{\rm max,L}^{({\mathcal S})}$ and $B_i < B_{\rm max}^{\rm (sub)}$ simultaneously~\cite{kubo_APL}: 
\begin{eqnarray}
B_{\rm max,L}= {\rm min} \{ \gamma_1^{-1} B_{\rm v}^{(\mathcal{S})}, \gamma_2^{-1} B_{\rm max}^{\rm (sub)} \} . 
\label{eq:multilayer_max_London}
\end{eqnarray}
To find the maximum value of $B_{\rm max,L}$, let us see the solid curves in Fig.~\ref{fig8}, corresponding to $\lambda_1 > \lambda_2$. 
While $\gamma_1^{-1}$ increases as $d_{\mathcal S}$ decreases, 
$\gamma_2^{-1}$ increases as $d_{\mathcal S}$ increases.  
$B_{\rm max}$ is maximized when the condition $\gamma_1^{-1} B_{\rm v}^{(\mathcal{S})} =\gamma_2^{-1} B_{\rm max}^{\rm (sub)}$ is satisfied. 
By substituting the definitions of $\gamma_1$ and $\gamma_2$, 
this condition becomes the quadratic equation $( 1+ \lambda_2/\lambda_1 ) u^2 -2ru - ( 1- \lambda_2/\lambda_1) = 0$, 
where $u\equiv e^{d_{\mathcal S}/\lambda_1}$ and $r\equiv B_{\rm v}^{(\mathcal{S})}/B_{\rm max}^{\rm (sub)}$. 
The solution is given by $u = (r + \sqrt{r^2+1 -\lambda_2^2/\lambda_1^2}\,)/ (1+ \lambda_2/\lambda_1)\equiv u_0$ or~\cite{gurevich_AIP} 
\begin{eqnarray}
\fl
d_{\mathcal S}^{\rm opt} 
= \lambda_1 \log u_0
= \lambda_1 \log \Biggl[ \frac{\lambda_1}{\lambda_1+\lambda_2}\frac{B_{\rm v}^{(\mathcal{S})}}{B_{\rm max}^{\rm (sub)}} 
+ \sqrt{ \Bigl(\frac{\lambda_1}{\lambda_1+\lambda_2} \frac{B_{\rm v}^{(\mathcal{S})}}{B_{\rm max}^{\rm (sub)}} \Bigr)^2 +\frac{\lambda_1-\lambda_2}{\lambda_1+\lambda_2} } \,\,\Biggr] .
\label{eq:multilayer_London_opt_ds}
\end{eqnarray}
Substituting $u=u_0$ into $\gamma_2^{-1}=(1/2) (u+u^{-1}) + (1/2)(\lambda_2/\lambda_1)(u-u^{-1})$, 
we find $\gamma_2^{-1}=\sqrt{r^2 +1 - (\lambda_2/\lambda_1)^2}$. 
Then the optimized $B_{\rm max,L}$ is given by~\cite{gurevich_AIP} 
\begin{eqnarray}
B_{\rm max,L}^{\rm opt} 
= \gamma_1^{-1} B_{\rm v}^{(\mathcal{S})} =\gamma_2^{-1} B_{\rm max}^{\rm (sub)}
= \sqrt{ (B_{\rm v}^{(\mathcal{S})})^2 + \biggl( 1-\frac{\lambda_2^2}{\lambda_1^2} \biggr) (B_{\rm max}^{\rm (sub)})^2 }, 
\label{eq:multilayer_London_opt_Bmax}
\end{eqnarray}
which is the main result in this subsection together with the optimum conditions given by  Eqs.~(\ref{eq:multilayer_condition}) and (\ref{eq:multilayer_London_opt_ds}).

So far, we have examined the S-I-S structure in the framework of the London theory, 
where the main results explicitly depends on the vortex penetration field of the ${\mathcal S}$ layer material, $B_{\rm v}^{(\mathcal{S})}$. 
As mentioned in the last section, however, 
the vortex penetration field defined in the London theory is unsatisfactory. 
The main results should be expressed by the superheating field of the GL or quasiclassical theories.

%%%%%%%%%%%%%%%%%%
%%%%%%%%%%%%%%%%%%
\subsection{S-I-S structure with a thin ${\mathcal I}$ layer at $T\simeq T_c$}
%%%%%%%%%%%%%%%%%%
%%%%%%%%%%%%%%%%%%

Next we investigate the same system as the above, the S-I-S structure with a negligibly thin ${\mathcal I}$ layer, in the framework of the GL theory~\cite{gurevich_AIP, posen_PRAppl} and rewrite the main results by using the GL superheating field. 
We follow the discussion in Ref.~\cite{gurevich_AIP}.

For simplicity, we assume the ${\mathcal S}$ layer and the substrate are made of materials with $\kappa \gg 1$. 
Then the GL equation is given by $A''=A-A^3$ in the usual dimensionless expression [see also the discussion below Eq.~(\ref{eq:superheating_GL_1})]. 
Restoring the dimensional units, we have
\begin{eqnarray}
A'' =\frac{1}{\lambda^2} A -\frac{4\pi^2\xi^2}{\phi_0^2 \lambda^2}A^3, 
\label{eq:GL_multi}
\end{eqnarray}
where $\lambda = \lambda_1$ and $\xi=\xi_1$ at $0 \le x \le d_{\mathcal S}$ and $\lambda = \lambda_2$ and $\xi=\xi_2$ at $x>d_{\mathcal S}$. 
Multiplying $A'$ and integrating $\lambda^2 A'A'' = AA' -(4\pi^2\xi^2/\phi_0^2) A^3A'$, we obtain
\begin{eqnarray}
\lambda^2 {A'}^2 -A^2 + \frac{2\pi^2\xi^2}{\phi_0^2}A^4=C 
\hspace{0.5cm} (0 < x < d_{\mathcal S}) , 
\label{eq:GL_multi_1}\\
\lambda^2 {A'}^2 -A^2 + \frac{2\pi^2\xi^2}{\phi_0^2}A^4=0
\hspace{0.5cm} (d_{\mathcal S} < x) ,
\label{eq:GL_multi_2}
\end{eqnarray}
where $C$ is a constant. 
The ${\mathcal S}$ layer and the substrate of the optimized S-I-S structure can achieve $A(0)=\phi_0/2\sqrt{3}\pi\xi_1$ and $A(d)=\phi_0/2\sqrt{3}\pi\xi_2$, respectively [see Eq.~(\ref{eq:superheating_GL_1})], 
when the applied field is $B_0=A'(0)=B_{\rm max,GL}^{\rm opt}$. 
Substituting $x=0$ into into Eq.~(\ref{eq:GL_multi_1}), 
$x=d_{\mathcal S}$ into Eq.~(\ref{eq:GL_multi_1}), 
and $x=d_{\mathcal S}$ into Eq.~(\ref{eq:GL_multi_2}), 
we have
\begin{eqnarray}
\lambda^2 (B_{\rm max,GL}^{\rm opt})^2 - \frac{5\phi_0^2}{72\pi^2\xi^2}=C , \\
\lambda^2 A'(d_{\mathcal S})^2  - \frac{\phi_0^2}{72\pi^2\xi_0^2} \biggl( 6-\frac{\xi^2}{\xi_0^2} \biggr) =C , \\
\lambda_0^2 A'(d_{\mathcal S})^2  - \frac{5\phi_0^2}{72\pi^2\xi_0^2} =0 . 
\end{eqnarray}
Solving these three equations, we find~\cite{gurevich_AIP}
\begin{eqnarray}
B_{\rm max,GL}^{\rm opt} = 
\sqrt{ \biggl( 1-\frac{\xi_1^2}{5\xi_2^2} + \frac{\xi_1^4}{5\xi_2^4} \biggr) ( B_{s,{\rm GL}}^{({\mathcal S})})^2 
+ \biggl( 1-\frac{\lambda_2^2}{\lambda_1^2} \biggr)  (B_{s,{\rm GL}}^{\rm (sub)})^2 } ,
\label{eq:Bmaxopt_multi_GL}
\end{eqnarray}
where $B_{s,{\rm GL}}^{({\mathcal S})}=(\sqrt{5}/3) (\phi_0/2\sqrt{2}\pi\lambda_1\xi_1)$ and $B_{s,{\rm GL}}^{({\rm sub})}=(\sqrt{5}/3) (\phi_0/2\sqrt{2}\pi\lambda_2\xi_2)$ are the GL superheating fields in the ${\mathcal S}$ layer and the substrate, respectively [see Eq.~(\ref{eq:superheating_GL_2})]. 
In the above calculation, we have assumed the substrate can withstand up to its superheating field $B_{s,{\rm GL}}^{\rm (sub)}$, 
but it can be replaced by an empirical field limit $B_{\rm max}^{\rm (sub)}$. 
Furthermore, when $\xi_1 \ll \xi_2$, the second and third terms in the first parenthesis are negligible. 
Then Eq.~(\ref{eq:Bmaxopt_multi_GL}) is reduced to~\cite{gurevich_AIP} 
\begin{eqnarray}
B_{\rm max,GL}^{\rm opt} = 
\sqrt{  ( B_{s,{\rm GL}}^{({\mathcal S})})^2 
+ \biggl( 1-\frac{\lambda_2^2}{\lambda_1^2} \biggr)  (B_{\rm max}^{\rm (sub)})^2 } ,
\label{eq:Bmaxopt_multi_GL_3}
\end{eqnarray}
which has the same form as Eq.~(\ref{eq:multilayer_London_opt_Bmax}) except for $B_{\rm v}^{({\mathcal S})}$ being replaced by $B_{s,{\rm GL}}^{({\mathcal S})}$.

Eq.~(\ref{eq:Bmaxopt_multi_GL_3}) can be obtained by an easier way as follows. 
We disregard the nonlinear term in Eq.~(\ref{eq:GL_multi}) and obtain the London equation, Eq.~(\ref{eq:London_multi}): 
we assume the magnetic field attenuation is well described by the London equation.  
Its solution is given by Eqs.~(\ref{eq:multi_London_A1}) and (\ref{eq:multi_London_A2}). 
Then the surface current density is given by Eq.~(\ref{eq:multi_London_j0}), 
and the magnetic field at the interface is by Eq.~(\ref{eq:multi_London_gamma2}). 
The surface current density must be smaller than the depairing limit $B_{s,{\rm GL}}^{\mathcal (S)}/\mu_0 \lambda_1$, namely, 
$\gamma_1 B_0/\mu_0 \lambda_1 < B_{s,{\rm GL}}^{\mathcal (S)}/\mu_0 \lambda_1$ or $B_0 < \gamma_1^{-1} B_{s,{\rm GL}}^{\mathcal (S)}$. 
Furthermore, the magnetic field at the interface must be smaller than the empirical field limit of the substrate: 
$B_i = \gamma_2 B_0 < B_{\rm max}^{\rm (sub)}$ or $B_0 < \gamma_2^{-1} B_{\rm max}^{\rm (sub)}$. 
Then the maximum $B_0$ is given by $B_{\rm max,GL}={\rm min} \{ \gamma_1^{-1} B_{s,{\rm GL}}^{\mathcal (S)}, \gamma_2^{-1} B_{\rm max}^{\rm (sub)} \}$, 
which corresponds with Eq.~(\ref{eq:multilayer_max_London}) except for $B_{\rm v}^{\mathcal (S)}$ being replaced by $B_{s,{\rm GL}}^{\mathcal (S)}$. 
$B_{\rm max,GL}$ is maximized when $\gamma_1^{-1} B_{s,{\rm GL}}^{\mathcal (S)}= \gamma_2^{-1} B_{\rm max}^{\rm (sub)}$, 
and finally we obtain Eq.~(\ref{eq:Bmaxopt_multi_GL_3}).

By using the same scheme as the above, 
the optimum conditions and the optimized field limit can be expressed by using the superheating field of the quasiclassical theory as shown below, 
which is valid at an arbitrary temperature $0<T<T_c$.

%%%%%%%%%%%%%%%%%%
%%%%%%%%%%%%%%%%%%
\subsection{S-I-S structure with a thin ${\mathcal I}$ layer at $0< T < T_c$}
%%%%%%%%%%%%%%%%%%
%%%%%%%%%%%%%%%%%%

We repeat the same scheme as the above. 
The only difference is the deparing limit:  
$B_{s,{\rm GL}}^{\mathcal (S)}/\mu_0 \lambda_1$ is replaced by $B_{s}^{\mathcal (S)}/\mu_0 \lambda_1$ that is obtained in the framework of the quasiclassical theory. 
Let us summarize results. 
The field limit for a given $d_{\mathcal S}$ is given by~\cite{kubo_APL}
\begin{eqnarray}
B_{\rm max}={\rm min} \{ \gamma_1^{-1} B_{s}^{\mathcal (S)}, \gamma_2^{-1} B_{\rm max}^{\rm (sub)} \} , 
\label{eq:Bmax_QC}
\end{eqnarray}
where $\gamma_1$ and $\gamma_2$ are given by Eqs.~(\ref{eq:multi_London_gamma}) and (\ref{eq:multi_London_gamma2}), respectively. 
When the conditions~\cite{kubo_APL} 
\begin{eqnarray}
\lambda_1 > \lambda_2, 
\label{eq:multilayer_QC_opt_lambda} 
\end{eqnarray}
and~\cite{gurevich_AIP} 
\begin{eqnarray}
d_{\mathcal S} 
= \lambda_1 \log \biggl[ \frac{\lambda_1}{\lambda_1+\lambda_2} \frac{B_{s}^{(\mathcal{S})}}{B_{\rm max}^{\rm (sub)}} + \sqrt{ \Bigl( \frac{\lambda_1}{\lambda_1+\lambda_2} \frac{B_{s}^{(\mathcal{S})}}{B_{\rm max}^{\rm (sub)}} \Bigr)^2 +\frac{\lambda_1-\lambda_2}{\lambda_1+\lambda_2} }
\,\,\Biggr] ,
\label{eq:multilayer_QC_opt_ds}
\end{eqnarray}
are satisfied, 
$B_{\rm max}$ is maximized and is given by~\cite{gurevich_AIP}
\begin{eqnarray}
B_{\rm max}^{\rm opt} = 
\sqrt{  ( B_{s}^{({\mathcal S})})^2 
+ \biggl( 1-\frac{\lambda_2^2}{\lambda_1^2} \biggr)  (B_{\rm max}^{\rm (sub)})^2 } . 
\label{eq:Bmaxopt_multi_QC}
\end{eqnarray}
Note that all $B_{\rm v}^{\mathcal (S)}$ or $B_{s,{\rm GL}}^{\mathcal (S)}$ have been replaced by those obtained in the quasiclassical theory, $B_s^{\mathcal (S)}$: 
the formulae are valid at an arbitrary temperature $0<T<T_c$. 
When the ${\mathcal S}$ layer material is a superconductor with $\kappa \gg 1$ and an accelerator is operated at $T \ll T_c$, 
$B_s^{\mathcal (S)}$ is approximately given by $B_s^{\mathcal (S)}=0.84 B_c^{\mathcal (S)}$, 
which is derived in the last section for a superconductor with $\kappa \to \infty$ at  $T\to 0$. 
The same $B_s^{\mathcal (S)}$ is available as an approximate value when non-magnetic impurities are included~\cite{lin}.

%%%%%%%%%%%%%%%%%%
%%%%%%%%%%%%%%%%%%
\subsection{S-I-S structure with a finite $d_{\mathcal I}$ at $0<T< T_c$}
%%%%%%%%%%%%%%%%%%
%%%%%%%%%%%%%%%%%%

\begin{figure}[tb]
   \begin{center}
   \includegraphics[width=0.5\linewidth]{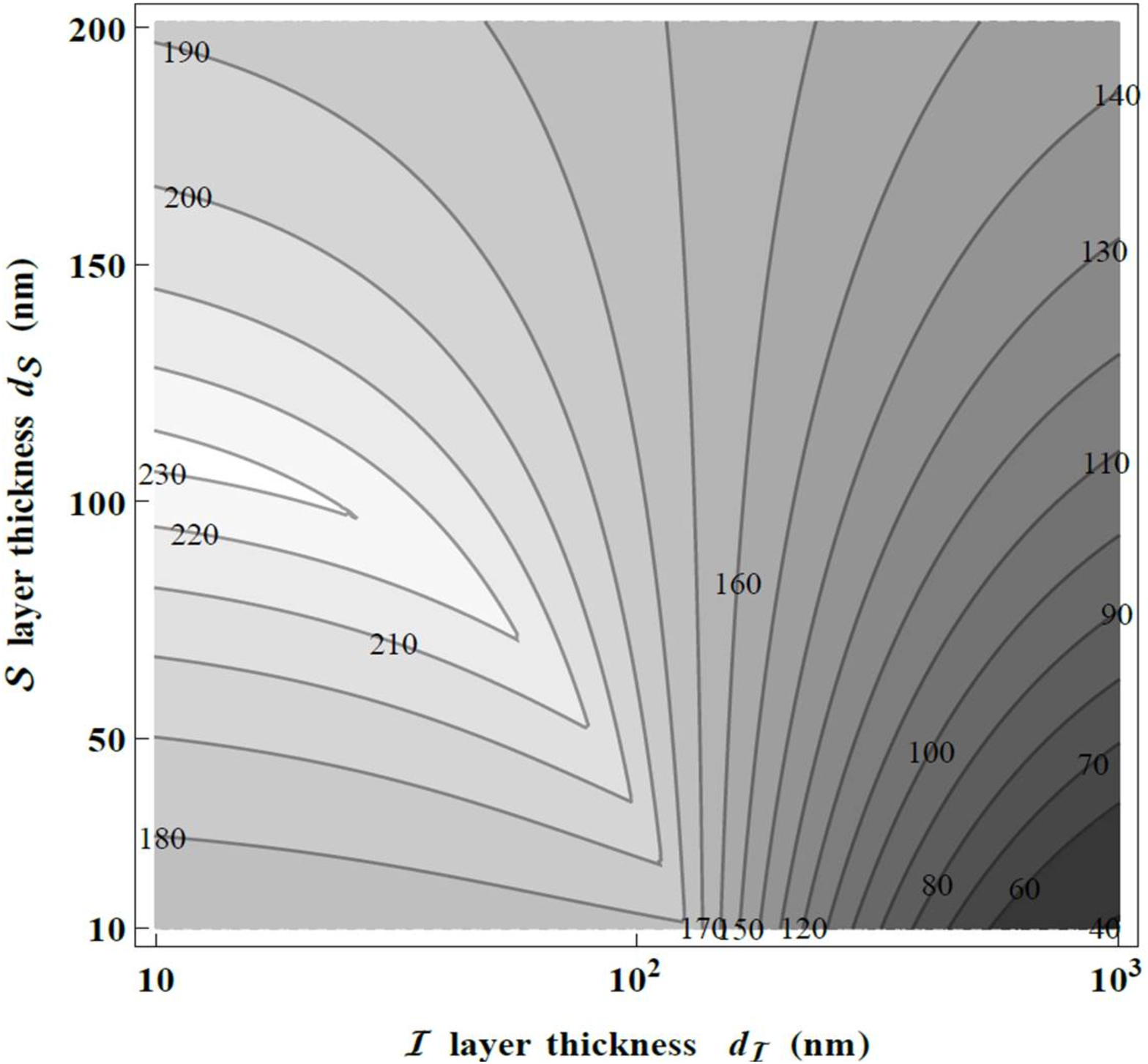}
   \end{center}\vspace{0cm}
   \caption{
$B_{\rm max}$ of dirty Nb-${\mathcal I}$-Nb system in unit of mT. 
Assumed parameters are $B_c^{\rm (Nb)}=200\,{\rm mT}$ and $\lambda_1= 180\,{\rm nm}$ for the ${\mathcal S}$ layer material, 
$B_{\rm max}^{\rm (sub)}=170\,{\rm mT}$ and $\lambda_2=40\,{\rm nm}$ for the substrate. 
See also Ref.~\cite{kubo_SRF2015}. 
   }\label{fig9}
\end{figure}

\begin{figure}[tb]
   \begin{center}
   \includegraphics[width=0.5\linewidth]{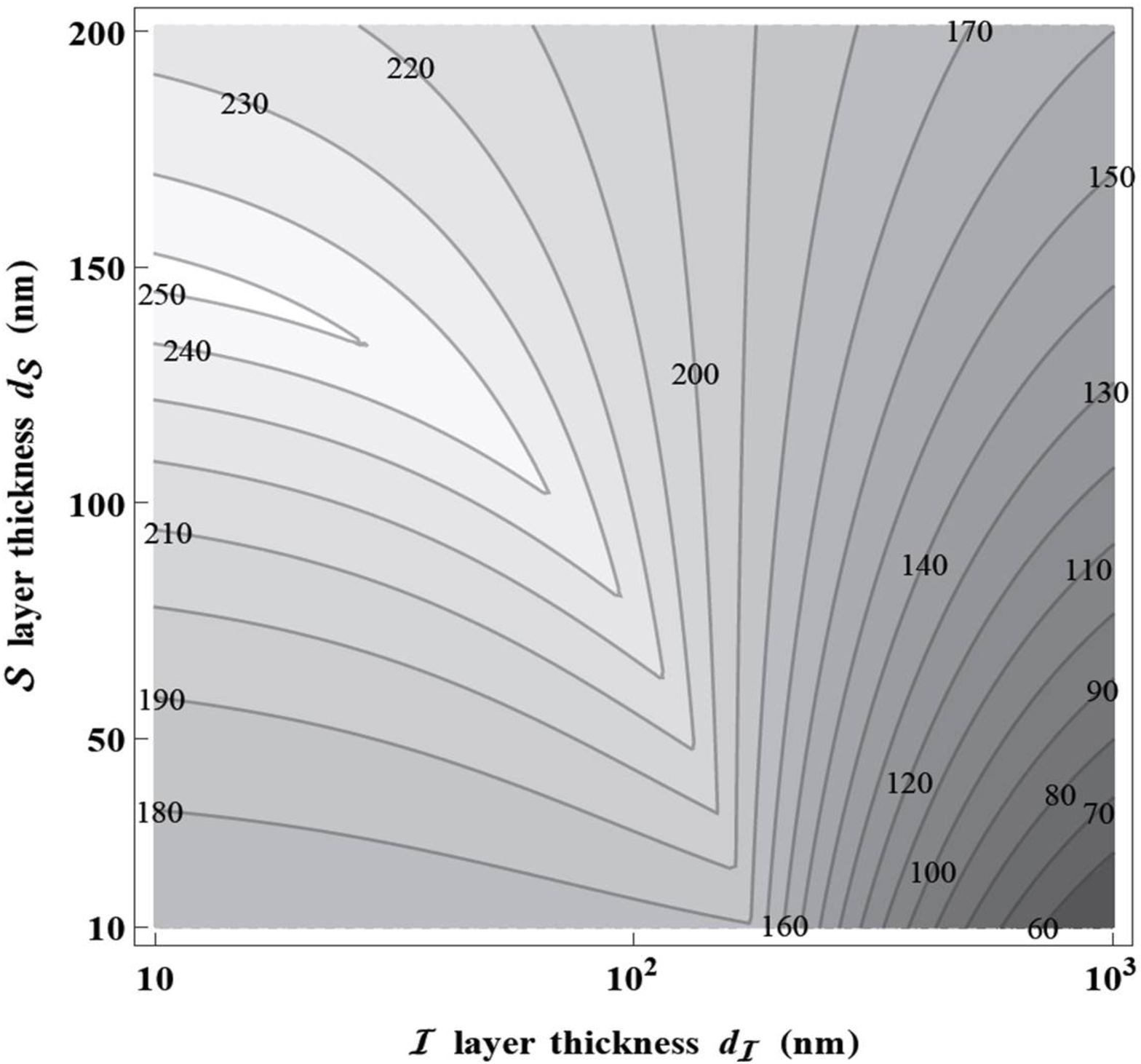}
   \end{center}\vspace{0cm}
   \caption{
$B_{\rm max}$ of NbN-${\mathcal I}$-Nb system in unit of mT. 
Assumed parameters are $B_c^{\rm (NbN)}=230\,{\rm mT}$ and $\lambda_1= 200\,{\rm nm}$ for the ${\mathcal S}$ layer material, 
$B_{\rm max}^{\rm (sub)}=170\,{\rm mT}$ and $\lambda_2=40\,{\rm nm}$ for the substrate. 
See also Ref.~\cite{kubo_SRF2015}. 
   }\label{fig10}
\end{figure}

\begin{figure}[tb]
   \begin{center}
   \includegraphics[width=0.5\linewidth]{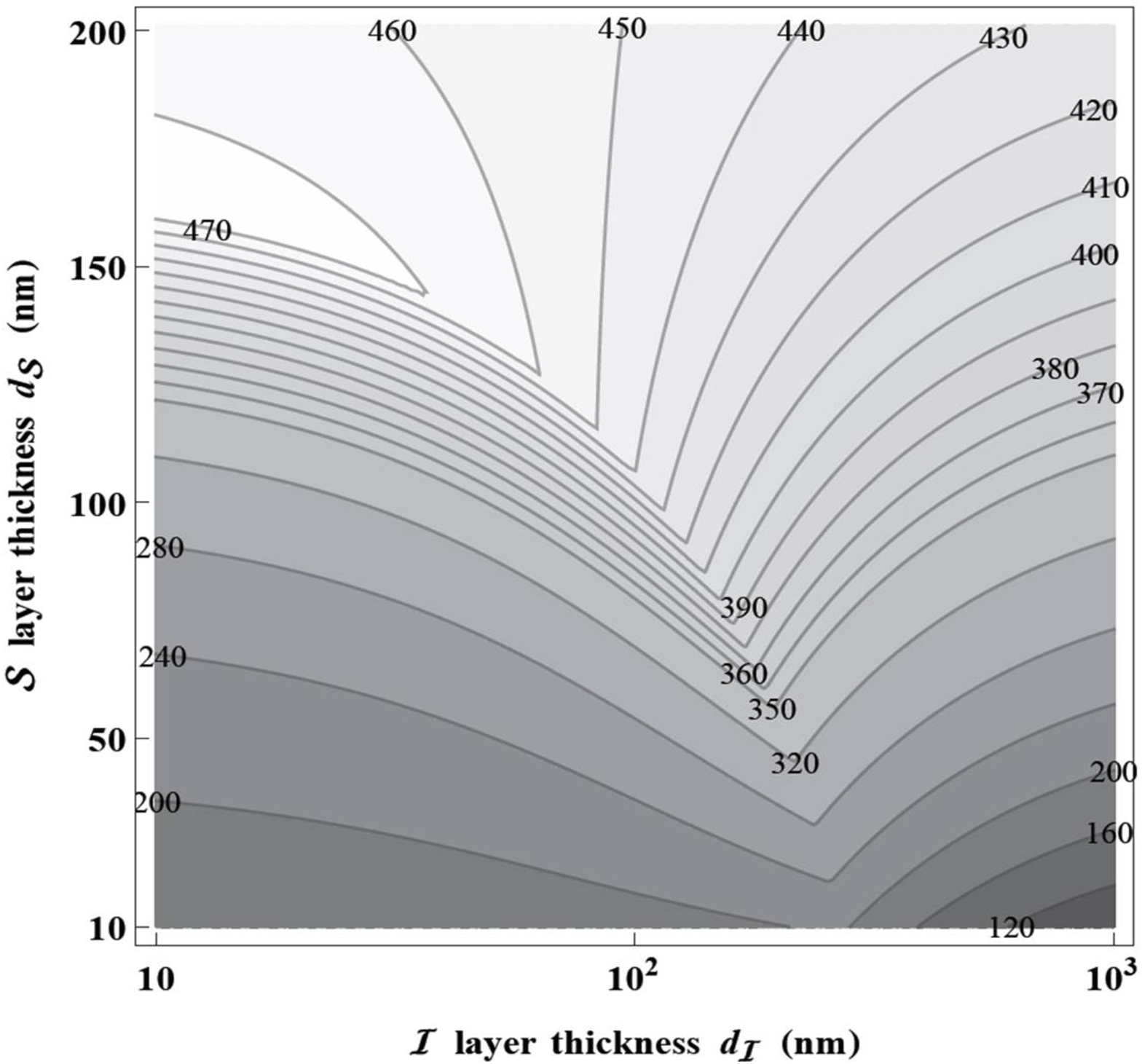}
   \end{center}\vspace{0cm}
   \caption{
$B_{\rm max}$ of ${\rm Nb_3 Sn}$-${\mathcal I}$-Nb system in unit of mT.   
Assumed parameters are $B_c^{\rm (Nb_3 Sn)}=540\,{\rm mT}$ and $\lambda_1= 120\,{\rm nm}$ for the ${\mathcal S}$ layer material, 
$B_{\rm max}^{\rm (sub)}=170\,{\rm mT}$ and $\lambda_2=40\,{\rm nm}$ for the substrate. 
See also Ref.~\cite{kubo_SRF2015}. 
   }\label{fig11}
\end{figure}

We have neglected the ${\mathcal I}$ layer thickness so far. 
Now we incorporate effects of a finite $d_{\mathcal I}$. 
When a frequency of the electromagnetic field is $\sim {\rm GHz}$ and $d_{\mathcal I}\ll 1\,{\rm cm}$, 
the magnetic field distribution in the S-I-S structure is given by (see \ref{appendix_electromagnetic_multi} and Ref.~\cite{kubo_APL, kubo_IPAC13}). 
\begin{eqnarray}
B &=& 
B_0
\frac{ \cosh \frac{d_{\mathcal{S}}-x}{\lambda_1} + \frac{\lambda_2+d_{\mathcal{I}}}{\lambda_1} \sinh \frac{d_{\mathcal{S}}-x}{\lambda_1}}
     { \cosh \frac{d_{\mathcal{S}}}{\lambda_1} + \frac{\lambda_2 + d_{\mathcal{I}}}{\lambda_1} \sinh \frac{d_{\mathcal{S}}}{\lambda_1}}  
\hspace{1cm} (0\le x \le d_{\mathcal S})  ,
\label{eq:B1} \\
B &=& 
B_0
\frac{ 1 }
     { \cosh \frac{d_{\mathcal{S}}}{\lambda_1} + \frac{\lambda_2+d_{\mathcal{I}}}{\lambda_1} \sinh \frac{d_{\mathcal{S}}}{\lambda_1}} 
= \widetilde{\gamma_2} B_0 
\hspace{0.5cm} (d_{\mathcal S} < x \le d_{\mathcal S} + d_{\mathcal I}) ,  
\label{eq:B2} \\
B &=& 
B_0 \frac{ e^{-\frac{x-d_{\mathcal{S}} -d_{\mathcal{I}}}{\lambda_2}}}
         { \cosh \frac{d_{\mathcal{S}}}{\lambda_1} + \frac{\lambda_2+d_{\mathcal{I}}}{\lambda_1} \sinh \frac{d_{\mathcal{S}}}{\lambda_1}} 
= \widetilde{\gamma_2} B_0 e^{-\frac{x-d_{\mathcal{S}} -d_{\mathcal{I}}}{\lambda_2}}
\hspace{0.5cm} (x \ge d_{\mathcal S} + d_{\mathcal I}),  
\label{eq:B3}
\end{eqnarray}
where 
\begin{eqnarray}
\widetilde{\gamma_2} \equiv \frac{ 1 }{ \cosh \frac{d_{\mathcal{S}}}{\lambda_1} + \frac{\lambda_2+d_{\mathcal{I}}}{\lambda_1} \sinh \frac{d_{\mathcal{S}}}{\lambda_1}} .
\label{eq:tildegamma2}
\end{eqnarray}
Then the surface current density is given by $J(0)=-B'(0)/\mu_0$ or~\cite{kubo_APL} 
\begin{eqnarray}
J(0) = \widetilde{\gamma_1} \frac{B_0}{\mu_0\lambda_1} , 
\hspace{2cm} 
\widetilde{\gamma_1} \equiv 
\frac{ \sinh \frac{d_{\mathcal{S}}}{\lambda_1} + \frac{\lambda_2+d_{\mathcal{I}}}{\lambda_1} \cosh \frac{d_{\mathcal{S}}}{\lambda_1}}
     { \cosh \frac{d_{\mathcal{S}}}{\lambda_1} + \frac{\lambda_2+d_{\mathcal{I}}}{\lambda_1} \sinh \frac{d_{\mathcal{S}}}{\lambda_1}} . 
\label{eq:j0_finite_dI}
\end{eqnarray}
Recall $\gamma_1$ defined by Eq.~(\ref{eq:multi_London_gamma}) is smaller than unity when $\lambda_1 > \lambda_2$. 
Then we find, when the condition
\begin{eqnarray}
\lambda_1 > \lambda_2 + d_{\mathcal I} 
\end{eqnarray}
is satisfied, 
$\widetilde{\gamma_1}$ is smaller than unity and the surface current is suppressed. 
The field limit can be evaluated by the same discussions as before:  
$J(0)$ must be smaller than the depairing limit $B_{s}^{\mathcal (S)}/\mu_0 \lambda_1$, 
and the magnetic field at the interface $\widetilde{\gamma}_2 B_0$ must be smaller than the empirical field limit of the substrate $B_{\rm max}^{\rm (sub)}$. 
Then we have~\cite{kubo_APL}
\begin{eqnarray}
B_{\rm max}={\rm min} \{ \widetilde{\gamma_1}^{-1} B_{s}^{\mathcal (S)}, \widetilde{\gamma_2}^{-1} B_{\rm max}^{\rm (sub)} \}. 
\label{eq:Bmax_finite_dI}
\end{eqnarray}
The factors $\gamma_1$ and $\gamma_2$ in Eq.~(\ref{eq:Bmax_QC}) have been replaced by $\widetilde{\gamma_1}$ and $\widetilde{\gamma_2}$, respectively. 
Figs.~\ref{fig9}-\ref{fig11} show contour plots of $B_{\rm max}$ calculated by using Eq.~(\ref{eq:Bmax_finite_dI}) for dirty Nb-${\mathcal I}$-Nb (proposed in Ref.~\cite{gurevich_AIP}), NbN-${\mathcal I}$-Nb, and ${\rm Nb_3 Sn}$-${\mathcal I}$-Nb systems. 
The abscissa and the ordinate represent $d_{\mathcal{I}}$ and $d_{\mathcal{S}}$, respectively. 
As seen in the contour plots, 
a large $d_{\mathcal{I}} \gtrsim {\mathcal O}(10^2)\,{\rm nm}$ leads to a reduction of the field limit. 
This can be understood as follows. 
Let us recall the Maxwell equation. 
The electric field decreases even in the ${\mathcal{I}}$ layer. 
As $d_{\mathcal{I}}$ increases, 
the electromagnetic field at the interface of the substrate decreases, 
and the surface current on the substrate also. 
This means the counterflow due to the substrate decreases, 
and the magnetic field attenuation in the ${\mathcal{S}}$ layer is promoted. 
This effect is self-consistently and automatically reflected to the solution of the Maxwell equation as seen in Eqs.~(\ref{eq:B1})-(\ref{eq:j0_finite_dI}). 
The rapid field attenuation in the ${\mathcal{S}}$ layer means the enhancement of the surface current density, 
which suppresses the field limit of the ${\mathcal{S}}$ layer. 
An extreme example is an S-I-S structure with $d_{\mathcal{I}}\to\infty$, 
which corresponds to an isolated thin film with a field applied on one side. 
Its field limit is strongly suppressed by a large current density due to the lack of the counterflow generated by the substrate. 
Aside from the above viewpoints, 
a small $d_{\mathcal{I}}$ is desirable taking into account the dielectric loss and the low thermal conductivity of the ${\mathcal{I}}$ layer. 
The dielectric loss is discussed in Section~\ref{section:SISresistance}.

The optimum conditions to maximize $B_{\rm max}$ are derived in much the same way as before and given by~\cite{gurevich_AIP, kubo_SRF2015}
\begin{eqnarray}
\fl
\lambda_1 > \lambda_2 + d_{\mathcal I} , \hspace{1cm}
d_{\mathcal I}\lesssim {\mathcal O}(10)\,{\rm nm} , \label{eq:multilayer_QC_opt_lambda_finite_dI}
\\
\fl
d_{\mathcal S} 
= \lambda_1 \log \biggl[ \frac{\lambda_1}{\lambda_1+\lambda_2+d_{\mathcal I}} \frac{B_{s}^{(\mathcal{S})}}{B_{\rm max}^{\rm (sub)}} + \sqrt{ \Bigl( \frac{\lambda_1}{\lambda_1+\lambda_2+d_{\mathcal I}} \frac{B_{s}^{(\mathcal{S})}}{B_{\rm max}^{\rm (sub)}} \Bigr)^2+\frac{\lambda_1-\lambda_2-d_{\mathcal I}}{\lambda_1+\lambda_2+d_{\mathcal I}} }
\,\,\Biggr] .
\label{eq:multilayer_QC_opt_ds_finite_dI}
\end{eqnarray}
The optimized $B_{\rm max}$ is given by~\cite{gurevich_AIP, kubo_SRF2015} 
\begin{eqnarray}
B_{\rm max}^{\rm opt}
= \sqrt{ (B_{s}^{(\mathcal{S})})^2 + \biggl[ 1-\biggl(\frac{\lambda_2 + d_{\mathcal I}}{\lambda_1} \biggr)^2 \biggr] (B_{\rm max}^{\rm (sub)})^2 } .
\label{eq:multilayer_QC_opt_Bmax_finite_dI}
\end{eqnarray}
When $d_{\mathcal I} \ll \lambda_2$, 
these formulae are reduced to Eqs.~(\ref{eq:multilayer_QC_opt_lambda})-(\ref{eq:Bmaxopt_multi_QC}).

%

%%%%%%%%%%%%%%%%%%
%%%%%%%%%%%%%%%%%%
\subsection{Incorporate effect of defects}
%%%%%%%%%%%%%%%%%%
%%%%%%%%%%%%%%%%%%

According to studies on surface topographies~\cite{xu, roach}, 
the material surface are covered by multi-scale structures characterized by the fractal nature~\cite{avnir, takayasu}. 
Nano-scale defects almost continuously exists on the surface, 
and $B_{s}$ is reduced at each defect (see Ref.~\cite{kubo_PTEP_nano} for example). 
Furthermore, precipitates or variation of chemical composition also reduce $B_{s}$. 
Then $B_{s}$ of the real surface is effectively reduced to $\eta B_{s}$, 
where $\eta$ is a suppression factor that contains effects of surface defects.

\begin{figure}[tb]
   \begin{center}
   \includegraphics[width=1\linewidth]{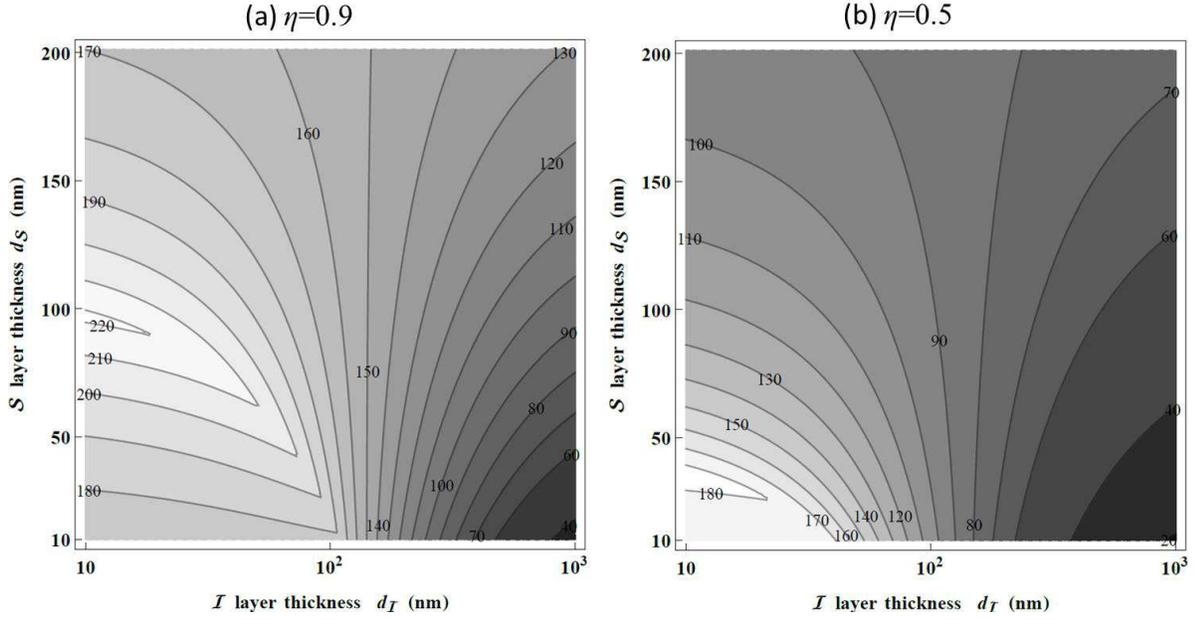}
   \end{center}\vspace{0cm}
   \caption{
$B_{\rm max}$ of dirty Nb-${\mathcal I}$-Nb system in unit of mT. 
Assumed parameters are $B_c^{\rm (Nb)}=200\,{\rm mT}$, $\lambda_1= 180\,{\rm nm}$, and (a) $\eta=0.9$ and (b) $\eta=0.5$ for the ${\mathcal S}$ layer material; 
$B_{\rm max}^{\rm (sub)}=170\,{\rm mT}$ and $\lambda_2=40\,{\rm nm}$ for the substrate. 
See also Ref.~\cite{kubo_SRF2015}. 
   }\label{fig12}
\end{figure}

\begin{figure}[tb]
   \begin{center}
   \includegraphics[width=1\linewidth]{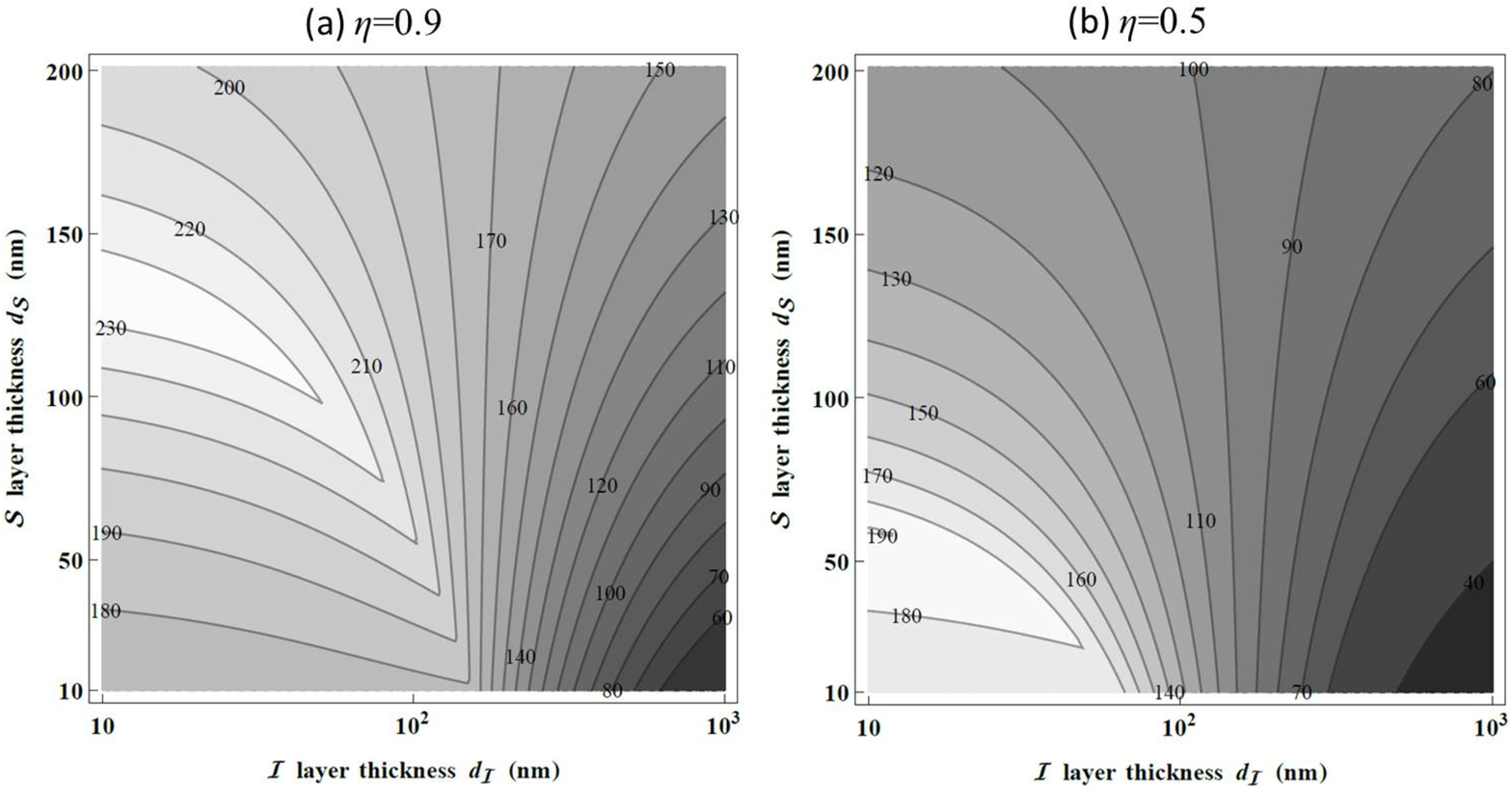}
   \end{center}\vspace{0cm}
   \caption{
$B_{\rm max}$ of NbN-${\mathcal I}$-Nb system in unit of mT. 
Assumed parameters are $B_c^{\rm (NbN)}=230\,{\rm mT}$, $\lambda_1= 200\,{\rm nm}$, and (a) $\eta=0.9$ and (b) $\eta=0.5$ for the ${\mathcal S}$ layer material; 
$B_{\rm max}^{\rm (sub)}=170\,{\rm mT}$ and $\lambda_2=40\,{\rm nm}$ for the substrate. 
See also Ref.~\cite{kubo_SRF2015}. 
   }\label{fig13}
\end{figure}

\begin{figure}[tb]
   \begin{center}
   \includegraphics[width=1\linewidth]{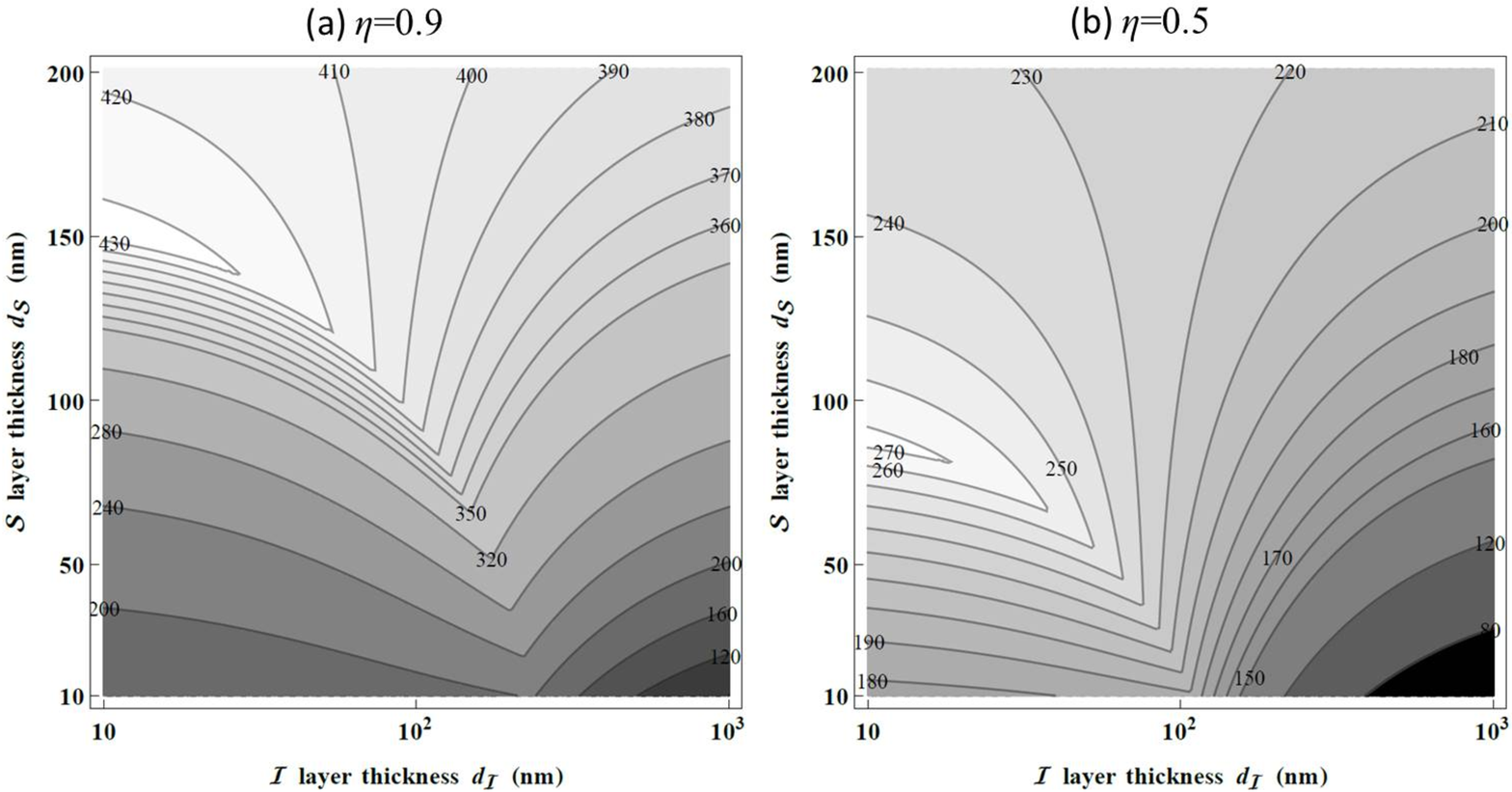}
   \end{center}\vspace{0cm}
   \caption{
$B_{\rm max}$ of ${\rm Nb_3 Sn}$-${\mathcal I}$-Nb system in unit of mT. 
Assumed parameters are $B_c^{\rm (Nb_3 Sn)}=540\,{\rm mT}$, $\lambda_1= 120\,{\rm nm}$, and (a) $\eta=0.9$ and (b) $\eta=0.5$ for the ${\mathcal S}$ layer material; 
$B_{\rm max}^{\rm (sub)}=170\,{\rm mT}$ and $\lambda_2=40\,{\rm nm}$ for the substrate. 
See also Ref.~\cite{kubo_SRF2015}. 
   }\label{fig14}
\end{figure}

In the context of the multilayer superconductor, 
the superheating field of the ${\mathcal S}$ layer would be reduced to $\eta B_{s}^{\mathcal (S)}$. 
This does not affect the field and current distributions: 
the field distribution is given by Eqs.~(\ref{eq:B1})-(\ref{eq:tildegamma2}), 
and the surface current is by Eq.~(\ref{eq:j0_finite_dI}). 
Then the field limit can be derived by replacing $B_{s}^{\mathcal (S)}$ by $\eta B_{s}^{\mathcal (S)}$: 
\begin{eqnarray}
B_{\rm max}={\rm min} \{ \widetilde{\gamma_1}^{-1} \eta B_{s}^{\mathcal (S)}, \widetilde{\gamma_2}^{-1} B_{\rm max}^{\rm (sub)} \}. 
\label{eq:Bmax_finite_dI_defect}
\end{eqnarray}
The optimum conditions and the optimized $B_{\rm max}$ are given by~\cite{gurevich_AIP, kubo_SRF2015}
\begin{eqnarray}
\fl
\lambda_1 > \lambda_2 + d_{\mathcal I}, \hspace{0.7cm} 
d_{\mathcal I}\lesssim {\mathcal O}(10)\,{\rm nm}, 
\label{eq:multilayer_condition_summary}\\
\fl
d_{\mathcal S} 
= \lambda_1 \log \biggl[ \frac{\lambda_1}{\lambda_1+\lambda_2+d_{\mathcal I}} \frac{\eta B_{s}^{(\mathcal{S})}}{B_{\rm max}^{\rm (sub)}} + \sqrt{ \Bigl( \frac{\lambda_1}{\lambda_1+\lambda_2+d_{\mathcal I}} \frac{\eta B_{s}^{(\mathcal{S})}}{B_{\rm max}^{\rm (sub)}} \Bigr)^2 +\frac{\lambda_1-\lambda_2-d_{\mathcal I}}{\lambda_1+\lambda_2+d_{\mathcal I}} }
\,\,\Biggr], 
\label{eq:multilayer_QC_opt_ds_summary}\\
\fl
B_{\rm max}^{\rm opt}
= \sqrt{ (\eta B_{s}^{(\mathcal{S})})^2 + \biggl[ 1-\biggl(\frac{\lambda_2 + d_{\mathcal I}}{\lambda_1} \biggr)^2 \biggr] (B_{\rm max}^{\rm (sub)})^2 } .
\label{eq:multilayer_QC_opt_Bmax_summary}
\end{eqnarray}
Assuming some concrete values of $\eta$, 
we can make the similar contour plots as in the last subsection. 
Figs.~\ref{fig12}, \ref{fig13}, and \ref{fig14} show contour plots of $B_{\rm max}$ for the cases of $\eta=0.9$ and $\eta=0.5$. 
Note here the optimum ${\mathcal S}$ layer thickness decreases as $\eta$ decreases (see Figs.~\ref{fig9} and \ref{fig12}, Figs.~\ref{fig10} and \ref{fig13}, and Figs.~\ref{fig11} and \ref{fig14}). 
This can be understood as follows. 
As $\eta$ decreases, the field limit of the ${\mathcal S}$ layer decreases. 
The decreased field limit can be compensated by suppressing the surface current, 
which is possible by reducing $d_{\mathcal S}$. 
However, a complete compensation leads to a too thin $d_{\mathcal S}$ to protect the substrate. 
As a result, the optimum $d_{\mathcal S}$ falls into a moderately reduced value that can partially compensate the decreased field limit.

%

%%%%%%%%%%%%%%%%%%
%%%%%%%%%%%%%%%%%%
\subsection{Surface resistance of multilayer superconductor}\label{section:SISresistance}
%%%%%%%%%%%%%%%%%%
%%%%%%%%%%%%%%%%%%

The surface resistance of the S-I-S structure can be obtained by calculating the total joule dissipation~\cite{gurevich_AIP}, 
which is given by (see \ref{appendix_electromagnetic_multi})
\begin{eqnarray}
R_s
= 2\lambda_1 \frac{\mu_0^2}{B_0^2} R_s^{\mathcal (S)} \int_0^{d_{\mathcal S}} \!\!\!\! J^2 dx
+ 2\lambda_2 \frac{\mu_0^2}{B_0^2} R_s^{\rm (sub)} \int_{d_{\mathcal S}+d_{\mathcal I}}^{\infty} \!\!\!\!\! J^2 dx
+ \frac{2\mu_0^2}{B_0^2} p_{\mathcal I} , 
\label{eq:SISresistance}
\end{eqnarray}
where $J$ is the screening current distribution derived from the London equation, 
$R_s^{\mathcal (S)}$ is the surface resistance of the semi-infinite superconductor made of the ${\mathcal S}$ layer material, 
$R_s^{\rm (sub)}$ is the surface resistance of the semi-infinite superconductor made of the substrate material, 
and $p_{\mathcal I}$ is the dielectric loss. 
The evaluation of Eq.~(\ref{eq:SISresistance}) is straightforward~\cite{gurevich_AIP}: 
\begin{eqnarray}
R_s
= \biggl[ \frac{1+r_{\lambda}^2}{2} \sinh\frac{2d_{\mathcal S}}{\lambda_1} + r_{\lambda} \biggl( \cosh\frac{2d_{\mathcal S}}{\lambda_1} -1 \biggr) - (1-r_{\lambda}^2)\frac{d_{\mathcal S}}{\lambda_1} \biggr] \widetilde{\gamma}_2^2 R_s^{\mathcal (S)} \nonumber \\
+ \widetilde{\gamma}_2^2 R_s^{\rm (sub)} 
+ \widetilde{\gamma}_2^2 \mu_0^2 \omega^3 \epsilon'' \lambda_2^2 d_{\mathcal I} ,
\label{eq:SISresistance2}
\end{eqnarray}
where $r_{\lambda}\equiv (\lambda_2+d_{\mathcal I})/\lambda_1$ and we used the fact that the electric field in the ${\mathcal I}$ layer is given by $-i\omega \lambda_2 \widetilde{\gamma}_2 B_0$. 
The first, second, and third terms correspond to a contribution from the ${\mathcal S}$ layer, substrate, and ${\mathcal I}$ layer, respectively.

Let us roughly evaluate the third term, the dielectric loss contribution. 
Substituting $\widetilde{\gamma}_2 \sim 1$, $\omega\sim 10^{10}\,{\rm s^{-1}}$, $\epsilon'' < \epsilon_0$, $\lambda_2 \sim 10^{-7}\,{\rm m}$, 
we find it is smaller than $(d_{\mathcal I}/{\rm nm})\times 10^{-7}\,{\rm n\Omega}$. 
For example, when $d_{\mathcal I}=100\,{\rm nm}$, the dielectric loss contribution is given by $< 10^{-5}\,{\rm n\Omega}$ and is negligible. 
This smallness can be understood by reminding that the electric field in the ${\mathcal I}$ layer is given by $|E| = \omega \lambda_2 \widetilde{\gamma}_2 B_0 \sim 10^{-5}\,{\rm V/m}$ for $B_0=10\,{\rm mT}$, 
which is much smaller than that of the plane wave in the vacuum $|E| \sim c B_0 \sim 1\,{\rm MV/m}$ for the same $B_0$.

See also Ref.~\cite{iwashita_normal, iwashita_normal_LINAC2010} for the multilayer normal conductor (N-I-N structure), 
where a reduction of power loss of a normal conducting RF cavity by using the N-I-N structure is proven theoretically and experimentally.

%%%%%%%%%%%%%%%%%%
%%%%%%%%%%%%%%%%%%
\subsection{Summary of Section~\ref{section:multilayer}}
\label{subsection:summary_SIS}
%%%%%%%%%%%%%%%%%%
%%%%%%%%%%%%%%%%%%

Let us summarize the main results of this section. 
\begin{enumerate}
\item We started with an investigation of the S-I-S structure with the ideal surface and a negligibly thin ${\mathcal I}$ layer in the framework of the London theory. 
Typical field and current distributions in the S-I-S structure are given by Fig.~\ref{fig7}. 
The field limit is given by Eq.~(\ref{eq:multilayer_max_London}). 
The optimum conditions to maximize the field limit and the optimized field limit are given by Eqs.~(\ref{eq:multilayer_condition}), (\ref{eq:multilayer_London_opt_ds}) and (\ref{eq:multilayer_London_opt_Bmax}).  
\item The same system was examined in the GL theory, which is valid only at $T\simeq T_c$. 
The optimized field limit is given by Eq.~(\ref{eq:Bmaxopt_multi_GL_3}) when the coherence length of the ${\mathcal S}$ layer is smaller than that of the substrate. 
\item At $0< T< T_c$, the field limit is given by Eq.~(\ref{eq:Bmax_QC}), and the optimum conditions and the optimized field limit are given by Eqs.~(\ref{eq:multilayer_QC_opt_lambda})-(\ref{eq:Bmaxopt_multi_QC}), 
which are expressed by using the superheating field derived in the quasiclassical theory. 
\item In much the same way, a generalized model with a finite $d_{\mathcal I}$ was studied. 
The field limit is given by Eq.~(\ref{eq:Bmax_finite_dI}), 
and the optimum conditions and the optimized field limit are given by Eqs.~(\ref{eq:multilayer_QC_opt_lambda_finite_dI})-(\ref{eq:multilayer_QC_opt_Bmax_finite_dI}), 
which depend on $d_{\mathcal I}$ (see also Figs.~\ref{fig9}-\ref{fig11}). 
\item Furthermore, effects of material and topographic defects were incorporated. 
The field limit is given by Eq.~(\ref{eq:Bmax_finite_dI_defect}),
and the optimum conditions and the optimized field limit are given by Eqs.~(\ref{eq:multilayer_condition_summary})-(\ref{eq:multilayer_QC_opt_Bmax_summary}), 
where the superheating field of the ${\mathcal S}$ layer material is reduced by a factor $\eta$ (see also Figs.~\ref{fig12}-\ref{fig14}). 
These are the most general formulae, 
which can be applied to the S-I-S structure with surface defects and a finite $d_{\mathcal I}$ under an arbitrary temperature $0< T< T_c$. 
\item Finally, the surface resistance formula was derived. See Eq.~(\ref{eq:SISresistance2}). 
\end{enumerate}
%

%%%%%%%%%%%%%%%%%%%%%%%%%%%%%%%%%%%%
%%%%%%%%%%%%%%%%%%%%%%%%%%%%%%%%%%%%
%%%%%%%%%%%%%%%%%%%%%%%%%%%%%%%%%%%%
%%%%%%%%%%%%%%%%%%%%%%%%%%%%%%%%%%%%
%%%%%%%%%%%%%%%%%%%%%%%%%%%%%%%%%%%%
%%%%%%%%%%%%%%%%%%%%%%%%%%%%%%%%%%%%
\section{Multilayer superconductor without insulator layer}\label{section:multilayer_without_I}
%%%%%%%%%%%%%%%%%%%%%%%%%%%%%%%%%%%%
%%%%%%%%%%%%%%%%%%%%%%%%%%%%%%%%%%%%
%%%%%%%%%%%%%%%%%%%%%%%%%%%%%%%%%%%%
%%%%%%%%%%%%%%%%%%%%%%%%%%%%%%%%%%%%
%%%%%%%%%%%%%%%%%%%%%%%%%%%%%%%%%%%%
%%%%%%%%%%%%%%%%%%%%%%%%%%%%%%%%%%%%

As mentioned in the introduction section, 
the role of the ${\mathcal I}$ layer is to intercept propagating vortex loops and to localize vortex dissipation in the ${\mathcal S}$ layer. 
The ${\mathcal I}$ layer is essential in the multilayer approach.  
Nonetheless, the multilayer superconductor without ${\mathcal I}$ layer is also an interesting system and worth studying. 
Here we summarize the two reasons mentioned in the introduction section again:  
(i) it can be regarded as a model of the surface of baked Nb, 
in which a penetration depth decreases in the first several tens of nm from the surface due to a depth-dependent mean free path~\cite{ciovati_bake, romanenko_bake}.  
The simplest model of the baked Nb is the S-S bilayer structure (see also the discussion section of Ref.~\cite{kubo_PTEP_nano}). 
Studying this system may help our understanding on how the low temperature baking works. 
(ii) some SRF researchers have made S-S bilayer structures such as ${\rm MgB_2}$-Nb or ${\rm Nb_3 Sn}$-Nb. 
The results of the sample tests~\cite{tan_SRF2015, laxdal_TTC2016} should be understood theoretically~\cite{kubo_LINAC14}. 
In this section, we review some features of the S-S bilayer structure that have already been revealed through studies on the S-I-S structure. 

%%%%%%%%%%%%%%%%%%%%%%%%%%%%%%%%%%%%%%%%%%%%%%%%%%%%%
\subsection{Theoretical field limit} 
%%%%%%%%%%%%%%%%%%%%%%%%%%%%%%%%%%%%%%%%%%%%%%%%%%%%%

%
\begin{figure}[tb]
   \begin{center}
   \includegraphics[width=0.7\linewidth]{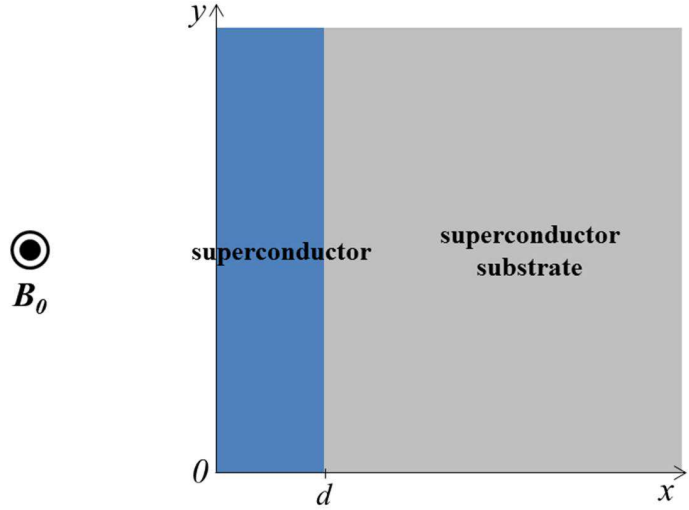}
   \end{center}\vspace{0cm}
   \caption{
Model of the S-S bilayer structure. 
The surface and boundary are parallel to the $y$-$z$ plane and then perpendicular to the $x$ axis. 
The thickness of the surface layer is given by $d$. 
The applied magnetic field is given by ${\bf B}_0 = (0,0,B_0)$. 
   }\label{fig15}
\end{figure}

\begin{figure}[tb]
   \begin{center}
   \includegraphics[width=0.6\linewidth]{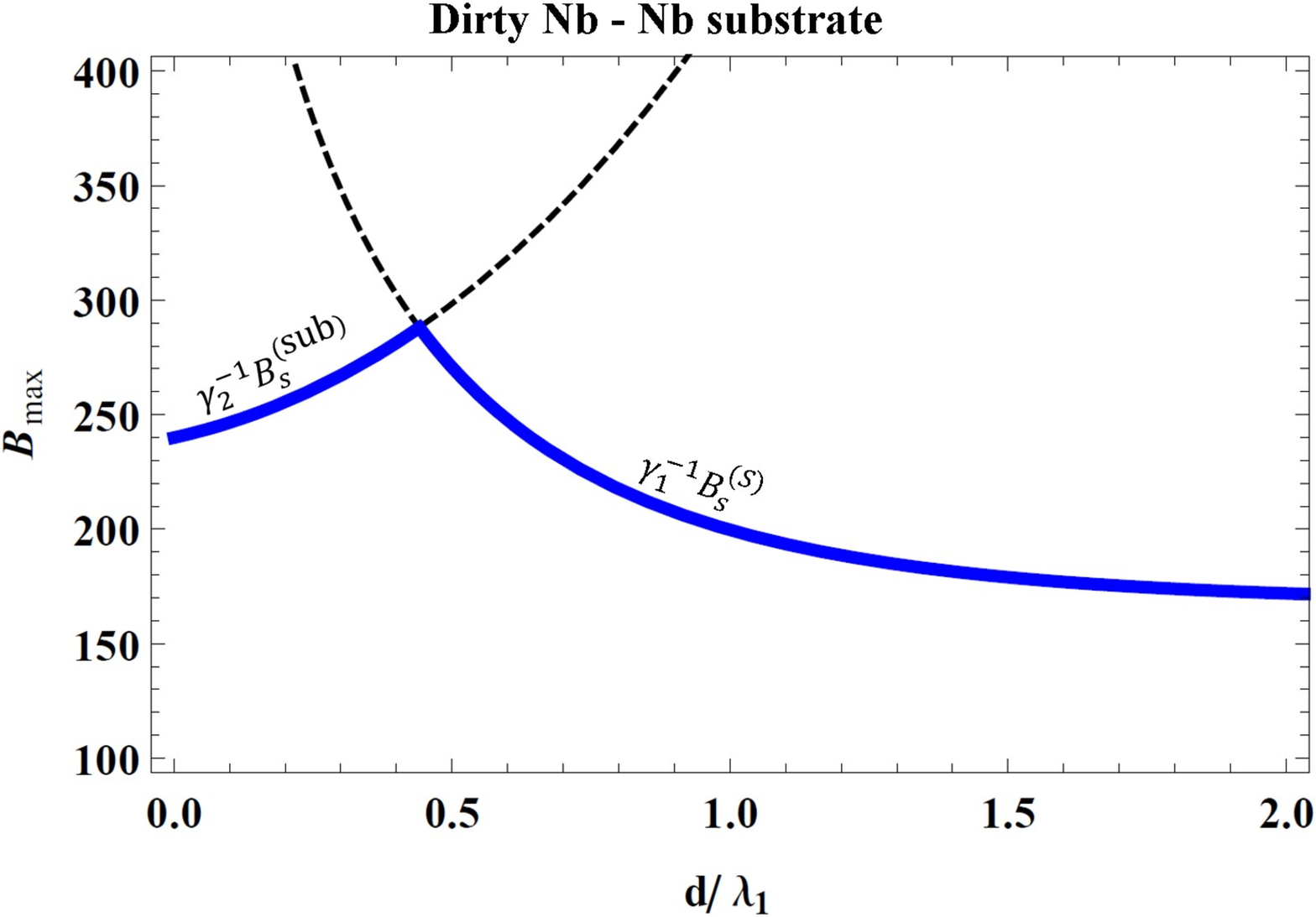}
   \end{center}\vspace{0cm}
   \caption{
Theoretical field limit of the S-S bilayer structure that consists of a dirty Nb layer and a clean Nb substrate as a function of the surface layer thickness. 
The assumed parameters are $\lambda_1=180\,{\rm nm}$, $B_s^{\mathcal (S)}=0.84 B_c^{\mathcal (S)}=170\,{\rm mT}$, $\lambda_2=40\,{\rm nm}$, and $B_s^{\rm (sub)}=240\,{\rm mT}$. 
   }\label{fig16}
\end{figure}

\begin{figure}[tb]
   \begin{center}
   \includegraphics[width=0.6\linewidth]{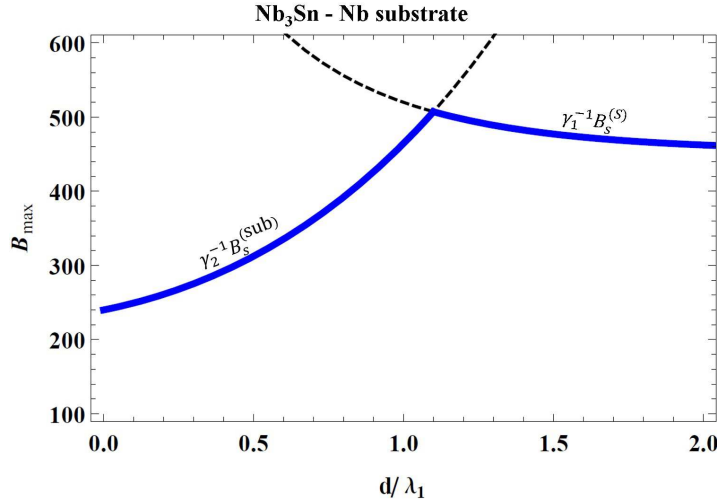}
   \end{center}\vspace{0cm}
   \caption{
Theoretical field limit of the S-S bilayer structure that consists of a ${\rm Nb_3 Sn}$ layer  and a clean Nb substrate as a function of the surface layer thickness. 
The assumed parameters are $\lambda_1=120\,{\rm nm}$, $B_s^{\mathcal (S)}=0.84 B_c^{\mathcal (S)}=450\,{\rm mT}$, $\lambda_2=40\,{\rm nm}$, and $B_s^{\rm (sub)}=240\,{\rm mT}$. 
   }\label{fig17}
\end{figure}

We consider the model shown in Fig.~\ref{fig15}. 
Materials of the surface layer and the substrate are assumed to be superconductors with $\lambda_1$ and $\lambda_2$, respectively. 
The theoretical field limit of the S-S bilayer structure~\cite{gurevich_AIP, kubo_LINAC14} can be derived by the absolutely same procedure as the S-I-S structure. 
To obtain the current and field distribution, we solve the London equation, 
$\lambda^2 A'' -A=0$, 
where $\lambda=\lambda_1$ at $0\le x\le d$ and $\lambda=\lambda_2$ at $x> d$. 
Its solution is given by the same one as Eqs.~(\ref{eq:multi_London_A1}) and (\ref{eq:multi_London_A2}). 
Then the current densities at the surface and S-S boundary can be obtained by using $J=-A/\mu_0 \lambda^2$ and are given by
\begin{eqnarray}
J(0) = \gamma_1 \frac{B_0}{\mu_0 \lambda_1} , 
\hspace{1.8cm}
\gamma_1 
\equiv \frac{  \sinh \frac{d}{\lambda_1} + \frac{\lambda_2}{\lambda_1} \cosh \frac{d}{\lambda_1}}{ \cosh \frac{d}{\lambda_1} + \frac{\lambda_2}{\lambda_1}  \sinh \frac{d}{\lambda_1}} , \\
J(d) = \gamma_2 \frac{B_0}{\mu_0 \lambda_2}, 
\hspace{2cm}
\gamma_2
\equiv \frac{1}{ \cosh \frac{d}{\lambda_1} + \frac{\lambda_2}{\lambda_1}  \sinh \frac{d}{\lambda_1}} . 
\end{eqnarray}
These current densities must be smaller than the depairing limit of the surface layer $B_{s}^{\mathcal (S)}/\mu_0 \lambda_1$ and that of the substrate $B_{s}^{\rm (sub)}/\mu_0 \lambda_2$, respectively, 
where $B_{s}^{\mathcal (S)}$ and $B_{s}^{\rm (sub)}$ are the superheating fields of the surface and substrate material for an arbitrary temperature derived by using the quasiclassical theory. 
Then we have~\cite{kubo_APL} 
\begin{eqnarray}
B_{\rm max}= {\rm min} \{ \gamma_1^{-1} B_s^{\mathcal (S)}, \gamma_2^{-1} B_{s}^{\rm (sub)} \}, 
\label{eq:Bmax_SS}
\end{eqnarray}
Note that $\gamma_i$ ($i=1,2$) are functions of $d$ as shown in Fig.~\ref{fig7}, 
and then $B_{\rm max}$ is also a function of $d$. 
The optimization of $d$ can also be carried out in much the same way as the S-I-S structure. 
$B_{\rm max}$ is maximized when $\gamma_1^{-1} B_{s}^{\mathcal (S)}= \gamma_2^{-1} B_{s}^{\rm (sub)}$, 
and finally we obtain the optimum conditions to maximize the field limit~\cite{kubo_APL, gurevich_AIP}, 
\begin{eqnarray}
\lambda_1 > \lambda_2, 
\label{eq:lambda_opt_SS} \\
d_{\mathcal S} 
= \lambda_1 \log \biggl[ \frac{\lambda_1}{\lambda_1+\lambda_2} \frac{B_{s}^{(\mathcal{S})}}{B_{s}^{\rm (sub)}} + \sqrt{ \Bigl( \frac{\lambda_1}{\lambda_1+\lambda_2} \frac{B_{s}^{(\mathcal{S})}}{B_{s}^{\rm (sub)}} \Bigr)^2 +\frac{\lambda_1-\lambda_2}{\lambda_1+\lambda_2} }
\,\,\Biggr] . 
\label{eq:dS_opt_SS}
\end{eqnarray}
The optimized field limit is given by~\cite{gurevich_AIP} 
\begin{eqnarray}
B_{\rm max}^{\rm opt} = 
\sqrt{  ( B_{s}^{({\mathcal S})})^2 
+ \biggl( 1-\frac{\lambda_2^2}{\lambda_1^2} \biggr)  (B_{s}^{\rm (sub)})^2 } , 
\label{eq:Bmax_opt_SS}
\end{eqnarray}
which is the same one as the S-I-S structure with negligibly thin $d_{\mathcal I}$ [see Eq.~(\ref{eq:Bmaxopt_multi_QC})]. 
Figs.~\ref{fig16} and \ref{fig17} show examples of $B_{\rm max}$ as functions of $d$. 
The peak values correspond to $B_{\rm max}^{\rm opt}$.


It should be noted that, even if the theoretical field limit is high, 
such a high field cannot be necessarily achieved actually. 
As mentioned in the introduction section, 
the Meissner state ceases to be stable at $B_0>B_{c1}$ (see Fig.~\ref{fig1}). 
While the surface barrier still protects the material against penetration of vortices, 
taking into account the surface barrier is reduced at material and topographic defects that cover the cavity surface, 
achieving a field much higher than $B_{c1}$ would not be easy without an additional mechanism to stabilize the Meissner state. 
In the S-I-S structure, 
the stability of the Meissner state at $B_0>B_{c1}$ is assured by the existence of the ${\mathcal I}$ layer, 
which stops penetration of vortices and suppresses vortex dissipation. 
In the S-S bilayer structure, however, the ${\mathcal I}$ layer is absent: 
we have only the S-S boundary. 
Is there any mechanism to stabilize the Meissner state in the S-S bilayer structure? 
Our next task is to examine a role of the S-S boundary.

%%%%%%%%%%%%%%%%%%%%%%%%%%%%%%%%%%%%%%%%%%%%%%%%%%%%%
\subsection{Interaction between a vortex and the S-S boundary} 
%%%%%%%%%%%%%%%%%%%%%%%%%%%%%%%%%%%%%%%%%%%%%%%%%%%%%

%%%%%%%%%%%%%
\subsubsection{Infinite superconductor with two regions} 
%%%%%%%%%%%%%

As an instructive exercise, first we consider an infinite superconductor that consists of two regions, $x<0$ with $\lambda=\lambda_1$ and $x\ge 0$ with $\lambda=\lambda_2$. 
We examine the interaction between a vortex and the boundary.  
Suppose there exists a vortex parallel to ${\bf \hat{z}}$ at $x=x_0=-|x_0|$, 
where $|x_0|$ is assumed to be smaller than $\lambda_1$ and $\lambda_2$ for simplicity. 
The force acting on the vortex can be evaluated by the method of images as usual. 
By using an analogy with an line charge embedded in a infinite dielectric with two regions, 
we find the current distribution for $x<0$ can be expressed by the superposition of the current circulating the vortex at $x=-|x_0|$ and an image vortex with flux $\phi_1=\tau \phi_0$ at $x=+|x_0|$, 
and the current distribution for $x>0$ can be expressed by an image vortex with $\phi_1' =\tau' \phi_0$ at $x=-|x_0|$. 
Imposing the continuity conditions of $j_x$ and $A_y$ at the boundary, 
we find~\cite{kubo_LINAC14} 
\begin{eqnarray}
\tau = \frac{\lambda_1^2-\lambda_2^2}{\lambda_1^2 + \lambda_2^2} , 
\hspace{1cm}
\tau'= 1-\tau . 
\label{eq:tau}
\end{eqnarray}
Then the force acting on the vortex ${\bf f}_{\rm B}$ is given by~\cite{kubo_LINAC14}
\begin{eqnarray}
{\bf f}_{\rm B} 
= {\bf j}_{\rm img} \times \phi_0 {\bf \hat{z}} 
= -\frac{\phi_0 \phi_1}{4\pi\mu_0\lambda_1^2 |x_0|} {\bf \hat{x}} , 
\label{eq:fB_infinite_two}
\end{eqnarray}
where ${\bf j}_{\rm img}$ is the current circulating the image vortex with flux $\phi_1$ at $x=|x_0|$. 
Thus the S-S boundary pushes the vortex to the direction of the material with larger penetration depth. 
Note that, instead of using the method of images, 
we can directly solve the London equation and obtain the same result as the above (see \ref{appendix_infinite_two}).

%%%%%%%%%%%%%%%%%%%%%%%%%%%%%%%%%%%%%%%%%%%%%%%%%%%%%
\subsubsection{Thin superconductor layer on a superconductor substrate} 
%%%%%%%%%%%%%%%%%%%%%%%%%%%%%%%%%%%%%%%%%%%%%%%%%%%%%

%
\begin{figure}[tb]
   \begin{center}
   \includegraphics[width=1\linewidth]{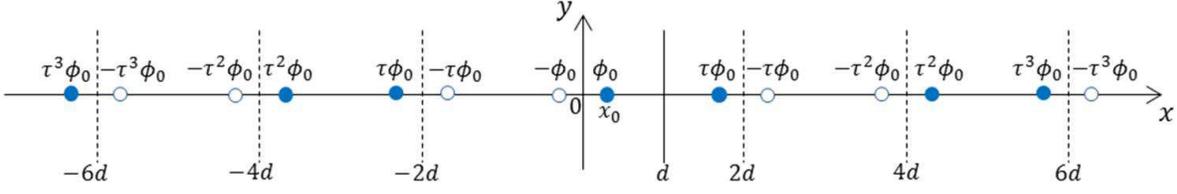}
   \end{center}\vspace{0cm}
   \caption{
Image vortices necessary for satisfying the boundary conditions at $x=0$ and $x=d$. 
The factor $\tau$ is given by Eq.~(\ref{eq:tau}). 
   }\label{fig18}
\end{figure}

Now we go back to the system shown in Fig.~\ref{fig15}. 
Suppose there exists a vortex parallel to ${\bf \hat{z}}$ at $x=x_0$ inside the surface layer. 
The easiest way to evaluate the force acting on the vortex is to use the method of images. 
In order to satisfy the boundary conditions at $x=0$ and $x=d$, 
an infinite number of image vortices are necessary in common with the multilayer suparconductor. 
We need 
(i) an antivortex at $x=-x_0$ to satisfy the condition at $x=0$, 
(ii) a vortex with flux $\tau \phi_0$ at $x=2d-x_0$ and an antivortex with flux $\tau \phi_0$ at $x=2d+x_0$ to satisfy the condition at $x=d$, 
which violate the condition at $x=0$, 
(iii) an antivortex with flux $\tau \phi_0$ at $x=-2d+x_0$ and a vortex with flux $\tau \phi_0$ at $x=-2d-x_0$ to satisfy the condition at $x=0$ again, 
which violate the condition at $x=d$, 
(iv) an antivortex with flux $\tau^2 \phi_0$ at $x=4d-x_0$ and a vortex with flux $\tau^2 \phi_0$ at $x=4d+x_0$ to satisfy the condition at $x=d$, and so on.
Finally an infinite number of image vortices are introduced (see Fig.~\ref{fig18}). 
The total force is given by (see \ref{appendix_summation_nano_layer})
\begin{eqnarray}
\fl
{\bf f}_{\rm B}
=  \frac{\phi_0^2}{4\pi\mu_0 \lambda_1^2} 
\biggl[ -\frac{1}{x_0}+\sum_{n=1}^{\infty} (-1)^n \tau^n \biggl( \frac{1}{n d-x_0} -\frac{1}{n d+x_0} \biggr) \biggr] {\bf \hat{x}}
\nonumber \\
\fl
= -\frac{\phi_0^2}{4 \pi \mu_0 \lambda_1^2 d} \biggl[ \frac{d}{x_0} F\Bigl(1,\frac{x_0}{d};1+\frac{x_0}{d};-\tau\Bigr) + \frac{\tau}{1-\frac{x_0}{d}} F\Bigl(1,1-\frac{x_0}{d};2-\frac{x_0}{d};-\tau\Bigr) \biggr] {\bf \hat{x}},  
\label{eq:infinite_fI_bi}
\end{eqnarray}
where $F(a,b;c;z)=[\Gamma(c)/\Gamma(b)\Gamma(c-b)] \int_0^1 dt (1-tz)^{-a} t^{b-1} (1-t)^{c-b-1}$ is the Gaussian hypergeometric function. 
Note that Eq.~(\ref{eq:infinite_fI_bi}) is reduced to Eq.~(\ref{eq:fB_semi_infinite}) as $x_0 \to 0$ and to Eq.~(\ref{eq:fB_infinite_two}) as $x_0\to d$. 
The same result can be obtained by directly solving the London equation (see \ref{appendix_nano_layer}).

\begin{figure}[tb]
   \begin{center}
   \includegraphics[width=0.7\linewidth]{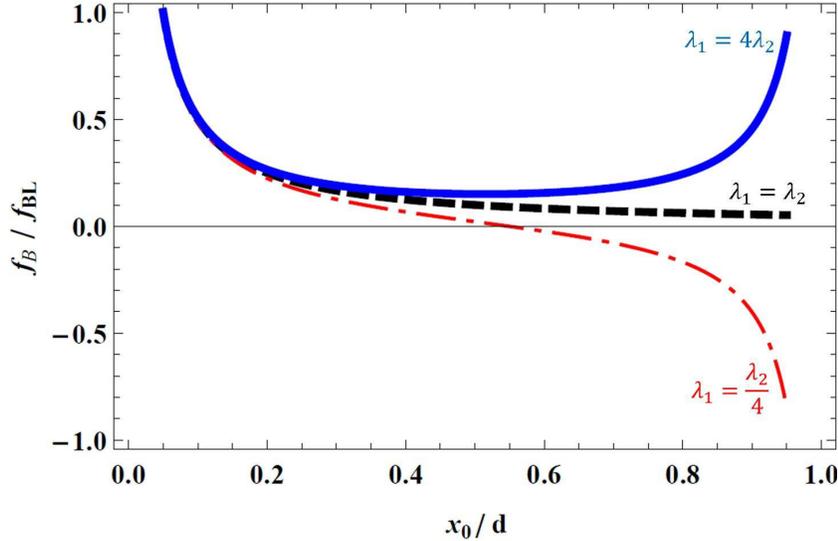}
   \end{center}\vspace{0cm}
   \caption{
The force acting on the vortex inside the surface layer calculated by using Eq.~(\ref{eq:infinite_fI_bi}). 
The sign is positive when the force directs the surface and then acts as a barrier.  
The short distance cutoff is assumed to be $d/20$. 
The force in the vicinity of $x_0\simeq 0$ corresponds to that of the Bean-Livingston barrier. 
The force near $x_0\simeq d$ is due to the S-S boundary, 
which is absent in a simple semi-infinite superconductor ($\lambda_1=\lambda_2$). 
When $\lambda_1>\lambda_2$, 
the force due to the S-S boundary acts as a barrier against penetration of vortices. 
   }\label{fig19}
\end{figure}

Fig.~\ref{fig19} shows $f_{\rm B}$ in unit of $f_{\rm BL}$ as functions of the vortex position $x_0/d$, 
where $f_{\rm BL}\equiv -\phi_0^2/4\pi \mu_0 \lambda_1^2 \xi_1$. 
Note that the sign of $f_{\rm B}/f_{\rm BL}$ is positive when it directs the surface and then acts as a barrier. 
When $\lambda_1=\lambda_2$, the present system is reduced to a simple semi-infinite superconductor, and only the Bean-Livingston barrier exists, 
which attenuates as $x_0$ increases (see the black dashed curve). 
On the other hand, when $\lambda_1 \ne \lambda_2$, 
the vortex feels not only the Bean-Livingston barrier but also the force due to the S-S boundary (see the blue solid curve and red dashed-dotted curve). 
In particular, when $\lambda_1 > \lambda_2$, 
the force due to the S-S boundary acts as a barrier to prevent penetration of vortices~\cite{kubo_LINAC14}.

As seen in the above, the S-S bilayer structure is protected by the double barriers: 
the Bean-Livingstone barrier and the barrier due to the S-S boundary. 
Both the barriers can be reduced by defects and have weak spots, 
but a vortex that penetrates from a weak spot of the Bean-Livingstone barrier may be stopped by the S-S boundary: 
there is a second chance to stop the vortex. 
While the S-S boundary is not as robust as the ${\mathcal I}$ layer in the S-I-S structure, 
it is also expected to contribute to preventing penetration of vortices. 
The low temperature baking~\cite{kako_bake, kneisel_40, ono_bake, lilje_bake} transforms the Nb surface from an simple semi-infinite clean Nb to a layered structure with $\lambda_1 > \lambda_2$ that consists of a dirty Nb layer and a clean Nb substrate~\cite{ciovati_bake, romanenko_bake},  
where the boundary of dirty and clean Nb plays a role of barrier and may be related to the cure of the high field Q drop~\cite{kubo_PTEP_nano} together with other factors that would significantly affect SRF performances at a high field such as the difference of the density of states between the dirty and clean Nb~\cite{gurevich_review}. 
The same would be true for the modified low temperature baking~\cite{grassellino_bake}. 
The S-S boundary in ${\rm MgB_2}$-Nb or ${\rm Nb_3 Sn}$-Nb also satisfies $\lambda_1 > \lambda_2$ and plays a role of barrier against penetration of vortices.

It should be noted that the ${\mathcal I}$ layer in the S-I-S structure plays a role not only in stopping penetration of vortices but also in suppressing vortex dissipation, 
because the dissipative vortex core disappears in the ${\mathcal I}$ layer. 
On the other hand, in the S-S bilayer structure,  
the double barrier would contribute to stopping vortex penetration, 
but the dissipative vortex core is conserved in contrast to the S-I-S structure: 
the whole length of an oscillating vortex inside the surface layer contributes to dissipation.

%%%%%%%%%%%%%%%%%%%%%%%%%%%%%%%%%%%%%%%%%%%%%%%%%%%%%
\subsection{Surface resistance of the S-S bilayer structure} 
%%%%%%%%%%%%%%%%%%%%%%%%%%%%%%%%%%%%%%%%%%%%%%%%%%%%%

%
\begin{figure}[tb]
   \begin{center}
   \includegraphics[width=0.6\linewidth]{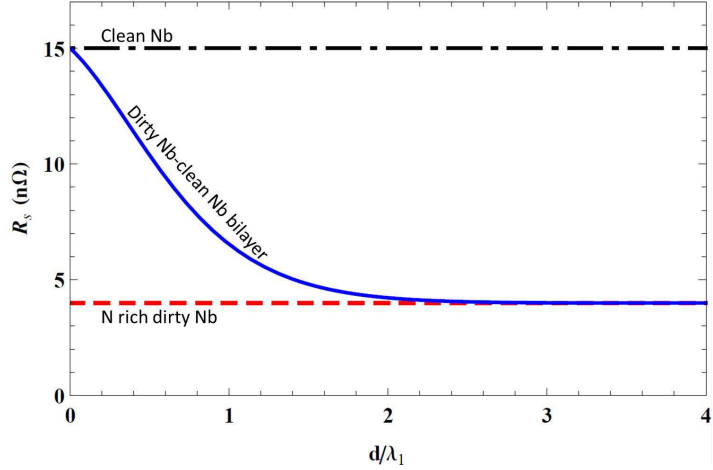}
   \end{center}\vspace{0cm}
   \caption{
Surface resistance of a bilayer structure that consists of a nitrogen rich dirty Nb layer and a clean Nb substrate as a function of the surface layer thickness. 
The assumed parameters are $\lambda_1=180\,{\rm nm}$, 
$\lambda_2=40\,{\rm nm}$, 
$R_s^{\mathcal (S)} = 4\,{\rm n\Omega}$, 
and $R_s^{\rm (sub)} = 15\,{\rm n\Omega}$. 
   }\label{fig20}
\end{figure}
\begin{figure}[tb]
   \begin{center}
   \includegraphics[width=0.6\linewidth]{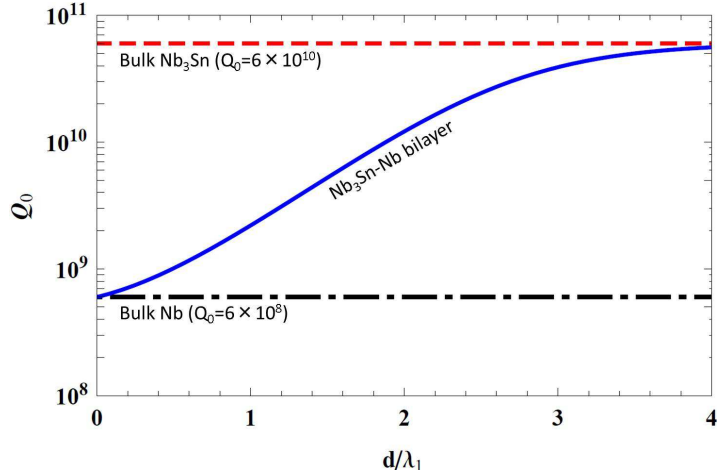}
   \end{center}\vspace{0cm}
   \caption{
Quality factor of a ${\rm Nb_3 Sn}$ cavity at $T=4.2\,{\rm K}$ as a function of ${\rm Nb_3 Sn}$ thickness. 
The assumed parameters are $\lambda_2/\lambda_1=1/3$, 
$R_s^{\mathcal (S)}= 4.5\,{\rm n\Omega}$, 
and $R_s^{\rm (sub)} = 450\,{\rm n\Omega}$. 
   }\label{fig21}
\end{figure}

The surface resistance of the S-S bilayer structure can be derived in much the same way as the S-I-S structure~\cite{gurevich_AIP} (see also \ref{appendix_electromagnetic_multi}). 
\begin{eqnarray}
\fl
R_s
= 2\lambda_1 \frac{\mu_0^2}{B_0^2} R_s^{\mathcal (S)} \int_0^d \!\!\!\! J(x)^2 dx
+ 2\lambda_2 \frac{\mu_0^2}{B_0^2} R_s^{\rm (sub)} \int_d^{\infty} \!\!\!\!\! J(x)^2 dx \nonumber \\
\fl
= \biggl[ \frac{1+(\frac{\lambda_2}{\lambda_1})^2}{2} \sinh\frac{2d}{\lambda_1} + \frac{\lambda_2}{\lambda_1} \biggl( \cosh\frac{2d}{\lambda_1} -1 \biggr) - \biggl\{ 1-\biggl(\frac{\lambda_2}{\lambda_1} \biggr)^2 \biggr\} \frac{d}{\lambda_1} \biggr] \gamma_2^2 R_s^{\mathcal (S)} 
+ \gamma_2^2 R_s^{\rm (sub)}  ,
\label{eq:SS_bilayer_resistance}
\end{eqnarray}
where $J(x)$ is the screening current density, 
$R_s^{\mathcal (S)}$ is the surface resistance of the semi-infinite superconductor made of the ${\mathcal S}$ layer material, and 
$R_s^{\rm (sub)}$ is the surface resistance of the substrate.

Fig.~\ref{fig20} shows an example of the surface resistance of a bilayer structure that consists of a nitrogen rich dirty Nb layer and a clean Nb substrate. 
When $d\to 0$, the system is reduced to a bulk clean Nb, and $R_s \to R_s^{\rm sub}$. 
Conversely, when $d \gg \lambda_1$, the system is reduced to a nitrogen rich bulk Nb, and $R_s \to R_s^{\mathcal (S)}$. 
The surface of Nb after the modified baking with $d\lesssim \lambda_1$ would have an intermediate value between the nitrogen rich bulk Nb and clean Nb.

Another example is shown in Fig.~\ref{fig21}: 
the quality factor $Q_0$ of ${\rm Nb_3 Sn}$ cavity. 
When $d \gg \lambda_1$, the cavity $Q_0$ is determined by the surface resistance of ${\rm Nb_3 Sn}$ and is larger than $10^{10}$ even at $T=4.2\,{\rm K}$~\cite{posen_Nb3Sn}. 
However, $Q_0$ rapidly decreases with $d$ and falls below $10^{10}$ at $d \simeq 2\lambda_1$ due to the large contribution from the surface resistance of the Nb substrate. 
If there is a large non-uniformity of ${\rm Nb_3 Sn}$ thickness and exists an area with $d \sim \lambda_1$, 
it can be a significant heat source and may cause a Q degradation or quench.

%%%%%%%%%%%%%%%%%%
%%%%%%%%%%%%%%%%%%
\subsection{Summary of Section~\ref{section:multilayer_without_I}}
\label{subsection:summary_SS}
%%%%%%%%%%%%%%%%%%
%%%%%%%%%%%%%%%%%%

Let us summarize the main results of this section. 
\begin{enumerate}
\item The theoretical field limit the S-S bilayer structure was examined in much the same way as the S-I-S structure. 
The field limit is given by Eq.~(\ref{eq:Bmax_SS}), which is maximized when Eqs.~(\ref{eq:lambda_opt_SS}) and (\ref{eq:dS_opt_SS}) are satisfied. 
The optimized field limit is given by Eq.~(\ref{eq:Bmax_opt_SS}). See Figs~\ref{fig16} and \ref{fig17}. 
It should be noted that, in order to achieve a theoretical field limit much higher than the lower critical field, 
a mechanism to stabilize the Meissner state, such as the ${\mathcal I}$ layer of the S-I-S structure, is necessary. 
\item The interaction among a vortex, the surface and the S-S boundary was examined. 
The force acting on a vortex inside the surface layer is given by Eq.~(\ref{eq:infinite_fI_bi}). 
See also Fig.~\ref{fig19}. 
The S-S boundary provides an additional barrier to prevent penetration of vortices. 
It would not be as robust as the ${\mathcal I}$ layer of the S-I-S structure, 
but it also contributes to pushing up the onset vortex penetration. 
\item Finally the surface resistance of the S-S bilayer structure was examined. 
The surface resistance formula is given by Eq.~(\ref{eq:SS_bilayer_resistance}). 
See also Figs.~\ref{fig20} and \ref{fig21}.
\end{enumerate}
%

%%%%%%%%%%%%%%%%%%
%%%%%%%%%%%%%%%%%%
\section{Summary}
%%%%%%%%%%%%%%%%%%
%%%%%%%%%%%%%%%%%%

We have reviewed recent progresses in theoretical understanding of the S-I-S structure and summarized important formulae that will be necessary for planning proof-of-concept experiments. 
Some results of the S-S bilayer structure obtained in studies of the S-I-S structure have also been introduced, 
which would be useful to study a system that can be modeled by the S-S bilayer structure such as ${\rm Nb_3 Sn}$-Nb, ${\rm MgB_2}$-Nb, and Nb surface after the low temperature baking. 
Important results are summarized in the end of each section: 
see Secs.~\ref{subsection:summary_SIS} and \ref{subsection:summary_SS} for the S-I-S and S-S structures, respectively.

%%%%%%%%%%%%%%%%%%
%%%%%%%%%%%%%%%%%%
\ack
%%%%%%%%%%%%%%%%%%
%%%%%%%%%%%%%%%%%%
This work was supported by
Photon and Quantum Basic Research Coordinated Development Program from MEXT
and JSPS KAKENHI Grant Numbers JP26800157 and JP26600142.

\appendix
%%%%%%%%%

%%%%%%%%%%%%%%%%%%%%%%%%%%%%%%%%%%%%%%%%%%%%%%%%%%%%%
%%%%%%%%%%%%%%%%%%%%%%%%%%%%%%%%%%%%%%%%%%%%%%%%%%%%%
\section{Vortex in an infinite superconductor}\label{appendix_infinite}
%%%%%%%%%%%%%%%%%%%%%%%%%%%%%%%%%%%%%%%%%%%%%%%%%%%%%
%%%%%%%%%%%%%%%%%%%%%%%%%%%%%%%%%%%%%%%%%%%%%%%%%%%%%

The magnetic field distribution in an infinite superconductor can be derived by solving the London equation 
$-\lambda^2 \nabla^2 {\bf B} + {\bf B} = \phi_0 \delta^{(2)}({\bf r}-{\bf r}_0)$ or 
\begin{eqnarray}
-\lambda^2 (\partial_x^2 +\partial_y^2 ) B(x,y) + B(x,y) = \phi_0 \delta(x-x_0) \delta(y) , \label{aeq:mod_London_1}
\end{eqnarray}
where ${\bf B}= B(x,y) {\bf \hat{z}}$ and ${\bf r}_0=(x_0,0)$. 
While we can treat this equation in the polar coordinate, 
here we use the Cartesian coordinate as an instructive exercise toward problems without the rotational symmetry. 
Eq.~(\ref{aeq:mod_London_1}) can be written as 
\begin{eqnarray}
B_k'' -p^2 B_k = -\frac{\phi_0}{\lambda^2} \delta(x-x_0) , \label{aeq:mod_London_2}
\end{eqnarray}
where 
\begin{eqnarray}
p\equiv \sqrt{k^2 + \frac{1}{\lambda^2}} ,
\end{eqnarray}
$B_k(x) = \int_{-\infty}^{\infty} \!\! dy B(x,y) e^{-iky}$, and the prime denote the derivative over $x$. 
By introducing the Fourier transformation $B_{k k'} = \int_{-\infty}^{\infty} \!\! dx B_k(x) e^{-ik'x}$, 
Eq.~(\ref{aeq:mod_London_2}) becomes an algebraic equation, whose solution can be inverse Fourier transformed on the complex $k'$-plane with poles at $x=\pm ip$. 
Then we find
\begin{eqnarray}
B_k(x) = \frac{\phi_0}{2\lambda^2} \frac{1}{p} e^{-p|x-x_0|}.  \label{aeq:mod_London_solution}
\end{eqnarray}
The self-energy of vortex is given by $\epsilon_v = (\phi_0/2\mu_0) B({\bf r}_0)$ or
\begin{eqnarray}
\epsilon_v 
= \frac{\phi_0^2}{4\mu_0 \lambda^2} \int_{-\infty}^{\infty} \frac{dk}{2\pi} \frac{1}{p}e^{-p\xi}
= \frac{\phi_0^2}{4\pi \mu_0 \lambda^2} K_0 \Bigl(\frac{\xi}{\lambda}\Bigr) , 
\label{aeq:self_energy_infinite}
\end{eqnarray}
where the standard prescription ${\bf r}_0 =(x_0,0) \to (x_0+\xi, 0)$ is used, 
and $K_0(z)=(1/2)\int_{-\infty}^{\infty}dt \exp(-z\cosh t)$ is the modified Bessel function. 
By using $K_0(z)\simeq \log(1/z)+\log 2 - \gamma +{\mathcal O}(z^2)$, where $\gamma=0.577$ is the Euler constant, 
Eq.~(\ref{aeq:self_energy_infinite}) is reduced to 
\begin{eqnarray}
\epsilon_v \simeq \frac{\phi_0^2}{4\pi \mu_0 \lambda^2} \log \frac{\lambda}{\xi} , 
\end{eqnarray}
for $\lambda/\xi \gg 1$. 
The current density can be derived by $J=-(1/\mu_0)B'=-(1/\mu_0)\int_{-\infty}^{+\infty} (dk/2\pi) B_k'(x) e^{iky}$. 
When we are interested in a scale smaller than $\lambda$, 
$p$ can be replaced by $|k|$, 
and the current density at a distance $r$ from the vortex core is given by
\begin{eqnarray}
J({\bf r})|_{|{\bf r}-{\bf r_0}|=r}=J(x_0+r,0) =\frac{\phi_0}{2\pi\mu_0\lambda^2} \int_0^{\infty} e^{-kr} = \frac{\phi_0}{2\pi\mu_0\lambda^2 r}. 
\end{eqnarray}
%

%%%%%%%%%%%%%%%%%%%%%%%%%%%%%%%%%%%%%%%%%%%%%%%%%%%%%
%%%%%%%%%%%%%%%%%%%%%%%%%%%%%%%%%%%%%%%%%%%%%%%%%%%%%
\section{Vortex in a semi-infinite superconductor}\label{appendix_semi_infinite}
%%%%%%%%%%%%%%%%%%%%%%%%%%%%%%%%%%%%%%%%%%%%%%%%%%%%%
%%%%%%%%%%%%%%%%%%%%%%%%%%%%%%%%%%%%%%%%%%%%%%%%%%%%%

A system with a single vortex in a semi-infinite superconductor can be treated in much the same way as in \ref{appendix_infinite}. 
The governing equation is Eq.~(\ref{aeq:mod_London_2}), 
and the general solution can be written as $B_k(x) = (\phi_0/2\lambda^2) (1/p) e^{-p |x-x_0|} + C e^{-px}$, where $C$ is a constant. 
Since $j_x(x,y) = \partial_y B(x,y) = \int (dk/2\pi) B_k(x) ik e^{iky}$, 
the boundary condition, $j_x=0$ at the surface, can be written as $B_k(0)=0$. 
Then we have $C = -(\phi_0/2\lambda^2) (1/p) e^{-p x_0}$, and 
\begin{eqnarray}
B_k(x) = \frac{\phi_0}{2\lambda^2} \frac{1}{p} \biggl( e^{-p |x-x_0|} -e^{-p(x+x_0)} \biggr).  \label{aeq:mod_London_solution_semi-infinite}
\end{eqnarray}
The self-energy of the vortex $\epsilon_v = (\phi_0/2\mu_0) B({\bf r}_0)$ depends on its position due to the existence of the surface, 
in contrast to that of the free vortex treated in \ref{appendix_infinite}. 
This means that the vortex is attracted to a direction that yields a smaller $\epsilon_v$ with a force given by $f_{\rm B}=-\partial_{x_0} \epsilon_v=-(\phi_0/2\mu_0) \int (dk/2\pi) \partial_{x_0} B_k(x_0)$ or 
\begin{eqnarray}
f_{\rm B}
= -\frac{\phi_0^2}{2 \pi \mu_0 \lambda^2} \int_{0}^{\infty}\!\!\! dk \, e^{-2p x_0}
= -\frac{\phi_0^2}{2 \pi \mu_0 \lambda^2} \frac{1}{\lambda} K_1\Bigl( \frac{2x_0}{\lambda}\Bigr). 
\label{aeq:force_semi_infinite}
\end{eqnarray}
where $K_{\nu}(z)=\int_0^{\infty}dt e^{-z\cosh t} \cosh\nu t$ is used. 
When the vortex is placed at the vicinity of the surface, $x_0/\lambda \ll 1$, 
Eq.~(\ref{aeq:force_semi_infinite}) is reduced to 
\begin{eqnarray}
f_{\rm B} = -\frac{\phi_0^2}{4 \pi \mu_0 \lambda^2 x_0} ,
\label{aeq:force_semi_infinite_2}
\end{eqnarray}
where the asymptotic behavior $\lim_{z\to 0} K_{\nu}(z)=(\nu-1)!\, 2^{\nu-1}z^{-\nu}$ is used. 
It should be noted that Eq.~(\ref{aeq:force_semi_infinite_2}) can be derived by an easier way. 
Since we are interested only in a scale much smaller than $\lambda$, 
we can replace $p$ by $|k|$. 
Then Eq.~(\ref{aeq:force_semi_infinite}) becomes 
\begin{eqnarray}
f_{\rm B}
= -\frac{\phi_0^2}{2 \pi \mu_0 \lambda^2} \int_{0}^{\infty}\!\!\! dk \, e^{-2|k| x_0}
= -\frac{\phi_0^2}{4 \pi \mu_0 \lambda^2 x_0} , 
\label{aeq:force_semi_infinite_3}
\end{eqnarray}
which corresponds with Eq.~(\ref{aeq:force_semi_infinite_2}). 

%%%%%%%%%%%%%%%%%%%%%%%%%%%%%%%%%%%%%
%%%%%%%%%%%%%%%%%%%%%%%%%%%%%%%%%%%%%
\section{The superheating field of a clean superconductor at $T\to 0$} \label{appendix_superheating}
%%%%%%%%%%%%%%%%%%%%%%%%%%%%%%%%%%%%%
%%%%%%%%%%%%%%%%%%%%%%%%%%%%%%%%%%%%%

We use the same unit as Ref.~\cite{catelani}: 
$\widetilde{\nabla}=\lambda_* \nabla$, 
$\widetilde{\bf A}=(2\pi\xi_* / \phi_0){\bf A}$, 
$\widetilde{\bf B}=\widetilde{\nabla} \times \widetilde{\bf A}=\sqrt{3/2\mu_0 N(0)}({\bf B}/\Delta_{00})$, 
$\widetilde{\Delta}= \Delta/\Delta_{00}$, 
$\widetilde{T}= k_B T/\Delta_{00}$, 
$\widetilde{\omega}_n = \hbar \omega_n/\Delta_{00}$, 
$\lambda_*^{-2}\equiv (4\mu_0/3) (2\pi\xi_*/ \phi_0)^2 \Delta_{00}^2 N(0)$, 
$\xi_*\equiv \hbar v_{\rm F}/2\Delta_{00}$, 
$\nu = \Delta_{00} N(0)$, 
$N(0)$ is the density of states per one spin at the Fermi surface, 
$\Delta_{00}$ is the zero-temperature and zero-field order parameter, 
$v_{\rm F}$ is the Fermi velocity, 
$k_B$ is the Boltzmann constant, 
and $\omega_n = (2\pi k_B T /\hbar) (n +1/2)$ is the Matsubara frequency~\cite{matsubara}. 
In the following, we omit all the tildes for brevity. 
Then the free energy in unit of $\Delta_{00}$ is given by
\begin{eqnarray}
\Omega = 
\nu \int\! d^3 r \biggl[ \frac{1}{3}(\nabla\times {\bf A} - {\bf B_a})^2 + \Delta^2 \log \frac{T}{T_c} \nonumber \\
+ 2\pi T \sum_{n} \biggl\{ \frac{\Delta^2}{\omega_n} -2\Delta \langle f \rangle  -2 \omega_n (\langle g \rangle -1)) -2i \langle g{\bf n}\cdot {\bf A} \rangle \biggr\} 
\biggr], 
\label{aeq:QC_free_energy}
\end{eqnarray}
where ${\bf n}$ is the unit vector normal to the Fermi surface, 
the angular brackets means the angular averaging over the Fermi surface,
and the quasiclassical Green functions are given by $f=\Delta/\sqrt{\Omega_n^2+\Delta^2}$ and $g=\Omega_n/\sqrt{\Omega_n^2+\Delta^2}$ with $\Omega_n \equiv \omega_n + i {\bf n}\cdot \bf A$, 
which satisfy the constraint $g^2+f^2=1$ and the Eilenberger equation $\Omega_n f = \Delta g$ for $\kappa \equiv \lambda_*/\xi_* \to \infty$. 
The self-consistency condition is given by 
\begin{eqnarray}
\Delta \log\frac{T}{T_c} + 2\pi T \sum_{n} \biggl( \frac{\Delta}{\omega_n} - \langle f \rangle \biggr) = 0 . 
\label{aeq:QC_gap}
\end{eqnarray}
In this unit, the energy density of the magnetic field $B^2/2\mu_0$ is reduced to $(\nu/3) B^2$, 
and the condensation energy is given by $-(\nu/3)B_c(T)^2=\nu [\Delta_{0T}^2\log(T/T_c) + 2\pi T \sum_n \{ (\Delta_{0T}^2/\omega_n) -2 \Delta_{0T} f_0 -2 \omega_n (g_0-1)\}]=2\pi T \nu \sum_n (-2\sqrt{\omega_n^2+\Delta_{0T}^2} + 2\omega_n + \Delta_{0T}^2/\sqrt{\omega_n^2+\Delta_{0T}^2})$, 
where Eq.~(\ref{aeq:QC_gap}) is used, $f_0$ and $g_0$ are the zero-field quasiclassical Green functions, and $\Delta_{0T}$ is the zero-field order parameter in a finite temperature. 
When $T=0$, we have $B_c(0) =\sqrt{3/2}$. 
Restoring the dimensional units, 
the well-known result $B_c(0)=\sqrt{\mu_0 N(0)} \Delta_{00}$ is reproduced.

In much the same way as the last subsection, 
we consider the second variation of $\Omega$ under small perturbations $\Delta +\delta\Delta$ and ${\bf A}+ \delta{\bf A}$, 
which is given by
\begin{eqnarray}
\fl
\delta^2 \Omega =\nu \int \!\!d^3 r \Biggl[ \frac{1}{3}(\nabla \times \delta{\bf A}^2) 
+2\pi T \sum_n \Biggl\langle \frac{\Delta^2 [\delta\Delta^2 + ({\bf n}\cdot\delta{\bf A})^2]+2i\Omega_n \Delta \delta \Delta ({\bf n} \cdot \delta {\bf A})}{(\Omega_n^2+\Delta^2)^{\frac{3}{2}}} \Biggr\rangle \Biggr],  
\label{aeq:QC_second_variation}
\end{eqnarray}
where Eq.~(\ref{aeq:QC_gap}) is used. 
Expanding the perturbations as $\delta \Delta (x,y) =  \widetilde{\delta \Delta}(x) \cos k y$, $\delta A_x (x,y) = \widetilde{\delta A_x}(x) \sin k y$, and $\delta A_y (x,y) = \widetilde{\delta A_y}(x) \cos k y$, 
we obtain $\delta^2\Omega \propto \int dx [(1/3)(\widetilde{\delta A}_y' -k \widetilde{\delta A}_x)^2 + F_0 \widetilde{\delta \Delta}^2 +F_x \widetilde{\delta A}_x^2 + F_y \widetilde{\delta A}_y^2 + 2G\widetilde{\delta \Delta} \widetilde{\delta A}_y]$, 
where $F_0 \equiv 2\pi T \sum_n \langle \Delta^2 /(\Omega_n^2 + \Delta^2)^{\frac{3}{2}} \rangle$, 
$F_i \equiv 2\pi T \sum_n \langle \Delta^2 n_i /(\Omega_n^2 + \Delta^2)^{\frac{3}{2}} \rangle$ ($i=x,y$), 
and $G \equiv 2\pi T \sum_n \langle i \Omega_n \Delta n_y /(\Omega_n^2 + \Delta^2)^{\frac{3}{2}} \rangle$. 
Minimizing $\delta^2\Omega$ with respect to $\widetilde{\delta\Delta}$ and $\widetilde{\delta A}_x$, 
we find $\widetilde{\delta\Delta}=-(G/F_0)\widetilde{\delta A}_y$ and $\widetilde{\delta A}_x = k/(3F_x+k^2) \widetilde{\delta A}_y'$. 
Substituting these into $\delta^2\Omega$, we find the $\delta^2 \Omega$ is positive definite as long as $F_0 F_y = G^2$. 
At the limit $T\to 0$, 
$F_0$, $F_y$, and $G$ are analytically calculable. 
Using the notation $b\equiv \Delta/A$, we obtain $(1-\sqrt{1-b^2}) (1/3) [1-(1+2b^2)\sqrt{1-b^2}]=(b\sqrt{1-b^2})^2$~\cite{catelani} or
\begin{eqnarray}
b=\sqrt{1-(2^{\frac{1}{3}}-1)^2} \equiv b_0. 
\label{aeq:QC_b}
\end{eqnarray}
Then we have $\sqrt{1-b_0^2}=2^{\frac{1}{3}}-1$.

When $A\to 0$, Eq.~(\ref{aeq:QC_gap}) is reduced to the zero-field self-consistency condition: 
$\log (T/T_c) + 2\pi T \sum_{n} (1/\omega_n - 1/\sqrt{\omega_n^2 +\Delta_{0T}^2}) = 0$. 
Combining this with Eq.~(\ref{aeq:QC_gap}), 
we obtain $2\pi T \sum_{n} (\Delta/\sqrt{\omega_n^2 +\Delta_{0T}^2} -\langle f \rangle) = 0$. 
At $T\to 0$, we find $\log[A(1+\sqrt{1-b^2})]=\sqrt{1-b^2}$~\cite{catelani} or $A=e^{\sqrt{1-b^2}}/(1+\sqrt{1-b^2})$. 
Substituting $b=b_0$ into this, we obtain $A_{\rm max}=2^{-\frac{1}{3}} \exp(2^{\frac{1}{3}}-1)$. 
Restoring the dimension, we have
\begin{eqnarray}
A_{\rm max}= \frac{\exp(2^{\frac{1}{3}}-1)}{2^{\frac{1}{3}}} \frac{\phi_0}{2\pi\xi_*} .
\label{aeq:QC_A}
\end{eqnarray}

The relation among $B_a$, $\Delta$ and $A$ for a superconductor with $\kappa \gg 1$ is given by 
$B_a^2 = B_c^2 + (3/\nu) \nu [ \Delta^2 \log (T/T_c) + 2\pi T \sum_n \{ \Delta^2/\omega_n -2 \Delta \langle f \rangle -2\omega_n (\langle g \rangle -1) -2i \langle g {\bf n}\cdot {\bf A} \rangle \}]$~\cite{lin}. 
By using Eq.~(\ref{aeq:QC_gap}), this becomes $B_a^2/B_c^2 =1 + B_c^{-2} 6\pi T \sum_n \{ 2\omega_n - \Delta \langle f \rangle -2\omega_n \langle g \rangle -2i \langle g {\bf n}\cdot {\bf A} \rangle \}$. 
The angular averaging are given by $\langle g \rangle = -ib (\cosh z_2-\cosh z_1)$, $\langle f \rangle =-ib(z_2 -z_1)$, and $2i\langle g {\bf n}\cdot {\bf A} \rangle = \omega_n (-\langle g \rangle +4/\langle g \rangle) -\Delta \langle f \rangle $, respectively, 
where $z_1\equiv (a-i)/b$, $z_2 \equiv (a+i)/b$, $a=\omega_n/A$, and $b=\Delta/A$. 
Then we have $B_a^2/B_c^2 =1 + B_c^{-2} 6\pi T \sum_n \{ 2\omega_n - (2/A){\rm Im}(\Omega_0 \sqrt{\Omega_0^2 + \Delta^2}) \}$, 
where $\Omega_0\equiv \omega_n + iA$. 
When $T\to 0$, 
substituting $b=b_0$ and $A=A_{\rm max}$ into $B_a^2/B_c(0)^2 = 1 - (2/3) A^2 \{ (1/2) -(3/2)(1-b^2) +(1-b^2)^{\frac{3}{2}} \} $, 
we find~\cite{galaiko, catelani, lin}
\begin{eqnarray}
B_s(0)
= \sqrt{1- (2^{\frac{5}{3}}-3)\exp(2^{\frac{4}{3}}-2) } \, B_c(0) 
\simeq 0.84 B_c(0) .
\label{aeq:QC_Bs}
\end{eqnarray}
%

%%%%%%%%%%%%%%%%%%%%%%%%%%%%%%%%%%%%%%%%%%%%%%%%%%%%%
%%%%%%%%%%%%%%%%%%%%%%%%%%%%%%%%%%%%%%%%%%%%%%%%%%%%%
\section{The summation in Eq.~(\ref{eq:infinite_fI_multi})} \label{appendix_infinite_fI_multi}
%%%%%%%%%%%%%%%%%%%%%%%%%%%%%%%%%%%%%%%%%%%%%%%%%%%%%
%%%%%%%%%%%%%%%%%%%%%%%%%%%%%%%%%%%%%%%%%%%%%%%%%%%%%

Let us evaluate the summation
\begin{eqnarray}
S \equiv \sum_{n=1}^{\infty} \biggl( \frac{1}{n d_{\mathcal S}-x_0} -\frac{1}{n d_{\mathcal S}+x_0} \biggr) 
= \frac{1}{d_{\mathcal S}} \sum_{n=1}^{\infty} \biggl( \frac{1}{n -a} -\frac{1}{n +a} \biggr) ,
\end{eqnarray}
where $a \equiv x_0/d_{\mathcal S}$. 
By using the difference equation of the digamma function, $\psi(z+N)-\psi(z)=\sum_{n=1}^N (n+z-1)^{-1}$, 
we have
\begin{eqnarray}
S = \frac{1}{d_{\mathcal S}} \lim_{N\to \infty} [ \psi(-a +1 +N) - \psi(-a +1) - \psi(a +1 +N) + \psi(a +1)] \nonumber \\
=\frac{1}{d_{\mathcal S}} \biggl[ \lim_{N\to \infty} \log \frac{-a +1 +N}{a +1 +N} + \psi(a +1) - \psi(1-a) \biggr] \nonumber \\
= \frac{1}{d_{\mathcal S}} \biggl[  \psi(a +1) - \psi(1-a) \biggr]
= \frac{1}{d_{\mathcal S}} \biggl[ \frac{1}{a} + \psi(a)- \psi(1-a) \biggr] ,
\end{eqnarray}
where the relation $\psi(a +1) = \psi(a) + 1/a$ is used. 
Then, using the reflection formula, $\psi(z)-\psi(1-z)=-\pi \cot\pi z$, we find
\begin{eqnarray}
S = \frac{1}{d_{\mathcal S}} \biggl[ \frac{1}{a} -\pi \cot\pi a \biggr]
= \frac{1}{x_0} - \frac{1}{d_{\mathcal S}} \pi \cot\frac{\pi x_0}{d_{\mathcal S}}. 
\end{eqnarray}
%

%%%%%%%%%%%%%%%%%%%%%%%%%%%%%%%%%%%%%%%%%%%%%%%%%%%%%
%%%%%%%%%%%%%%%%%%%%%%%%%%%%%%%%%%%%%%%%%%%%%%%%%%%%%
\section{Vortex in a thin film}\label{appendix_thin_film}
%%%%%%%%%%%%%%%%%%%%%%%%%%%%%%%%%%%%%%%%%%%%%%%%%%%%%
%%%%%%%%%%%%%%%%%%%%%%%%%%%%%%%%%%%%%%%%%%%%%%%%%%%%%

Next we tackle a system with a single vortex in a thin film with a thickness $d\ll \lambda$ by using the same technique as \ref{appendix_semi_infinite} (see Ref.~\cite{stejic} for more detailed discussions). 
The general solution of Eq.~(\ref{aeq:mod_London_2}) can be written as $B_k(x) = (\phi_0/2\lambda^2) (1/p) e^{-p |x-x_0|} + C_1 e^{px} + C_2 e^{-px}$, 
where $C_1$ and $C_2$ are constants. 
The boundary conditions are given by $j_x(0)=j_x(d)=0$ or $B_k(0)=B_k(d)=0$. 
Then we find $C_1=-(\phi_0/2\lambda^2p)[e^{-p(d-x_0)}-e^{-p(d+x_0)}]/(e^{pd}-e^{-pd})$, 
$C_2=(\phi_0/2\lambda^2p)[-e^{p(d-x_0)}+e^{-p(d-x_0)}]/(e^{pd}-e^{-pd})$, and 
\begin{eqnarray}
B_k(x) = \frac{\phi_0}{2\lambda^2 p \sinh pd} \Bigl[ \cosh p(d- |x-x_0|) - \cosh p(x+x_0-d) \Bigr],  \label{aeq:mod_London_solution_thin_film}
\end{eqnarray}
which corresponds with that given in Ref.~\cite{stejic} if we translate the cordinate as $x\to x+d/2$ and $x_0\to x_0+d/2$. 
In much the same way as the above, the self-energy of the vortex $\epsilon_v({\bf r}_0) = (\phi_0/2\mu_0) B({\bf r}_0)$ depends on its position due to the existance of the surfaces, 
and the vortex is exerted a force given by $f_{\rm B}=-\partial_{x_0} \epsilon_v=-(\phi_0/2\mu_0) \int (dk/2\pi) \partial_{x_0} B_k(x_0)$ or 
\begin{eqnarray}
f_{\rm B}
= -\frac{\phi_0^2}{2 \pi \mu_0 \lambda^2} \int_{0}^{\infty}\!\!\! dk \, \frac{\sinh [pd(1-2a)]}{\sinh pd} , 
\label{aeq:force_thin_film_1}
\end{eqnarray}
where $a \equiv x_0/d$. 
Since we are interested in a scale much smaller than $\lambda$, we may replace $p$ by $|k|$. 
Substituting $t= e^{-2kd}$, the integral becomes $(1/2d)\int_0^1 dt [t^{a-1}-t^{(1-a)-1}]/(1-t)=[-\psi(a)+\psi(1-a)]/2d$, 
where $\psi(z)=-\int_0^1 [(1/\log t) + t^{z-1}/(1-t)]$ is the digamma function. 
Then, using the reflection formula, $\psi(z)-\psi(1-z)=-\pi \cot\pi z$, 
Eq.~(\ref{aeq:force_thin_film_1}) becomes
\begin{eqnarray}
f_{\rm B}
= -\frac{\phi_0^2}{4 \pi \mu_0 \lambda^2 d} \pi \cot \frac{\pi x_0}{d} .
\label{aeq:force_thin_film_2}
\end{eqnarray}
When the vortex is at the edge of the film $x_0/d \ll 1$, 
Eq.~(\ref{aeq:force_thin_film_2}) is reduced to 
\begin{eqnarray}
f_{\rm B}
= -\frac{\phi_0^2}{4 \pi \mu_0 \lambda^2 x_0} ,
\label{aeq:force_thin_film_3}
\end{eqnarray}
where $\cot(\pi x_0/d) \simeq d/\pi x_0$ is used. 
Note that Eq.~(\ref{aeq:force_thin_film_3}) is equal to the force acting on the vortex at the edge of a semi-infinite superconductor.

%%%%%%%%%%%%%%%%%%%%%%%%%%%%%%%%%%%%%%%%%%%%%%%%%%%%%
%%%%%%%%%%%%%%%%%%%%%%%%%%%%%%%%%%%%%%%%%%%%%%%%%%%%%
\section{Electromagnetic field in a superconductor} \label{appendix_electromagnetic_multi}
%%%%%%%%%%%%%%%%%%%%%%%%%%%%%%%%%%%%%%%%%%%%%%%%%%%%%
%%%%%%%%%%%%%%%%%%%%%%%%%%%%%%%%%%%%%%%%%%%%%%%%%%%%%

We briefly summarize some results necessary for calculations of the electromagnetic field distribution and the surface resistance in the S-I-S structure. 
Let us introduce the complex conductivity 
\begin{eqnarray}
\sigma=\sigma' + i\sigma'' .
\end{eqnarray}
Then the current density can be written as ${\bf j}=\sigma {\bf E}$. 
Then starting from the maxwell equation $\nabla\times {\bf E}=-\partial_t {\bf B}$, 
we obtain $-\triangle {\bf E}=-\partial_t \nabla\times{\bf B}=i \mu_0 \omega {\bf j}= i\mu_0 \sigma \omega {\bf E}$, 
where the displacement current term is always negligible. 
In much the same way as the above, starting from $\nabla\times {\bf B}=\mu_0 {\bf j}$, 
we obtain $-\triangle {\bf B}= i\mu_0 \sigma \omega {\bf B}$. 
Then the London equations for the electromagnetic field are given by~\cite{hasan}
\begin{eqnarray}\label{aeq:MaxwellLondon1}
\triangle {\bf E} = \frac{1}{\ell^2} {\bf E} \,, \hspace{2cm} 
\triangle {\bf B} = \frac{1}{\ell^2} {\bf B}\, , 
\end{eqnarray}
where 
\begin{eqnarray}
\frac{1}{\ell^2} \equiv \frac{\mu_0 \sigma \omega}{i} = \mu_0 \omega \sigma'' \biggl( 1-i \frac{\sigma'}{\sigma''} \biggr) 
= \frac{1}{\lambda^2} ( 1-i \mu_0 \omega \sigma' \lambda^2) \,. 
\label{aeq:ell}
\end{eqnarray}
As $\omega \to 0$, the second term approaches zero, and $\ell \to \lambda$.

In calculations of the electromagnetic field distribution, 
we can replace $\ell$ by $\lambda$. 
For example, using $\omega \sim 10^{9}\,{\rm Hz}$, $\lambda\sim 10^{-7}\,{\rm m}$ and $\sigma' \sim 10^{7}\,{\rm S/m}$, 
we obtain $\mu_0 \omega \sigma' \lambda \simeq 10^{-4}$, 
and the second term of Eq.~(\ref{aeq:ell}) is negligible.  
Then Eq.~(\ref{aeq:MaxwellLondon1}) are reduced to 
\begin{eqnarray}\label{aeq:MaxwellLondon2}
\triangle {\bf E} = \frac{1}{\lambda^2} {\bf E} \, , \hspace{2cm} 
\triangle {\bf B} = \frac{1}{\lambda^2} {\bf B} \, .
\end{eqnarray}
By using Eq.~(\ref{aeq:MaxwellLondon2}), we can calculate the electromagnetic field distribution in the S-I-S structure. 
The result and its derivation processes are shown in Ref.~\cite{kubo_APL, kubo_IPAC13}.

On the other hand, in calculation of the surface resistance, the second term of Eq.~(\ref{aeq:ell}) is essential. 
For example, let us evaluate the surface resistance of the semi-infinite superconductor. 
The surface resistance is determined by the total joule dissipation:  
$(1/2)R_s H_0^2 = (1/2)\int_0^{\infty}dx {\rm Re}( E J^*) = (1/2) \int_0^{\infty}dx |J|^2 {\rm Re} (1/\sigma)$ or 
\begin{eqnarray}
R_s =  \frac{\mu_0^2}{B_0^2} \int_0^{\infty}  dx |J|^2 {\rm Re} \biggl( \frac{1}{\sigma} \biggr) . 
\label{aeq:Rs}
\end{eqnarray}
When we neglect terms with ${\mathcal O}(\sigma'^2/\sigma''^2)$, 
Eq.~(\ref{aeq:Rs}) is reduced to a simple form.  
Since $1/\sigma \simeq -i/\sigma'' + \sigma'/\sigma''^2$, 
we obtain ${\rm Re}(1/\sigma)=(1/\sigma'')(\sigma'/\sigma'')$. 
Then we may consider only the zeroth order for the contribution from the factor $|J|^2$ and we can regard $|J|^2$ as $J^2|_{\ell = \lambda}$. 
Thus we have
\begin{eqnarray}
R_s =  \frac{\mu_0^2}{B_0^2} \int_0^{\infty}  dx J|_{\ell=\lambda}^2 
\frac{\sigma'}{\sigma''^2} 
=  \frac{\mu_0^2}{B_0^2} \int_0^{\infty}  dx J|_{\ell=\lambda}^2 
\sigma' \mu_0^2 \omega^2 \lambda^4 , 
\label{aeq:Rs2}
\end{eqnarray}
where $\lambda^{-2}=\mu_0 \omega \sigma''$ is used. 
Substituting $B|_{\ell=\lambda}=B_0 e^{-\frac{x}{\lambda}}$ or $J|_{\ell=\lambda}=-B'/\mu_0=(B_0/\mu_0\lambda) e^{-\frac{x}{\lambda}}$ into Eq.~(\ref{aeq:Rs2}), 
we obtain 
\begin{eqnarray}
R_s = \frac{1}{2} \sigma' \mu_0^2 \omega^2 \lambda^3 , 
\end{eqnarray}
By using Eq.~(\ref{aeq:Rs}), the surface resistance of the S-I-S structure and the S-S bilayer can also be evaluated.

%%%%%%%%%%%%%%%%%%%%%%%%%%%%%%%%%%%%%%%%%%%%%%%%%%%%%
%%%%%%%%%%%%%%%%%%%%%%%%%%%%%%%%%%%%%%%%%%%%%%%%%%%%%
\section{Vortex in an infinite superconductor with two regions}\label{appendix_infinite_two}
%%%%%%%%%%%%%%%%%%%%%%%%%%%%%%%%%%%%%%%%%%%%%%%%%%%%%
%%%%%%%%%%%%%%%%%%%%%%%%%%%%%%%%%%%%%%%%%%%%%%%%%%%%%

Suppose the vortex is at $x=x_0=-|x_0|$. 
Then the magnetic field distribution in this system can be obtained by solving the set of equations
\begin{eqnarray}
B_k'' -p_1^2 B_k = -\frac{\phi_0}{\lambda_1^2} \delta(x-x_0) \hspace{1.5cm} (x\le 0), \label{aeq:London_infinite_two_left} \\
B_k'' -p_2^2 B_k = 0  \hspace{3.8cm} (x\ge 0) , \label{aeq:London_infinite_two_right}
\end{eqnarray}
where $p_1=\sqrt{k^2 + \lambda_1^{-2}}$ and $p_2=\sqrt{k^2 + \lambda_2^{-2}}$. 
The general solution of Eq.~(\ref{aeq:London_infinite_two_left}) can be written as $B_k(x) = (\phi_0/2\lambda_1^2) (1/p_1) e^{-p_1 |x-x_0|} + C_1 e^{p_1 x}$, 
and that of Eq.~(\ref{aeq:London_infinite_two_right}) is given by $B_k(x) = C_2 e^{-p_2 x}$, 
where $C_1$ and $C_2$ are constants.  
The boundary conditions are given by $j_x(-0)=j_x(+0)$ and $A_y(-0)=A_y(+0)$, 
which reduce to $B_k(-0)=B_k(+0)$ and $\lambda_1^2 B_k'(-0)=\lambda_2^2 B_k'(+0)$, respectively. 
Then we find $C_1=(\phi_0/2p_1\lambda_1^2) [(p_1\lambda_1^2-p_2\lambda_2^2)/(p_1\lambda_1^2+p_2\lambda_2^2)]e^{p_1 x_0}$, 
$C_2=[\phi_0/(p_1\lambda_1^2+p_2\lambda_2^2)]e^{p_1 x_0}$, and
\begin{eqnarray}
B_k(x) = \frac{\phi_0}{2p_1\lambda_1^2}\biggl( 
e^{-p_1 |x-x_0|} + \frac{p_1\lambda_1^2-p_2\lambda_2^2}{p_1\lambda_1^2+p_2\lambda_2^2} e^{p_1(x+x_0)}
\biggr)  \hspace{0.7cm} (x\le 0), \label{aeq:London_infinite_two_solution_left} \\
B_k(x) = \frac{\phi_0}{p_1\lambda_1^2+p_2\lambda_2^2} e^{-p_2 x +p_1 x_0}  \hspace{4.1cm} (x\ge 0) . \label{aeq:London_infinite_two_solution_right}
\end{eqnarray}
The force acting on the vortex is given by $f_{\rm B}=-(\phi_0/2\mu_0) \int (dk/2\pi) \partial_{x_0} B_k(x_0)$ or
\begin{eqnarray}
f_{\rm B}
= -\frac{\phi_0^2}{2 \pi \mu_0 \lambda_1^2} \int_{0}^{\infty}\!\!\! dk \, \frac{p_1\lambda_1^2-p_2\lambda_2^2}{p_1\lambda_1^2+p_2\lambda_2^2}e^{2p_1 x_0} .
\label{aeq:force_infinite_two_1}
\end{eqnarray}
When we focus attention on a scale much smaller than $\lambda_1$ and $\lambda_2$, 
we may replace $p_1$ and $p_2$ by $|k|$. 
Then the force is given by
\begin{eqnarray}
f_{\rm B}
= -\frac{\tau \phi_0^2}{2 \pi \mu_0 \lambda_1^2} \int_{0}^{\infty}\!\!\! dk \, e^{2 k x_0} 
= -\frac{\phi_0 \phi_1}{4 \pi \mu_0 \lambda_1^2 |x_0|} .
\label{aeq:force_infinite_two_2}
\end{eqnarray}
where 
\begin{eqnarray}
\phi_1 \equiv \tau \phi_0 , \hspace{2cm} 
\tau \equiv \frac{\lambda_1^2-\lambda_2^2}{\lambda_1^2+\lambda_2^2} .
\end{eqnarray}
Eq.~(\ref{aeq:force_infinite_two_2}) can be interpreted as a force due to the image with flux $\phi_1 = \tau \phi_0$ at $x=|x_0|$ (see also Ref.~\cite{kubo_LINAC14}).

%%%%%%%%%%%%%%%%%%%%%%%%%%%%%%%%%%%%%%%%%%%%%%%%%%%%%
%%%%%%%%%%%%%%%%%%%%%%%%%%%%%%%%%%%%%%%%%%%%%%%%%%%%%
\section{Summation in Eq.~(\ref{eq:infinite_fI_bi})}\label{appendix_summation_nano_layer}
%%%%%%%%%%%%%%%%%%%%%%%%%%%%%%%%%%%%%%%%%%%%%%%%%%%%%
%%%%%%%%%%%%%%%%%%%%%%%%%%%%%%%%%%%%%%%%%%%%%%%%%%%%%

Let us evaluate the summation in Eq.~(\ref{eq:infinite_fI_bi}),  
\begin{eqnarray}
\fl
S' 
\equiv \frac{1}{x_0} -\sum_{n=1}^{\infty} (-1)^{n} \tau^n \biggl( \frac{1}{nd-x_0} - \frac{1}{nd+x_0} \biggr) 
= \frac{1}{d} \biggl( \frac{1}{a} + \sum_{n=1}^{\infty} \frac{(-\tau)^n}{n+a} - \sum_{n=1}^{\infty} \frac{(-\tau)^n}{n-a} \biggr) 
\end{eqnarray}
where $a\equiv x_0/d$. 
We can rewrite $S'$ as 
\begin{eqnarray}
\fl
S'= \frac{1}{d} \biggl( \sum_{n=0}^{\infty} \frac{(-\tau)^n}{n+a} +\tau \sum_{n=0}^{\infty} \frac{(-\tau)^n}{n+1-a} \biggr)
= \frac{1}{d} [ \Phi(-\tau,1,a) + \tau \Phi(-\tau,1,1-a) ], 
\end{eqnarray}
where $\Phi(z,s,a)=\sum_{n=0}^{\infty} z^n/(n+a)^s$ is the Lerch transcendent. 
Through its integral representation, $\Phi(z,s,a)=\Gamma(s)^{-1}\int_0^{\infty} t^{s-1} e^{-at}/(1-ze^{-t})dt$, 
we arrive at 
\begin{eqnarray}
\fl
S'= \frac{1}{d} \biggl[ \frac{1}{a}F(1,a;1+a;-\tau) + \frac{\tau}{1-a} F(1,1-a;2-a;-\tau) \biggr] , 
\end{eqnarray}
where $F$ is the Gaussian hypergeometric function. 

%%%%%%%%%%%%%%%%%%%%%%%%%%%%%%%%%%%%%%%%%%%%%%%%%%%%%
%%%%%%%%%%%%%%%%%%%%%%%%%%%%%%%%%%%%%%%%%%%%%%%%%%%%%
\section{Vortex in a thin layer formed on a semi-infinite superconductor}\label{appendix_nano_layer}
%%%%%%%%%%%%%%%%%%%%%%%%%%%%%%%%%%%%%%%%%%%%%%%%%%%%%
%%%%%%%%%%%%%%%%%%%%%%%%%%%%%%%%%%%%%%%%%%%%%%%%%%%%%

Suppose the vortex is at $x=x_0$ ($0<x_0< d$). 
Then the magnetic field distribution in the surface layer can be obtained by solving
\begin{eqnarray}
B_k'' -p_1^2 B_k = -\frac{\phi_0}{\lambda_1^2} \delta(x-x_0) \hspace{1.5cm} (0< x <d), \label{aeq:London_nano_layer_left} \\
B_k'' -p_2^2 B_k = 0  \hspace{3.8cm} (x\ge d) , \label{aeq:London_nano_layer_right}
\end{eqnarray}
where $p_1=\sqrt{k^2 + \lambda_1^{-2}}$ and $p_2=\sqrt{k^2 + \lambda_2^{-2}}$. 
The general solution of Eq.~(\ref{aeq:London_nano_layer_left}) can be written as $B_k(x) = (\phi_0/2\lambda_1^2) (1/p_1) e^{-p_1 |x-x_0|} + C_1 e^{p_1 x} + C_2 e^{-p_1 x}$, 
and that of Eq.~(\ref{aeq:London_nano_layer_right}) is given by $B_k(x) = C_3 e^{-p_2 (x-d)}$, 
where $C_1$, $C_2$, and $C_3$ are constants.  
The boundary conditions are given by $j_x(0)=0$, $j_x(d-0)=j_x(d+0)$, and $A_y(d-0)=A_y(d+0)$, 
which reduce to $B_k(0)=0$, $B_k(d-0)=B_k(d+0)$ and $\lambda_1^2 B_k'(d-0)=\lambda_2^2 B_k'(d+0)$, respectively. 
Then we find $C_1=(\phi_0/2p_1\lambda_1^2) (e^{p_1 x_0}-e^{-p_1 x_0})/(1+\tilde{\tau}^{-1}e^{2p_1 d})$, 
$C_2=-(\phi_0/2p_1\lambda_1^2) (e^{p_1 x_0}+\tilde{\tau}^{-1}e^{2p_1 d - p_1 x_0})/(1+\tilde{\tau}^{-1}e^{2p_1 d})$, and
$C_3=(\phi_0/2p_2\lambda_2^2) e^{p_1 d} (-1+\tilde{\tau}^{-1})(e^{p_1 x_0}-e^{-p_1 x_0}) /(1+\tilde{\tau}^{-1}e^{2p_1 d})$, 
where $\tilde{\tau} \equiv (p_1\lambda_1^2-p_2\lambda_2^2)/(p_1\lambda_1^2+p_2\lambda_2^2)$. 
Then $B_k$ at $0< x <d$ is given by
\begin{eqnarray}
\fl
B_k(x) = \frac{\phi_0}{2p_1\lambda_1^2} \frac{e^{-p_1 |x-x_0|} -e^{-p_1(x+x_0)} + \tilde{\tau} e^{-2p_1 d} (e^{+p_1(x+x_0)}-e^{+p_1 |x-x_0|} ) }{1+\tilde{\tau}e^{-2p_1 d}} 
\end{eqnarray}
The force acting on the vortex is given by $f=-(\phi_0/2\mu_0) \int (dk/2\pi) \partial_{x_0} B_k(x_0)$ or
\begin{eqnarray}
f_{\rm B}
= -\frac{\phi_0^2}{2 \pi \mu_0 \lambda_1^2} \int_{0}^{\infty}\!\!\! dk  
\frac{e^{-2p_1x_0} + \tilde{\tau} e^{-2p_1 d} e^{2p_1 x_0} }{1+\tilde{\tau}e^{-2p_1 d}} .
\end{eqnarray}
Since we are focusing on a scale smaller than $\lambda_1$ and $\lambda_2$, 
$p_1$ and $p_2$ can be replaced by $|k|$, and $\tilde{\tau}$ by $\tau=(\lambda_1^2-\lambda_2^2)/(\lambda_1^2+\lambda_2^2)$. 
Then, substituting $t\equiv e^{-2kd}$ and $a\equiv x_0/d$, we have
\begin{eqnarray}
f_{\rm B}
= -\frac{\phi_0^2}{2 \pi \mu_0 \lambda_1^2} \int_{0}^{\infty}\!\!\! dk  
\frac{e^{-2k x_0} + \tau e^{-2k d} e^{2k x_0} }{1+\tau e^{-2k d}} \\
= -\frac{\phi_0^2}{4 \pi \mu_0 \lambda_1^2 d} \biggl[ \frac{1}{a} F(1,a;1+a;-\tau) + \frac{\tau}{1-a} F(1,1-a;2-a;-\tau) \biggr],  
\end{eqnarray}
where $F$ is the Gaussian hypergeometric function. 
When the vortex is at the surface $a=x_0/d \ll 1$, the contribution from the first term becomes dominant, and we have
\begin{eqnarray}
f_{\rm B}
= -\frac{\phi_0^2}{4 \pi \mu_0 \lambda_1^2 d} \frac{1}{a} F(1,0;1;-\tau) 
= -\frac{\phi_0^2}{4 \pi \mu_0 \lambda_1^2 x_0},  
\label{qeq:force_nano_layer_0}
\end{eqnarray}
where $F(1,0;1;-\tau)=1$ is used. 
Eq.~(\ref{qeq:force_nano_layer_0}) corresponds with the force acting on the vortex at the edge of the semi-infinite superconductor. 
On the other hand, when the vortex is at the boundary of two superconductors, $1-a=1-x_0/d \ll 1$, 
the contribution from the second term becomes dominant, and we have
\begin{eqnarray}
f_{\rm B}
= -\frac{\phi_0^2}{4 \pi \mu_0 \lambda_1^2 d} \frac{\tau}{1-a} F(1,0;1;-\tau) 
= -\frac{\phi_0 \phi_1}{4 \pi \mu_0 \lambda_1^2 (d-x_0)}, 
\label{qeq:force_nano_layer_1}
\end{eqnarray}
where $\phi_1 \equiv \tau \phi_0$. 
Eq.~(\ref{qeq:force_nano_layer_1}) corresponds with the force acting on the vortex near the boundary of two infinite superconductors given by Eq.~(\ref{aeq:force_infinite_two_2}).


\begin{thebibliography}{99}

%introduction

\bibitem{hasan}
H. Padamsee, J. Knobloch, and T. Hays, {\it RF Superconductivity for Accelerators} (John Wiley, New York, 1998). 

\bibitem{gurevich_review}
A. Gurevich, 
Rev. Accel. Sci. Technol. {\bf 5}, 119 (2012). 

\bibitem{saitoEP}
K. Saito et al., 
in {\it Proceedings of SRF1989, KEK, Tsukuba, Japan} (1989), p. 635, SRF89G18. 

\bibitem{furuyaEP}
T. Furuya, 
in {\it Proceedings of SRF1989, KEK, Tsukuba, Japan} (1989), p. 305, SRF89D02. 

\bibitem{bernerdHPR}
Ph. Bernard, D. Bloess, T. Flynn, C. Hauviller, W. Weingarten, P. Bosland, and J. Martignac, 
in {\it Proceedings of EPAC1992, Berlin, Germany} (1992), p. 1269. 

\bibitem{saitoHPR}
K. Saito, H. Miwa, K. Kurosawa, P. Kneisel, S. Noguchi, E. Kako, M. Ono, T. Shishido and T. Suzuki, 
in {\it Proceedings of SRF1993, CEBAF, Newport News, Virginia, USA} (1993), p. 1151, SRF93J03. 

\bibitem{kneiselHPR}
P. Kneisel, B. Lewis and L. Turlington, 
in {\it Proceedings of SRF1993, CEBAF, Newport News, Virginia, USA} (1993), p. 628, SRF93I09. 

\bibitem{kojimaCA}
Y. Kojima et al., 
in {\it Proceedings of SRF1989, KEK, Tsukuba, Japan} (1989), p. 85, SRF89A07. 

\bibitem{kako_bake}
E. Kako et al., %S. Noguchi, M.Ono, K. Saito, T. Shishido, T. Fujino, Y. Funahashi, H. Inoue, M. Matsuoka, T. Higuchi, T. Suzuki, and H. Umezawa, 
in {\it Proceedings of SRF1997, Gif-sur-Yvette, France} (1995), p. 425, SRF95C12. 

\bibitem{kneisel_40}
P. Kneisel, R. W. R${\rm \ddot{o}}$th and H. - G. Kiirschner,  
in {\it Proceedings of SRF1997, Gif-sur-Yvette, France} (1995), p. 449, SRF95C17. 

\bibitem{ono_bake}
M. Ono et al., 
in {\it Proceedings of SRF1997, Abano Terme (Padova), Italy} (1997), p. 472, SRF97C08. 

\bibitem{lilje_bake}
L. Lilje et al., 
in {\it Proceedings of SRF1999, La Fonda Hotel, Santa Fe, New Mexico, USA} (1999), p. 74, TUA001. 

\bibitem{iwashita}
Y. Iwashita, Y. Tajima, and H. Hayano, 
Phys. Rev. ST Accel. Beams {\bf 11}, 093501 (2008). 

\bibitem{champion}
M. S. Champion, L. D. Cooley, C. M. Ginsburg, D. A. Sergatskov, R. L. Geng, H. Hayano, Y. Iwashita, and Y. Tajima, 
IEEE Trans. Appl. Supercond. {\bf 19}, 3 (2009). 

\bibitem{yamamoto2010}
Y. Yamamoto,   
Nucl. Instrum. Methods Phys. Res. A {\bf 623}, 579 (2010). 

\bibitem{ge}
M. Ge, G. Wu, D. Burk, J. Ozelis, E. Harms, D. Sergatskov, D. Hicks, and L. D. Cooley, 
Supercond. Sci. Technol. {\bf 24}, 035002 (2011). 

\bibitem{yamamoto2013}
Y. Yamamoto,  H. Hayano, E. Kako, S. Noguchi, T. Shishido, and K. Watanabe,  
Nucl. Instrum. Methods Phys. Res. A {\bf 729}, 589 (2013). 

\bibitem{kubo_PTEP_pit}
T. Kubo, 
Prog. Theor. Exp. Phys. {\bf 2015}, 073G01 (2015).  

\bibitem{geng}
R. L. Geng, G. V. Eremeev, H. Padamsee, and V. D. Shemelin, 
in {\it Proceedings of PAC07, Albuquerque, New Mexico, USA} (2007), p. 2337, WEPMS006. 

\bibitem{watanabe}
K. Watanabe, S. Noguchi, E. Kako, K. Umemori, and T. Shishido, 
Nucl. Instrum. Methods Phys. Res. A {\bf 714}, 67 (2013). 

%introduction
%superheating

\bibitem{galaiko}
V. P. Galaiko, 
Sov. Phys. JETP {\bf 23}, 475 (1966). 

\bibitem{kramer}
L. Kramer, 
Phys. Rev. {\bf 170}, 475 (1968). 

\bibitem{bean}
C. P. Bean and J. D. Livingston, 
Phys. Rev. Lett. {\bf 12}, 14 (1964). 


\bibitem{christiansen}
P. V. Christiansen, 
Solid State Commun. {\bf 7}, 727 (1969). 

\bibitem{chapman}
S. J. Chapman, 
SIAM J. Appl. Math. {\bf 55}, 1233 (1995). 

\bibitem{transtrum}
M. K. Transtrum, G. Catelani, and J. P. Sethna, 
Phys. Rev. B {\bf 83}, 094505 (2011). 

\bibitem{catelani}
G. Catelani and J. P. Sethna, 
Phys. Rev. B {\bf 78}, 224509 (2008). 

\bibitem{lin}
F. P. Lin and A. Gurevich, 
Phys. Rev. B {\bf 85}, 054513 (2012). 



%introduction
%avalanche

\bibitem{aranson2001}
I. Aranson, A. Gurevich, and V. Vinokur, 
Phys. Rev. Lett. {\bf 87}, 067003 (2001). 

\bibitem{aranson2005}
I. Aranson, A. Gurevich, M. S. Welling, R. J. Wijngaarden, V. K. Vlasko-Vlasov, V. M. Vinokur, and U. Welp, 
Phys. Rev. Lett. {\bf 94}, 037002 (2005). 

\bibitem{duran_Nb}
C. A. Duran, P. L. Gammel, R. E. Miller, and D. J. Bishop, 
Phys. Rev. B {\bf 52}, 75 (1995). 

\bibitem{rudnev_NbN}
I. A. Rudnev, D. V. Shantsev, T. H. Johansen, and A. E. Primenko, 
Appl. Phys. Lett. {\bf 87}, 042502 (2005). 

\bibitem{rudnev_Nb3Sn}
I. A. Rudnev, S. V. Antonenko, D. V. Shantsev, T. H. Johansen, A. E. Primenko, 
Cryogenics {\bf 43}, 663 (2003). 

\bibitem{johansen_MgB2}
T. H. Johansen, M. Baziljevich, D. V. Shantsev, P. E. Goa, Y. M. Galperin, W. N. Kang, H. J. Kim, E. M. Choi, M.-S. Kim and S. I. Lee, 
Europhys. Lett. {\bf 59}, 599 (2002). 

%introduction
%multilayer

\bibitem{gurevich_APL}
A. Gurevich, 
Appl. Phys. Lett. {\bf 88}, 012511 (2006).  

\bibitem{kubo_APL}
T. Kubo, Y. Iwashita, and T. Saeki, 
Appl. Phys. Lett. {\bf 104}, 032603 (2014).  

\bibitem{gurevich_AIP}
A. Gurevich, 
AIP Advance {\bf 5}, 017112 (2015). 

\bibitem{posen_PRAppl}
S. Posen, M. K. Transtrum, G. Catelani, M. U. Liepe, and J. P. Sethna, 
Phys. Rev. Applied {\bf 4}, 044019 (2015). 

\bibitem{kubo_SRF2015}
T. Kubo, 
``Theory of multilayer coating for proof-of-concept experiments", 
SRF2015, Whistler, Canada (2015), TUBA07. 

\bibitem{kubo_PTEP_nano}
T. Kubo, 
Prog. Theor. Exp. Phys. {\bf 2015}, 063G01 (2015).  

\bibitem{ciovati_bake}
G. Ciovati, 
J. Appl. Phys. {\bf 96}, 1591 (2004). 

\bibitem{romanenko_bake}
A. Romanenko, A. Grassellino, F. Barkov, A. Suter, Z. Salman, and T. Prokscha, 
Appl. Phys. Lett. {\bf 104}, 072601 (2014).  

\bibitem{kubo_LINAC14}
T. Kubo, 
in {\it Proceedings of LINAC14, Geneva, Switzerland} (2014), p. 1026, THPP074. 

\bibitem{grassellino_bake}
A. Grassellino and S. Aderhold,  
``New Low T Nitrogen Treatments Cavity Results with Record Gradients and Q", 
TESLA Technology Collaboration (TTC) meeting, Saclay, France (2016). 

\bibitem{checchin}
M. Checchin, A. Grassellino, M. Martinello, S. Posen, A. Romanenko, and J. F. Zasadzinski, 
in {\it Proceedings of IPAC2016, Busan, Korea} (2016), p. 2254, WEPMR002. 

\bibitem{tan_SRF2015}
T. Tan, M. A. Wolak, X. Xi, L. Civale, and T. Tajima, 
in {\it Proceedings of SRF2015, Whistler, Canada} (2015), p. 512, TUBA06. 

\bibitem{laxdal_TTC2016}
R. Laxdal, 
``New insights for reaching higher gradients from muSR samples studies", 
TESLA Technology Collaboration (TTC) meeting, Saclay, France (2016). 


 
%superheating field (London)


\bibitem{kubo_SRF2013}
T. Kubo, Y. Iwashita, and T. Saeki, 
in {\it Proceedings of SRF2013, Paris, France} (2013), p. 427, TUP007. 

\bibitem{kubo_IPAC14}
T. Kubo, Y. Iwashita, and T. Saeki, 
in {\it Proceedings of IPAC2014, Doresden, Germany} (2014), p. 2522, WEPRI023.   


%superheating field (T~Tc)





%superheating field (T<<Tc)


\bibitem{eilenberger}
G. Eilenberger, 
Z. Phys. {\bf 214}, 195 (1968). 


\bibitem{matsubara}
T. Matsubara, 
Prog. Theor. Phys. {\bf 14}, 351 (1955).  




%multilayer London

\bibitem{stejic}
G. Stejic, A. Gurevich, E. Kadyrov, D. Christen, R. Joynt, and D. C. Larbalestier, 
Phys. Rev. B {\bf 49}, 1274 (1994). 

\bibitem{kubo_IPAC13}
T. Kubo, Y. Iwashita, and T. Saeki, 
in {\it Proceedings of IPAC13, Shanghai, China}, p. 2343, WEPWO014. 

%multilayer defect

\bibitem{xu}
C. Xu, H. Tian, C. E. Reece, and M. J. Kelley, 
Phys. Rev. ST Accel. Beams {\bf 14}, 123501 (2011). 

\bibitem{roach}
W. M. Roach, J. R. Skuza, D. B. Beringer, Z. Li, C. Clavero, and R. A. Lukaszew, 
Supercond. Sci. Technol. {\bf 25}, 125016 (2012). 

\bibitem{avnir}
D. Avnir, D. Farin, and P. Pfeifer, 
Nature {\bf 308}, 261 (1984).  

\bibitem{takayasu}
H. Takayasu, 
 {\it Fractals in the Physical Sciences} (Manchester University Press, New York, 1990). 


%multilayer surface resistance

\bibitem{iwashita_normal}
Y. Tajima, Y. Iwashita, H. Fujisawa, M. Ichikawa, and H. Tongu, 
Jpn. J. Appl. Phys. {\bf 46}, 4765 (2008). 

\bibitem{iwashita_normal_LINAC2010}
Y. Iwashita, 
in {\it Proceedings of LINAC2010, Tsukuba, Japan} (2014), p. 310, MOP106. 

\bibitem{posen_Nb3Sn}
S. Posen, M. Liepe, and D. L. Hall, 
Appl. Phys. Lett. {\bf 106}, 082601 (2015). 






\end{thebibliography}
\end{document}